\documentclass[10pt,preprint]{aastex}
\usepackage{color}

\newcommand{\av}{$A_V$}

\newcommand{\eg}{{\it e.g.}}
\newcommand{\etal}{et~al.}

\newcommand{\ic}{$I_{\rm C}$}

\newcommand{\jhk}{$JHK_{\rm s}$}

\newcommand{\ks}{$K_{\rm s}$}

\newcommand{\msun}{M$_{\sun}$}

\newcommand{\mum}{$\mu$m}

\begin{document}

\title{The North American and Pelican Nebulae II. MIPS Observations
and Analysis}

\slugcomment{Version from \today}

\author{L.\ M.\ Rebull\altaffilmark{1}, 
S.\ Guieu\altaffilmark{1},
J.\ R.\ Stauffer\altaffilmark{1},
L.\ A.\ Hillenbrand\altaffilmark{2},
A.\  Noriega-Crespo\altaffilmark{1},
K.\ R.\ Stapelfeldt\altaffilmark{3}, 
S.\ J.\ Carey\altaffilmark{1},
J.\ M.\ Carpenter\altaffilmark{2},
D.\ M.\ Cole\altaffilmark{3}, 
D.\ L.\ Padgett\altaffilmark{1},
S.\ E.\ Strom\altaffilmark{4}
S.\ C.\ Wolff\altaffilmark{4}
}

\altaffiltext{1}{Spitzer Science Center/Caltech, M/S 220-6, 1200
E.\ California Blvd., Pasadena, CA  91125
(luisa.rebull@jpl.nasa.gov)}
\altaffiltext{2}{Department of Astronomy, California Institute of
Technology, Pasadena, CA 91125}
\altaffiltext{3}{Jet Propulsion Laboratory, MS 183-900, California Institute
of Technology, Pasadena, CA 91109}
\altaffiltext{4}{NOAO, Tucson, AZ}

\begin{abstract}

We present observations of $\sim$7 square degrees of the North
American and Pelican Nebulae region at 24, 70, and 160 \mum\ with the
{\it Spitzer Space Telescope} Multiband Imaging Photometer for Spitzer
(MIPS).  We incorporate the MIPS observations with earlier {\em
Spitzer} Infrared Array Camera (IRAC) observations, as well as
archival near-infrared (IR) and optical data. We use the MIPS data to
identify 1286 young stellar object (YSO) candidates.  IRAC data alone
can identify 806 more YSO candidates, for a total of 2076 YSO
candidates.  Prior to the {\em Spitzer} observations, there were only
$\sim$200 YSOs known in this region. Three subregions within the
complex are highlighted as clusters: the Gulf of Mexico, the Pelican,
and the Pelican's Hat. The Gulf of Mexico cluster is subject to the
highest extinction (\av\ at least $\sim$30) and has the widest range
of infrared colors of the three clusters, including the largest
excesses and by far the most point-source detections at 70 \mum. Just
3\% of the cluster members were previously identified; we have
redefined this cluster as about 10-100 times larger (in projected
area) than was previously realized.

\end{abstract}


\section{Introduction}
\label{sec:intro}

Much of our current knowledge regarding star-forming patterns and
circumstellar disk evolution derives from study of molecular cloud
complexes within a few hundred parsecs of the Sun.  Among this group
are a large number of lower-mass clouds such as Taurus (140 pc; Torres
\etal\ 2007), populated with $\sim$300 low-mass ($<$2 M$_{\odot}$)
young stars (see, e.g., Kenyon \etal\ 2008). Also among them but more
infrequent are dense clusters like the Orion Nebula Cluster (ONC; 400
pc; see, e.g., Menten et al.\ 2007 or Muench \etal\ 2008), the
prototypical high mass and high density star forming region.  The
nearby cloud complexes (e.g., Rebull \etal\ 2010, Megeath \etal\ 2005)
have served as our primary empirical guide to understanding the
formation and early evolution of stars. However, they provide
snapshots of only certain kinds of star formation. Because environment
appears to play a significant role in defining stellar properties such
as stellar rotation rates (e.g., Clarke \& Bouvier 2000, Briggs et
al.\ 2007), disk evolution timescales (e.g., Robberto et al.\ 2004),
and multiplicity (e.g., Kraus \& Hillenbrand 2007), for a
comprehensive understanding of star formation, it is important that we
study more than just the nearest examples of star formation. 

The North American (NGC 7000) and Pelican (IC 5070) Nebulae (the
entire complex is refered to hereafter as ``NAN") is a complicated
mixed-mode star formation region. NGC 7000 and IC 5070 appear to lie
on either side of Lynds Dark Nebula (LDN) 935, but all three (NGC
7000, IC 5070, L935) are generally regarded as part of the same large
\ion{H}{2} complex (see, e.g., Wendker \etal\ 1983).  We take the
distance to the NAN to be 520 pc based on Laugalys \etal\ (2006) and
Strai\v{z}ys \etal\ (1989), though Laugalys \& Strai\v{z}ys (2002)
claim a distance of 600 pc; see distance discussion in Guieu \etal\
(2009).  Our results here are not strongly dependent on the distance
assumed to the complex.

The NAN has a mass in molecular gas of order $10^5$ M$_{\odot}$ (Feldt
\& Wendker 1993). There are at least 50 Herbig AeBe stars scattered
through the complex (e.g., Herbig 1958), at least nine embedded or
partially embedded young clusters (Cambr\'esy et al.\ 2002), and a
variety of other signposts of current star formation, such as HH
objects, jets, FU Ori stars, and strong H$\alpha$ emission line stars
(e.g., Bally \& Reipurth 2003, Ogura et al.\ 2002, Comer\'on \&
Pasquali 2005, Marcy 1980; see Reipurth \& Schneider 2008 for a
review). The few hundred largely unobscured young stellar objects
(YSOs) that have been previously identified as members of the complex
likely range in age from $<$1 Myr to several Myr.  If one extrapolates
using a typical initial mass function (IMF) from the
already-identified high mass NAN population, the predicted mass in
stars (to 0.2 M$_{\odot}$) should be at least of order $10^4$
M$_{\odot}$ and, given the age of the region, there should be
thousands of low mass classical T Tauri (CTT) and weak-lined T Tauri
(WTT) members of the complex.  In addition to the unobscured
population, an embedded population was suspected (e.g., Herbig 1958,
Osterbrock 1957). Using more than a million 2MASS sources in a
$2.5^{\circ}\times2.5^{\circ}$ region, Cambr\'esy et al.\ (2002)
mapped extinction and reddening toward the NAN complex, finding visual
extinction up to $\sim$30 magnitudes, as well as embedded clusters. 
Thus, star formation is ongoing.

The line of sight to the NAN ($l,b$=84.8$\arcdeg$,$-$0.53$\arcdeg$) is
heavily populated with foreground and background stars; the
contamination is estimated at 1900$\pm$500 stars deg$^{-2}$ to the
depth of the 2MASS survey (Cambr\'esy et al.\ 2002). One way of
identifying members of the NAN as distinct from the Galaxy is by
looking for those that have the prominent infrared (IR) signatures
indicative of circumstellar disks.  This region of sky was in the few
percent missed by the all-sky survey conducted in 1983 by the Infrared
Astronomical Satellite (IRAS), but it has been observed by both the
Midcourse Space Experiment (MSX) and AKARI. Given the high source
density and high contamination rate, both relatively high spatial
resolution and relatively deep observations are crucial for extracting
candidate cluster members. We have obtained deep {\em Spitzer Space
Telescope} (Werner \etal\ 2004a) Infrared Array Camera (IRAC; Fazio
\etal\ 2004) and Multiband Imaging Photometer for Spitzer (MIPS; Rieke
\etal\ 2004) maps of a $\sim$7 square degree region covering most of
the complex. This paper is the second in our series on the NAN
complex; Guieu \etal\ (2009; hereafter G09) have presented the IRAC
observations, and this paper presents the MIPS observations and
analysis. 

MIPS observations at 24, 70, and 160 \mum\ can elucidate many aspects
of the ongoing star formation in the NAN.  Although the emission from
stellar photospheres is falling rapidly at 24 \mum, emission from
circumstellar material makes many of the young cluster members  quite
bright at 24 \mum, so MIPS easily distinguishes the young stars from
most of the background Galactic population.   Diffuse emission from
the cloud itself gives us the location of much of the dust and
molecular material. This paper presents the NAN MIPS observations,
calling out properties unique to the MIPS bands, and using data from
other wavelengths as required in order to obtain a more complete list
of YSO members. MIPS reveals complex extended emission throughout the
NAN region at all three of its wavelengths, in addition to a rich
stellar population. 

In order to obtain a complete YSO membership list, including
confirmation of Spitzer-selected objects and identification of those
members that have lost their disks, one needs additional observations,
such as optical photometry.  As it pertains to the present discussion,
IRAC and optical data from G09 are included below.  A forthcoming
paper will include optical and/or near-IR spectroscopy of many of
these candidates to confirm or refute their YSO status.  Several
investigators (e.g., the Palomar Transient Factory [PTF], see, e.g.,
Covey \etal\ 2011 or Miller \etal\ 2011) are monitoring this region
for variability, which can also be used to confirm (or refute) a given
object as a YSO.

We first present the observational details for the MIPS observations
of the NAN complex (\S\ref{sec:obs}), with special attention to the
extended emission (\S\ref{sec:extemiss}) and  inclusion of ancillary
data (\S\ref{sec:ancildata}). Then we discuss the overall properties
of the point source catalog, including issues of contamination 
(\S\ref{sec:sourcecounts}).  We select young stellar object (YSO)
candidates in Section~\ref{sec:pickysos}, primarily using MIPS, but
including ancillary data where possible. We describe the ensemble of
YSO candidates in \S\ref{sec:ensemble} and define three new clusters
in \S\ref{sec:clusters}.  We have a few brief words on the source(s)
potentially powering the nebula in \S\ref{sec:ionizing}. Finally, we
summarize our main points in \S\ref{sec:concl}.  We compare our
results to those obtained from other mid-IR large survey missions (for
point sources and extended emission) in an Appendix. 


\section{Observations, Data Reduction, and Ancillary Data}
\label{sec:obs}

\begin{figure*}[tbp]
\epsscale{0.75}
\plotone{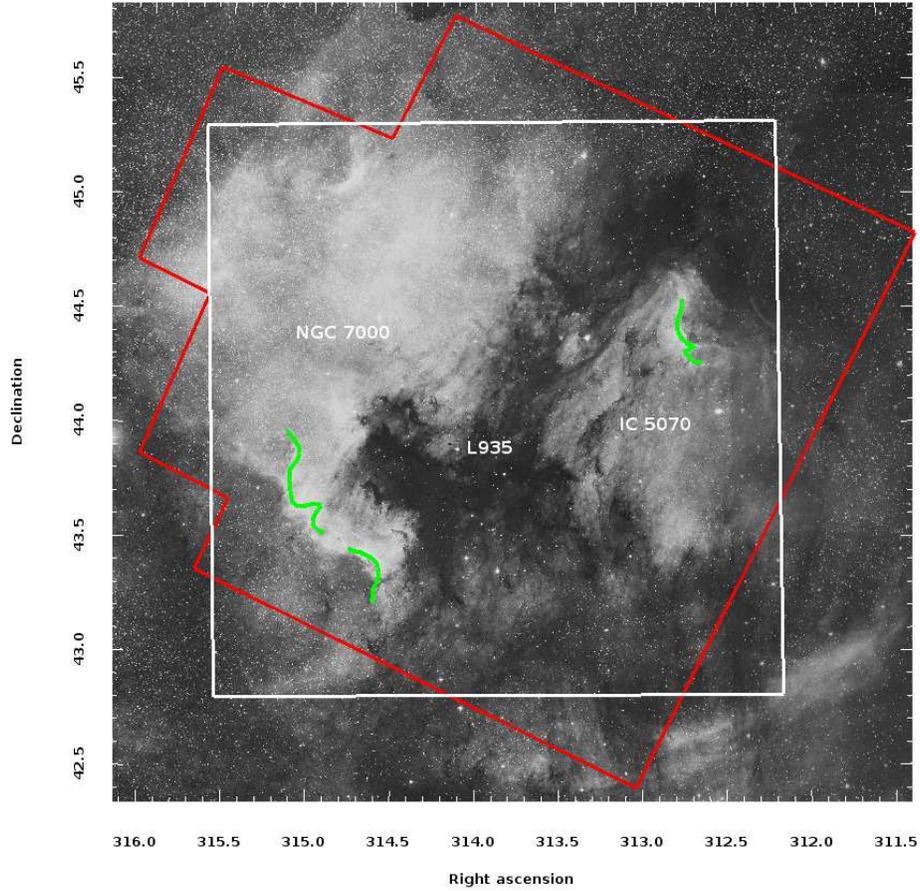}
\caption{Approximate location of MIPS (red, irregular polygon) and
IRAC (white rectangle) coverage, superimposed on a Palomar Observatory
Sky Survey (POSS) red image.  Three regions are indicated: NGC 7000,
IC 5070, and L935, as well as some of the brightest edges (upper right
and lower left, irregular contours) seen in the infrared images below
(Figures~\ref{fig:24mosaic}--\ref{fig:3color}) to guide the eye. The
coordinates are J2000 decimal degrees. }
\label{fig:where}
\end{figure*}

\begin{deluxetable}{lllll}
\tablecaption{Summary of MIPS observations (from programs 20015 and 462)
\label{tab:observations}}
\tablewidth{0pt}
\tablehead{
\colhead{field} & \colhead{map center} & \colhead{map
center} & \colhead{map offset} & \colhead{AORKEY}\\
 & \colhead{(RA, Dec in sexagesimal) } & \colhead{ (RA, Dec in degrees)} &
\colhead{(cross-scan, arcsec)} & }

\startdata
map 1  & 20:54:18,+44:05:00 & 313.575, 44.0833 & $-$3300 &   16793600\\ 
map 2  & 20:54:18,+44:05:00 & 313.575, 44.0833 & $-$2200 &   16793856\\ 
map 3  & 20:54:18,+44:05:00 & 313.575, 44.0833 & $-$1100 &   16794112\\ 
map 4  & 20:54:18,+44:05:00 & 313.575, 44.0833 & 0	 &   16794368\\ 
map 5  & 20:54:18,+44:05:00 & 313.575, 44.0833 & $+$1100 &   16794624\\ 
map 6  & 20:54:18,+44:05:00 & 313.575, 44.0833 & $+$2200 &   16794880\\ 
map 7  & 20:54:18,+44:05:00 & 313.575, 44.0833 & $+$3300 &   16795136\\ 
patch 1& 21:00:52,+44:32:33 & 313.217, 44.5425 & \ldots  & 24251136\\
patch 2& 21:01:56,+45:00:06 & 315.483, 45.0017 & \ldots  & 24253440\\
\enddata
\end{deluxetable}

\subsection{Observations}

The MIPS observations of the NAN complex were conducted in two epochs.
The first (and largest) component took place on 10-11 June 2006 and
covered $\sim$6 square degrees in the region of highest extinction
(see additional discussion in G09).   These observations were part of
Spitzer program id 20015 (P.\ I.: L.\ Rebull); the seven
AORKEYs\footnote{AORKEYs are the unique 8-digit identifier for the
AOR, which can be used to retrieve these data from the Spitzer
Archive.} appear in Table~\ref{tab:observations}.  Additional
observations were obtained on 24 June 2008 as part of program 462 (P.\
I.: L.\ Rebull) in an effort to increase the legacy value of this
dataset.  They were designed to cover the corner of the map to the
same integration time as the main map, and added about a square degree
to the total coverage. These two additional AORKEYs also appear in
Table~\ref{tab:observations}, and are referred to as the ``patch
data.''

The center of the final $\sim$7 square degree map is (J2000) Right
Ascension (RA; $\alpha$), Declination (Dec; $\delta$) =
20:54:18,+44:05:00 (313.575$\arcdeg$, 44.0833$\arcdeg$), or galactic
coordinates $l,b$=84.8$\arcdeg$, $-$0.53$\arcdeg$, or ecliptic
coordinates (J2000) 337.9$\arcdeg$, +57.7$\arcdeg$. Since this region
is at a high ecliptic latitude, the field of view rotates by about a
degree per day. In order to make AOR coverage as independent as possible
without constraints, map offsets (see Table~\ref{tab:observations})
were used along with a group-within constraint to ensure even coverage
and minimal overlap between tiles.   Also as a result of the high
ecliptic latitude, asteroids are not a significant concern; we have
just one epoch of observations per field. Figure~\ref{fig:where} shows
the region of 3-band coverage with MIPS, and the 4-band coverage with
IRAC (discussed in G09).  There are about 6 square degrees with both
MIPS and IRAC coverage. 

Medium-rate MIPS scan maps were obtained with the spacing between
adjacent scan legs set at 160$\arcsec$, 50\% of the 24 \mum\ detector
width.  The total integration times were 84 s, 42 s, and $\sim$4 s at
each point in the map, respectively, for 24, 70, and 160 \mum.  These
observations therefore can detect photospheres at 24 \mum\ down to
about 3 \msun\ at the distance of the NAN complex.  Much lower-mass
protostars with infrared excesses due to a circumstellar disk can also
be detected; typically, we can detect down to $\sim$0.3 \msun\ (mid-M)
stars with IR excesses. IRAS data, though incomplete in this region
(see Appendix~\ref{sec:irasmsx}), shows 100 \mum\ emission peaks up to
1800 MJy/sr, so we did not expect to retrieve valid (unsaturated) 160
\mum\ data, and thus did not plan on complete coverage in that band. 
However, these data (both the original and the secondary `patch' data)
were taken in cold MIPS campaigns and, although undersampled and
saturated in places, the smoothed 160 \mum\ data are still valid
representations of the emission; see below.  

The IRAC observations and data reduction are discussed in G09.  We
note that since the publication of that paper, we identified a minor
error in our analysis where some objects were erroneously identified
as saturated at 8 \mum, and the total size of the map was
over-estimated.  In the discussion below, when the G09 selection
methodology is included, it is meant to be the G09 selection
methodology as imposed on the corrected catalog.

We note for completeness that the four channels of IRAC are 3.6, 4.5,
5.8, and 8 microns, and that the three channels of MIPS are 24, 70,
and 160 microns. These bands can be referred to equivalently by their
channel number or wavelength; the bracket notation, e.g., [24],
denotes the measurement in magnitudes rather than flux density units.
Further discussion of the bandpasses can be found in, e.g., the
Instrument Handbooks, available from the Spitzer Science Center (SSC)
website.

\subsection{Data Reduction}

\begin{figure*}[tbp]
\epsscale{1.0}
\plotone{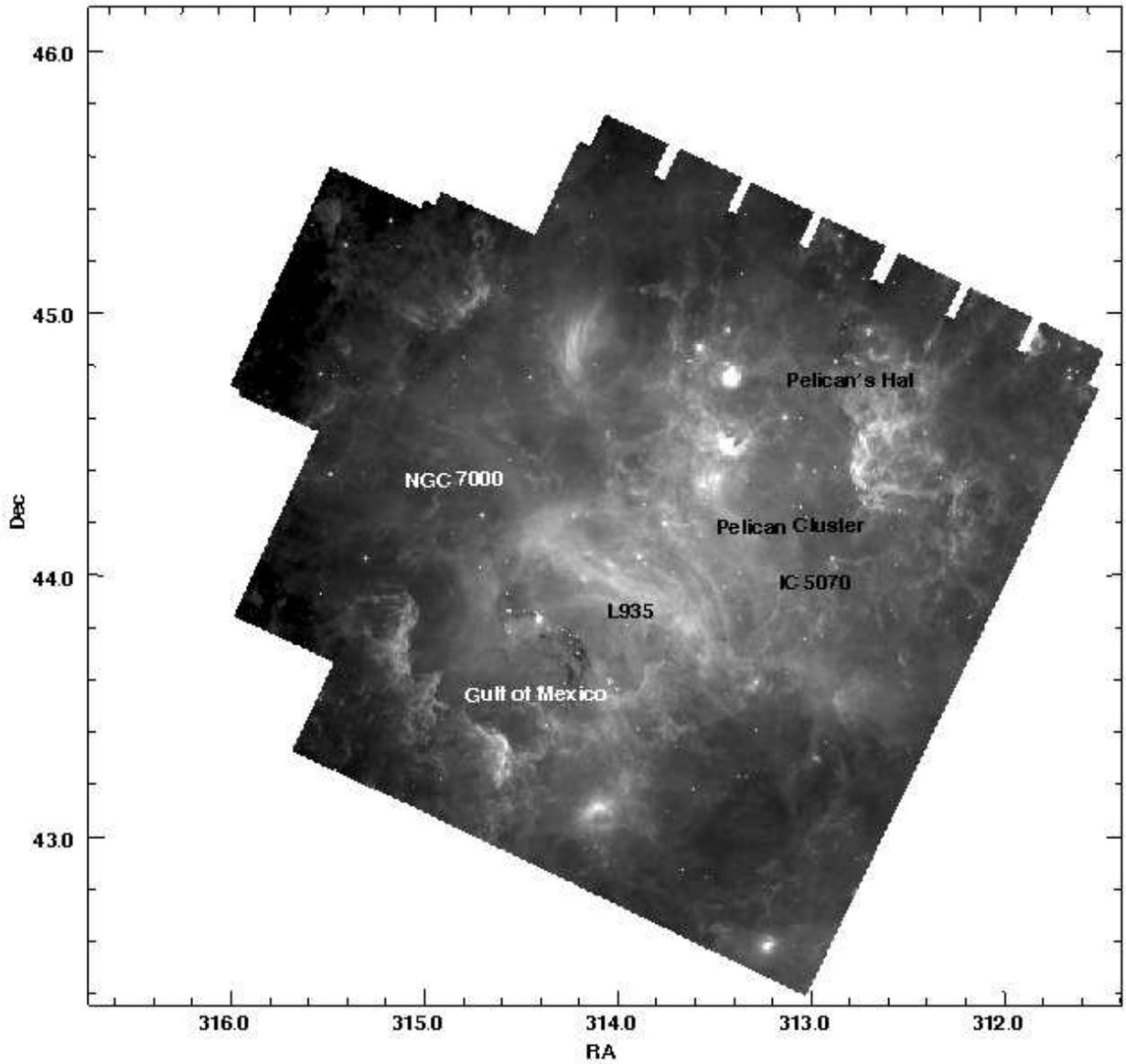}
\caption{Mosaic of the NAN at 24 \mum.  The greyscale colors
correspond to a histogram-equalized stretch of surface brightnesses.
The labels identifying IC 5070, L935, and NGC 7000 are in the same
locations as Fig.~\ref{fig:where}.  Three of the clusters which
will be discussed below are indicated as well: the Gulf of Mexico, the
Pelican's Hat, and the Pelican Cluster.}
\label{fig:24mosaic}
\end{figure*}

\begin{figure*}[tbp]
\epsscale{1.0}
\plotone{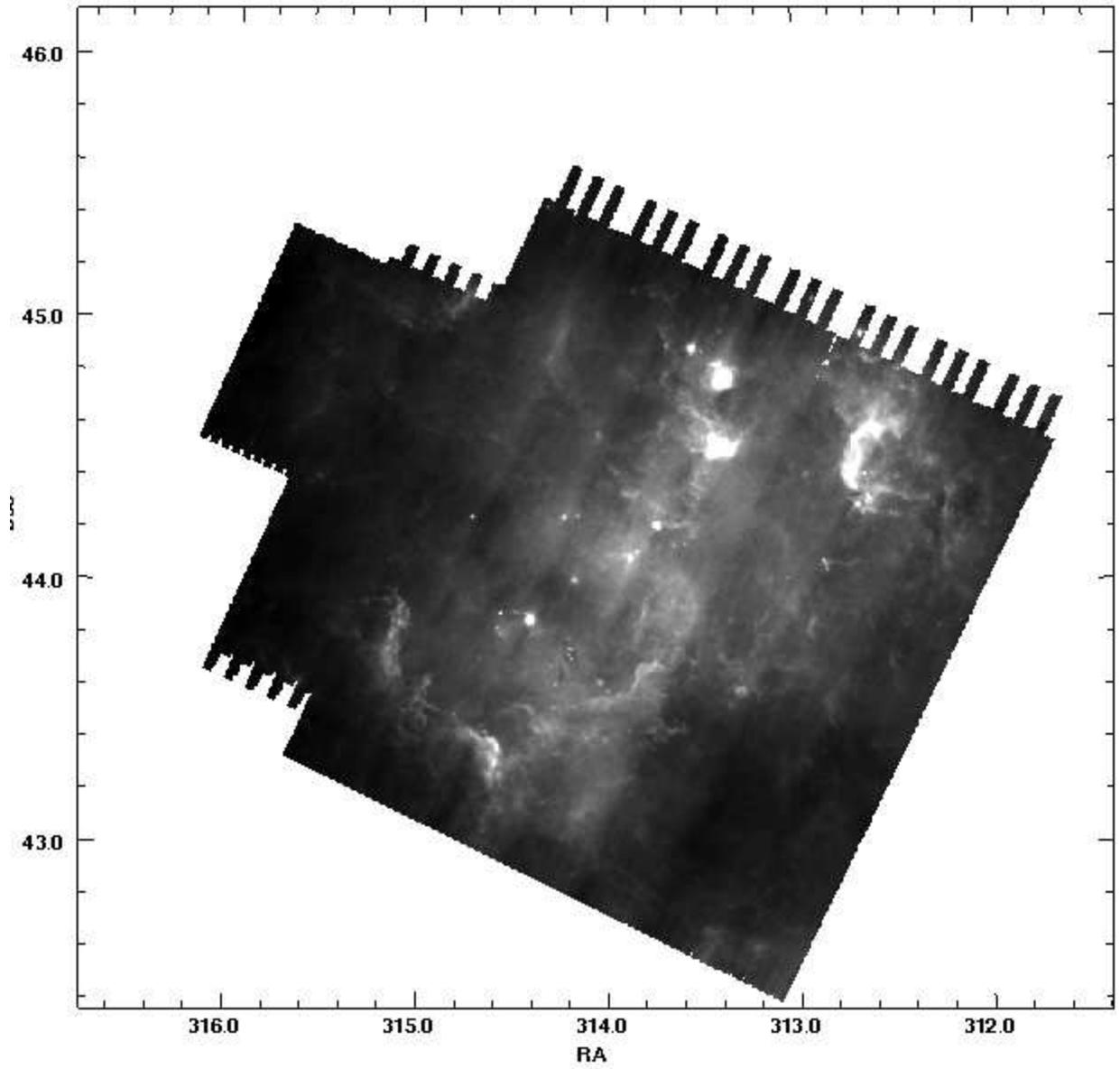}
\caption{Mosaic of the NAN at 70 \mum.  The greyscale colors correspond to
a histogram-equalized stretch of surface brightnesses. }
\label{fig:70mosaic}
\end{figure*}

\begin{figure*}[tbp]
\epsscale{1.0}
\plotone{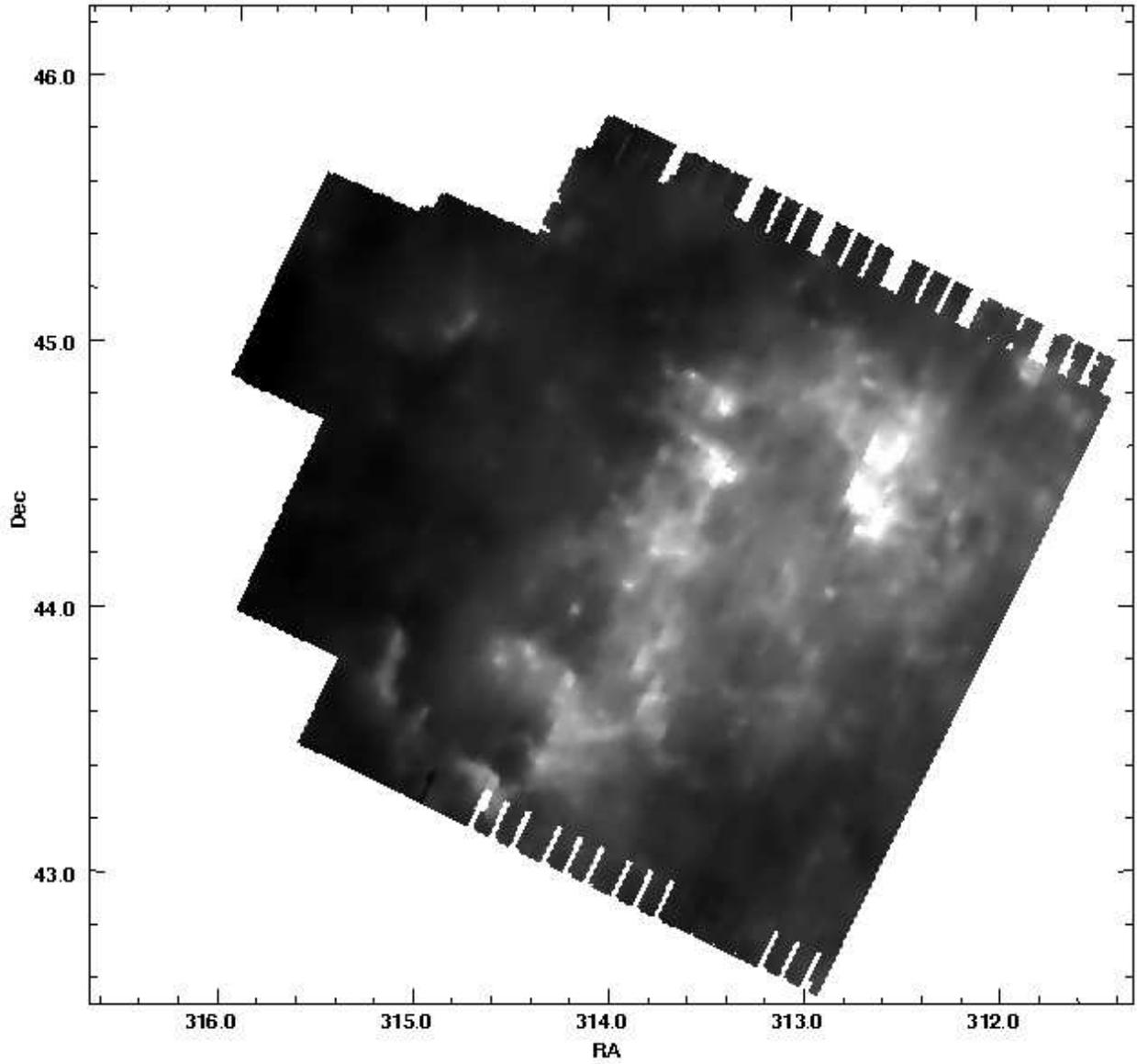}
\caption{Mosaic of the NAN at 160 \mum.  The greyscale colors correspond
to a histogram-equalized stretch of surface brightnesses. Note that
this image is highly smoothed; see text. }
\label{fig:160mosaic}
\end{figure*}

\begin{figure*}[tbp]
\epsscale{1.0}
\plotone{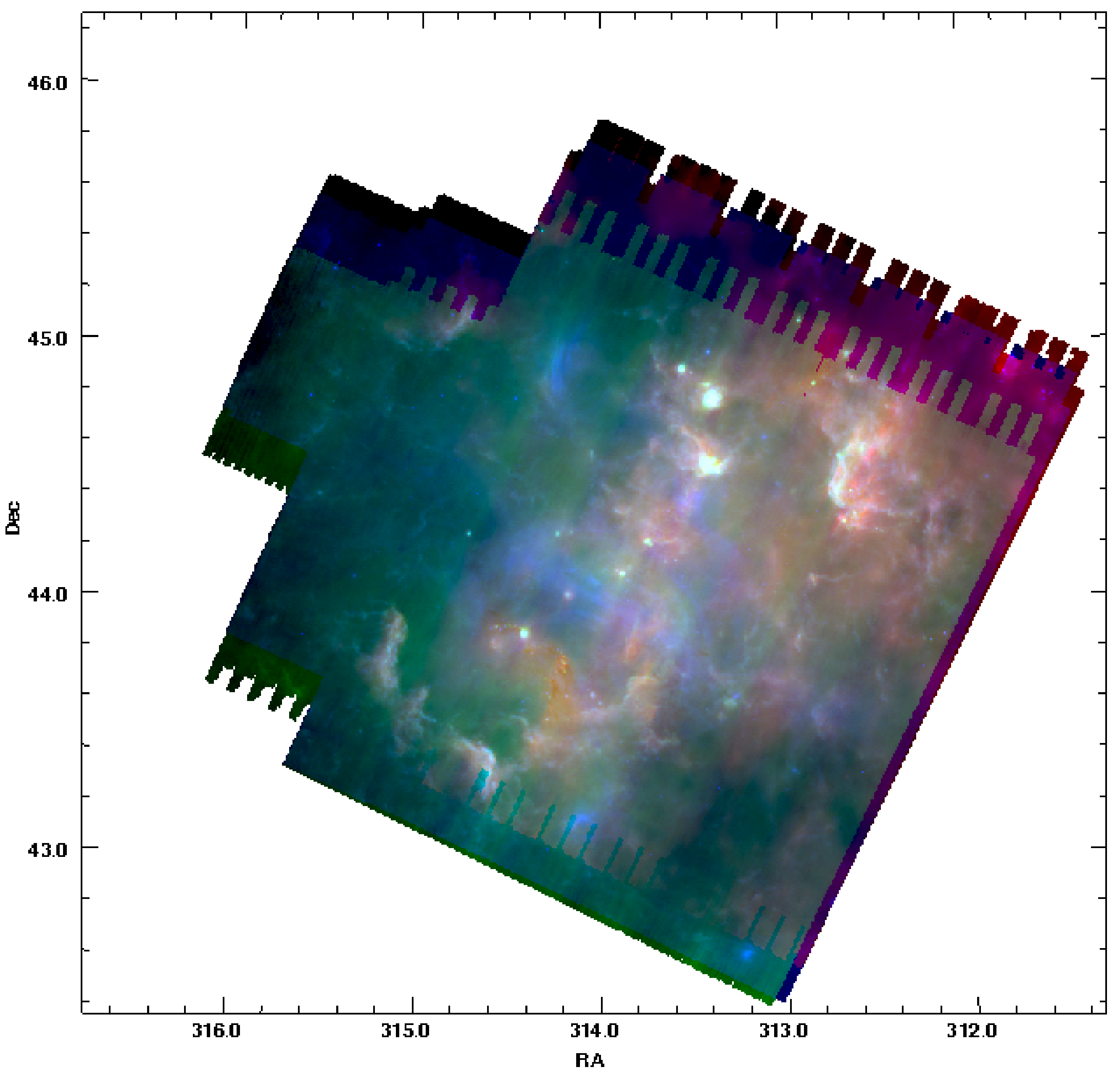}
\caption{Three-color mosaic of the NAN including all three MIPS bands. }
\label{fig:3color}
\end{figure*}

Figures~\ref{fig:24mosaic}, \ref{fig:70mosaic}, and
\ref{fig:160mosaic} show the individual mosaics by channel, and
Figure~\ref{fig:3color} shows a 3-color image with all three MIPS
channels included.  To guide the eye, the labels for IC 5070, L935,
and NGC 7000 appear in the same location in Figures~\ref{fig:where}
and \ref{fig:24mosaic} (optical through 160 \mum); the bright emission
contours on the upper left and lower right seen in all three MIPS
bands are indicated in Figure~\ref{fig:where}.  There is substantial
extended emission in all three MIPS channels throughout the MIPS maps;
see \S\ref{sec:extemiss}.  

\subsubsection{24 micron mosaics and point-source photometry}

We started with the SSC-pipeline-produced basic calibrated data
(BCDs); for the main portion of the map, the data were processed with
pipeline version S16.1, and the additional BCDs from the `patch'
AORs were processed with pipeline version S18.0.2.  Functionally,
there is no significant difference between these pipeline versions.
For a description of the MIPS pipeline, see Gordon \etal\ (2005).  We
constructed a mosaic from the BCDs and then performed point-source
extraction on the mosaics using the SSC mosaicking and point-source
extraction (MOPEX) software (Makovoz \& Marleau 2005), specifically
the APEX 1-Frame routines.    

Constructing the mosaics at 24 \mum\ was straightforward.  We used
MOPEX's point-response function (PRF) fitting to  obtain photometry
for all of the point sources in the map in a two-step process in order
to optimize parameters in regions of high nebulosity and/or source
density, e.g., the Gulf of Mexico.  Since the structure in the
nebulosity is very complex, often of width comparable to the PRF, it
can fool the point source detection algorithm into finding false
``chains'' of sources along the nebulosity.  As for the MIPSGAL
project (Carey \etal\ 2009), we used a very small median filter box
over which to extract the sources (APEX parameter
``extract\_medfilter'' set to 11 pixels, where the pixels are
2.5$\arcsec$, very close to the native pixel scale) to limit the
influence of the nebulosity.  While this does indeed limit the noise
from the nebulosity, it requires a 3\% correction to the derived flux
densities to match flux densities obtained with a PRF fitting using a
larger median filter box.  We visually inspected each of the sources,
dropping those along nebulous filaments that were clearly not true
point sources.  Imposing a signal-to-noise ratio cut of $\sim$10
removes most such sources while leaving the true sources, but
additional hand-inspection is required to remove false sources and add
real sources with a calculated signal-to-noise comparable to but below
10.  In very crowded regions such as the Gulf of Mexico, a larger
median filter box (and therefore no flux density correction) is
required.  We reprocessed the mosaic using these different parameters,
visually inspected the detected sources, and merged the source list
from this reduction with that from above.  

We estimate internal (statistical) 24 \mum\ flux density errors from
the MOPEX-reported signal-to-noise ratio (SNR) divided into the
measured flux density.   The systematic uncertainty on the absolute
flux densities at 24 \mum\ is estimated to be 4\% (Engelbracht \etal\
2007); we added this error in quadrature.

There were $\sim$4270 total real point sources automatically detected
at 24 \mum, ranging from 1.4 mJy to 4 Jy (beyond the generally quoted
saturation limit; see below).  In addition to those sources
automatically detected by MOPEX using the multi-step process described
above, we manually added several sources to the list.  For the
brightest end (brighter than $\sim$ 1 Jy), MOPEX has a tendency to
detect multiple sources when in reality there is only one very bright
source. For those, we did aperture photometry and allowed  recentering
on the true peak of the object, taking those flux densities over the
PRF-fitted flux densities.  At the faint end, in the regions of the
Gulf of Mexico and the Pelican cluster (see below for cluster
definitions), there were several sources that were obvious by eye, but
were not automatically extracted by MOPEX due to local brightness
variations or source density. Flux density measurements for those
objects were also manually obtained and added to the source list. 


There are 4334 sources in the complete 24 \mum\ point source catalog. 
The 24 \mum\ source surface density is about 600 sources per square
degree.  The zero point\footnote{Zero points are used in the following
standard formula to convert between flux densities and magnitudes: $m
= 2.5 \times \log \frac{F_{\rm zero point}}{F}$ where $m$ is magnitude
and $F$ is flux density.} used to convert these flux densities to
magnitudes was 7.14 Jy, based on the extrapolation from the Vega
spectrum as published in the MIPS Data Handbook.  We note for
completeness that there are also several hundred detections that are
apparently resolved and/or confused with bright localized nebulosity
(knots), and thus do not appear in our point source catalog.  A
detailed analysis of these objects and an assessment of whether or not
they are true point sources embedded in nebulous material (and, if so,
whether they are associated with the NAN complex) is beyond the scope
of this paper.

The faint limit of the catalog of 24 \mum\ sources is a function of
the nebular brightness, which varies across the field. Similarly, the
saturation limit for point sources with MIPS-24 is also a function of
location in the cloud because the total flux density registered by the
detector is that due to the point source itself plus any surrounding
extended nebular emission. For objects that are very bright, the MIPS
pipeline attempts to reconstruct the brightest portions of the mosaic
from the initial 1.0 s of exposure time, and therefore very bright
sources can have reasonably accurate flux densities (although less
accurate than fainter objects).  The extended emission at 24 \mum\
varies from $\sim$74 MJy sr$^{-1}$ in the brightest part of the
Pelican Nebula to $\sim$52 MJy sr$^{-1}$ in the ``streaks'' in the
center of the image (see Figure~\ref{fig:24mosaic}) to $\lesssim$26
MJy sr$^{-1}$ in the darker parts of the cloud.   An additional issue
when considering completeness is the resolution; the resolution of
MIPS-24 ($\sim7\arcsec$, 2.5$\arcsec$ pixel size) is poorer than IRAC
or 2MASS ($\sim2\arcsec$). Source multiplicity and confusion will also
affect the completness of the catalog, particularly in dense
regions.

\subsubsection{70 micron mosaics and point-source photometry}

We started with the SSC-pipeline-produced basic calibrated data (BCDs)
processed with pipeline version S14.4 or later; as for 24 \mum,
functionally, there is no significant difference among these
pipeline versions.  

For the germanium channels (both 70 and 160 \mum), the data must be
treated carefully due to the time-dependent response of the arrays,
the stimulator latencies, and bright source stimulator artifacts. The
steps to improve the Ge:Ga data are described in the MIPS Data
Handbook, available at the SSC website. 

Some additional steps have been applied to the 70 \mum\ data that have
been used and tested in the processing of the MIPSGAL Legacy data
(Paladini \etal\ 2011 in preparation). These steps include (i) a
`delta flat' between  stimulator flashes to correct time dependent
gain variations, (ii) a stim outlier rejection to correct corrupted
stims during the calibration process, and (iii) de-striping using a
Fourier algorithm using `ridgelets' to remove the stripes left by
`rowdy pixels' during the mapping procedure.  (For more information on
ridgelets or rowdy pixels, as well as the reduction procedure, 
see Paladini \etal\ 2011 in preparation.)
The pixel scale of this mosaic is the native pixel scale of
9.8$\arcsec$ px$^{-1}$.  This cleaned mosaic is shown in
Figure~\ref{fig:70mosaic}.

We also used MOPEX to construct mosaics from the filtered and
unfiltered BCDs generated by the SSC pipeline. We resampled these
pixels to be 4$\arcsec$ square, which can better enable source
extraction.   We used a PRF defined from clean and bright point
sources selected from other identically-produced mosaics of similar
regions, and performed point source detection and extraction using
MOPEX (specifically APEX 1-Frame) on the mosaic.  The
initially-produced source list was cleaned for instrumental artifacts
via manual inspection of the 70 \mum\ image and comparison to the 24
\mum\ image; e.g., if there was some question as to whether a faint
object seen at 70 \mum\ was real or an instrumental artifact, and
comparison to the 24 \mum\ image revealed a 24 \mum\ source, then the
70 \mum\ object was retained as a real source.   This catalog has the
same limitations as was found at 24 \mum; the brightness of the
nebulosity can drown out the faintest objects, and contribute to
saturation of the brightest point-source objects.  The resolution at
70 \mum\ ($\sim20\arcsec$) is coarser than at 24 \mum\
($\sim7\arcsec$), which complicates source matching to shorter bands
and source extraction in confused regions. For all of these reasons,
the 70 \mum\ catalog is neither complete nor unbiased, particularly in
the regions of bright interstellar medium (ISM) or the faintest end.
The estimated uncertainty on the 70 \mum\ point source flux densities
is 20\%.  No color corrections were applied.

There were 97 total point sources detected at 70 \mum, ranging from
$\sim$0.15 Jy to $\sim$90 Jy (beyond saturation); most  (95\%) of the
sources detected are $\sim$9 Jy or less, far below saturation.   The
surface density is about 15 sources per square degree.  All of the 70
\mum\ objects had identifiable counterparts at 24 \mum.  (Four
additional objects were flagged as potential point sources in the 70
\mum\ map but were resolved into highly nebulous bright clumps at 24
\mum, and thus were dropped from the final catalog.) The zero point
used to convert the flux densities to magnitudes was 0.775 Jy, based
on the extrapolation from the Vega spectrum as published in the MIPS
Data Handbook.

\subsubsection{160 micron mosaics}

For the 160 \mum\ data, we used the same MIPSGAL techniques described
above (from Paladini \etal\ 2011 in preparation) to create a clean
mosaic. An initial mosaic was created with 16$\arcsec$ pixels from the
S18.12 data, then it was de-striped and wavelet-de-noised. As
described above, the 160 \mum\ map is actually unfilled; 28\% of the
map is missing (``NaN'' pixels) due to incomplete coverage, the
missing readout in the array (see MIPS Instrument Handbook), or
saturation. In order to ``fill in'' these gaps in the map, we used a
two-dimensional 7$\times$7 (native) pixel boxcar median to interpolate
across missing (``NaN'') pixels. Although this methodology
substantially redistributes the flux density such that point-source
photometry is no longer possible, morphological information is
preserved.  Some regions appear as if they might have point sources,
but it is impossible to tell whether or not they are resolved given
that the image is smoothed. Note that several regions are saturated.

\subsection{Large-Scale Extended Emission and Texture in the Maps}
\label{sec:extemiss}

\subsubsection{Morphology}

\begin{figure*}[tbp]
\epsscale{1.0}
\plotone{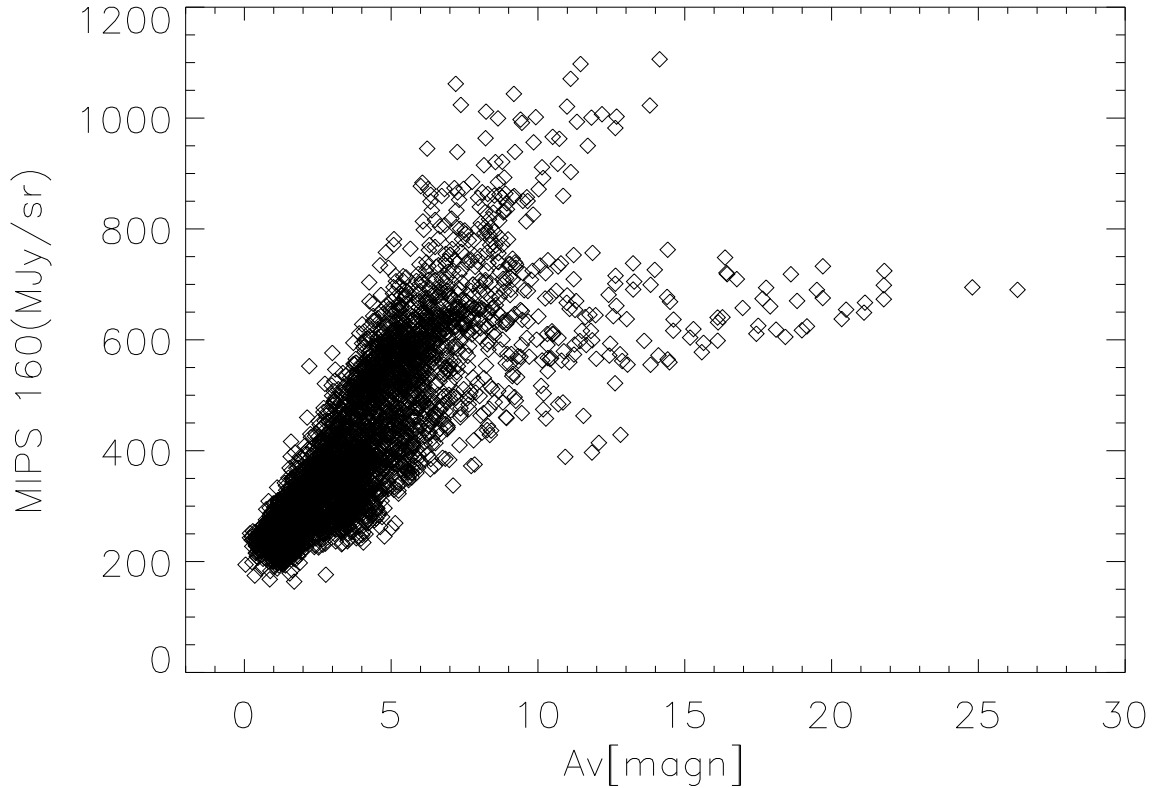}
\caption{Plot of \av\ (from Cambr\'esy \etal\ 2002) against 160 \mum\
surface brightess, in bins of 20 px.  As in other star-forming
regions, the one-to-one correlation holds to \av$\sim$5 mags. The 160
\mum\ emission closely follows the \av\ in space as well.
}
\label{fig:160ext}
\end{figure*}

Bright, complex, extended emission is present at all three wavelengths
throughout the MIPS NAN mosaics (Figs.~\ref{fig:24mosaic}-\ref{fig:3color}).  

At 24 \mum, the diversity of textures found in the ISM is stunning.
There are dark lanes in the Gulf of Mexico and north of the brightest
part of the Pelican (see \S\ref{sec:clusters}); there are broad arcs
near 314.25$\arcdeg$, 44.93$\arcdeg$ (20:57:00 +44:55:48) that
somewhat resemble the ISM in the Pleiades; there are striations near
the center of the map that seem to indicate large arcs centered on the
Gulf of Mexico; there are highly localized nebulous very bright blobs;
and, finally, there is overall turbulent ISM structure that appears to
be illuminated by a source near the center of the map. The brightest
portions so illuminated are to the northwest (forming the neck of the
Pelican), and the southeast (forming the west coast of Mexico, the
``Mexican Riviera''); these edges are indicated in
Figure~\ref{fig:where} for comparison.

Much of the brightest structure seen at 24 \mum\ is also seen at 70
\mum. Several of the highly localized nebulous very bright blobs are
still resolved at 70 \mum.  The brightest portions of the ISM at 70
\mum\ are not only the Pelican's neck and the Mexican Riviera, but
also the region near several ``Carribbean islands."  The 24 \mum\
striations seen in the middle of the map and to the northeast are not
apparent in the 70 \mum\ map.

At 160 \mum, the brightest parts of this map are the Pelican's neck
and two of the highly localized nebulous very bright blobs.  The
Mexican Riviera and the Carribbean islands are still discernible, even
in the highly smoothed map.  As in other star-forming regions, the 160
\mum\ emission closely follows the \av\ exinction contours, as derived
by Cambr\'esy \etal\ (2002).  Plotting \av\ values against 160 \mum\
surface brightness, Figure~\ref{fig:160ext}, yields a result similar
to other star-forming regions.  As in molecular clouds such as Taurus
(Flagey \etal\ 2009), the one-to-one correlation holds to \av$\sim$5
mags, perhaps higher. For large \av, we see the `saturation effect' of
the near-IR measurements, i.e., the near-IR does not accurately
portray the dusty/cold ISM  beyond  \av$\sim$10.


In all three MIPS maps, the darkest regions are to the southwest and
far east, where little ISM and few point sources at 24 \mum\ (none at
the other MIPS bands) are apparent.  To the southwest, since the star
counts at IRAC bands are not dramatically higher in this region, we
suspect that we are not seeing the edge of the cloud, but perhaps a
part of the cloud that has not yet begun star formation in earnest. 
The farthest east, darkest region seen in MIPS does not have IRAC
coverage, though we suspect that the map extends off the edge of the
cloud here.

\subsubsection{Comparison to NGC 7023}

\begin{figure*}[tbp]
\epsscale{1.0}
\plotone{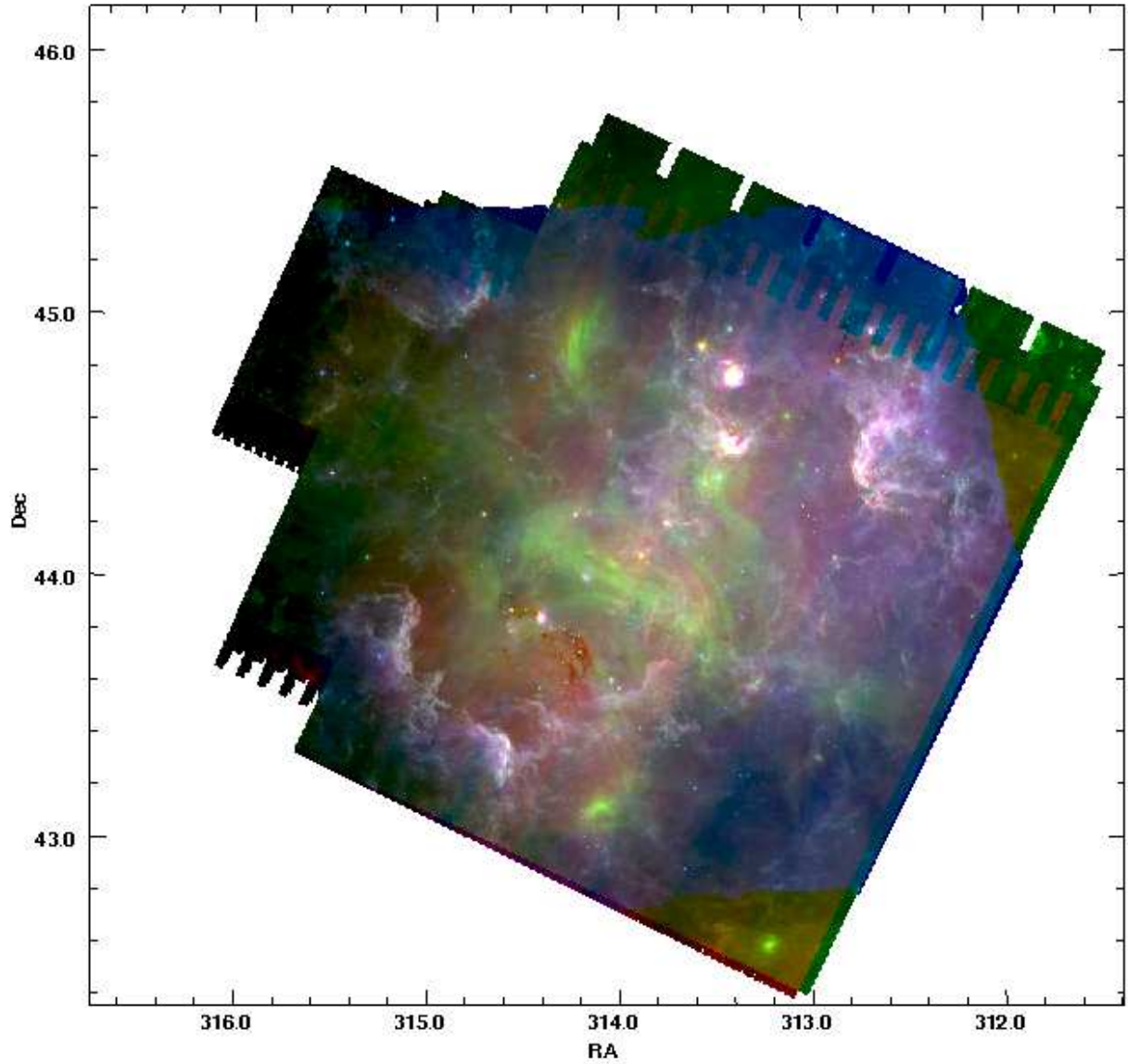}
\caption{Mosaic of IRAC-4 (8 \mum), MIPS-24, and MIPS-70 data.  Labels
are as in earlier similar figures.  Note that the 8 and 24 \mum\
emission are very similar in the large-scale
PDR-like features.}
\label{fig:82470}
\end{figure*}

We can compare the colors of the entire NAN region with those of the
classical reflection nebula/photo-dissociation region (PDR) NGC 7023
(e.g., Werner \etal\ 2004b).  


We assume $\sim$5\% uncertainties in the measurements themselves,
dominated by the absolute calibration.  The overall 8:24 \mum\ ratio
is quite similar for the NAN (0.414) and NGC 7023 (0.444), as is the
24:70 \mum\ ratio (NAN, 0.280; NGC 7023, 0.234).  We interpret this as
that the relative number of polycyclic aromatic hydrocarbons (PAHs) to
``Very Small Grains'' (VSG; $\sim$1.2-15 nm)\footnote{The nomenclature
describing the contribution of the different grain sizes to the
mid/far infrared emission is widely used and originates from the early
models applied to explain the spectral energy distribution and
extinction properties of the galactic ISM (D\'esert,  Boulanger, \&
Puget 1990), using empirical dust models. Models and observations have
grown in sophistication since then, but the overall picture holds,
with a strong contribution from PAH broad emission features between 3
to 12 \mum\ (see e.g., Tielens 2008), Very Small Grains (VSG; sizes
$\sim$1.2 to 15 nm) at mid-IR wavelengths and Big Grains (BG), mostly
silicates (sizes $\sim$ 15 to 100 nm), at the longer wavelengths.} is
similar in the two regions and that the NAN's properties are similar
to that of a PDR (consistent with the image morphology). 

The 70:160 \mum\ ratio, however, is much different -- 0.320 for the
NAN and 0.546 for NGC 7023. Even if the uncertainties are $\sim$20\%,
these values are still sigificantly different.  This suggests that
colder dust (inferred from the stronger 160 \mum\ continuum) arising
from ``Big Grains'' in the NAN brings down the ratio. This seems to be
consistent with the ``voids" of emission in the images of the region,
which could indicate even colder dust, particularly on the southwest
corner of the image. This kind of behavior can be seen in Herschel
images (e.g., Paradis \etal\ 2010).

Figure~\ref{fig:82470} compares the IRAC 8 \mum\ data with the MIPS 24
and 70 \mum\ data.  In this composite, similar structures are seen at
8 and 24 \mum, as in a PDR.  One of the most remarkable things of this
composite image is the ``veil" or wispy structure at 24 \mum\ in the
middle of the image without a counterpart from the dust tracers.
Perhaps this due to fine structure emission lines (e.g. [\ion{Fe}{2}]
at 26 \mum). This idea is reinforced by the small bowshock at
RA$\sim$314.25$\arcdeg$ and Dec$\sim$43.1$\arcdeg$ (20:57,+43:06).


\subsubsection{FCRAO Data}

\begin{figure*}[tbp]
\epsscale{1.0}
\plotone{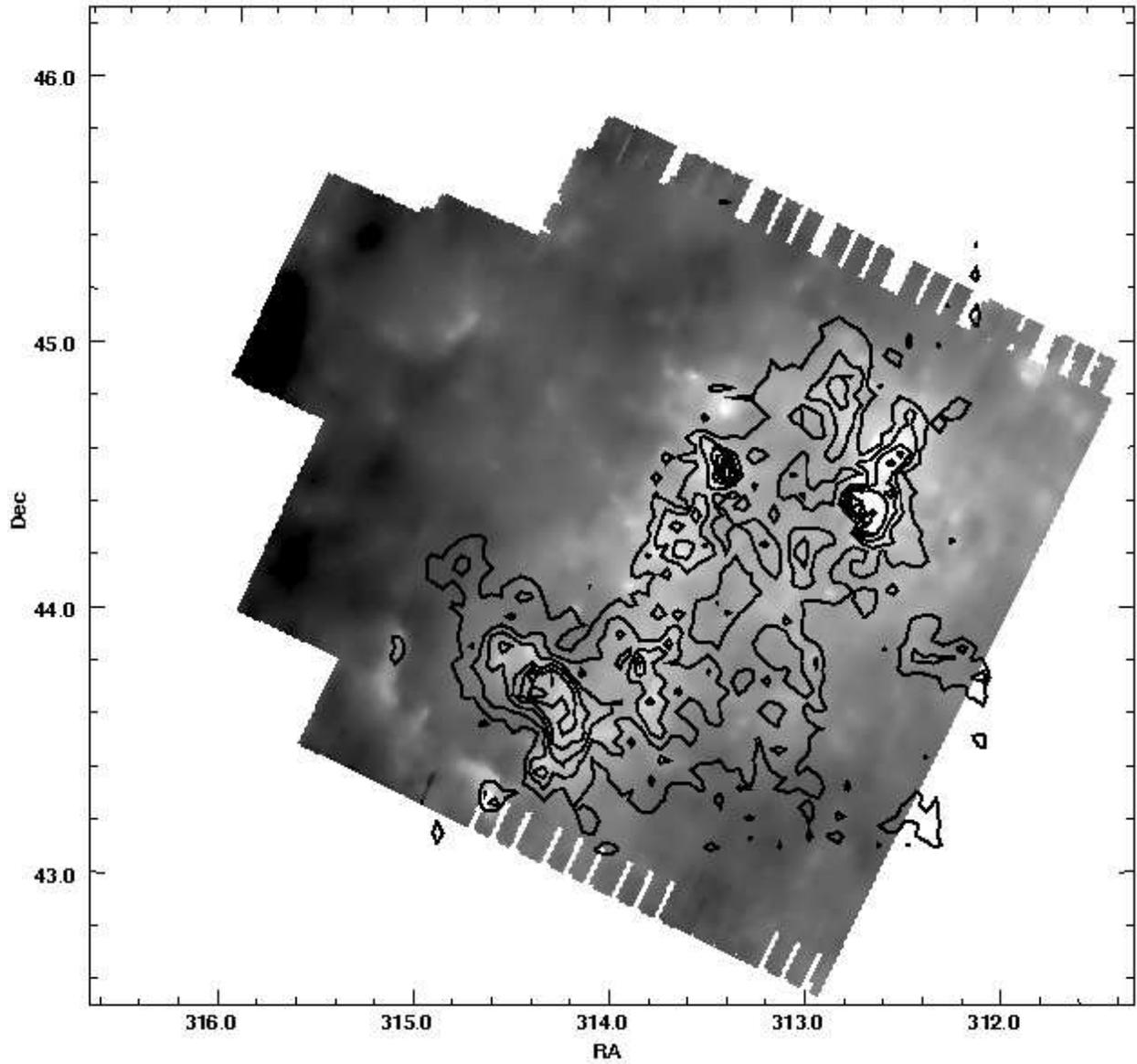}
\caption{MIPS-160 image (greyscale) and the FCRAO $^{13}$CO
radio data (contours, levels 2 to 16).  The 160 \mum\ emission is
closely correlated with the $^{13}$CO emission. }
\label{fig:fcrao}
\end{figure*}

Co-authors JMC and LAH obtained Five College Radio Astronomical
Observatory (FCRAO) data of this region in 1998. A complete discussion
of these data is beyond the scope of the present paper;
Figure~\ref{fig:fcrao} shows that the 160 \mum\ emission and the
$^{13}$CO emission are well-correlated.

\subsection{Bandmerging and Multi-wavelength Point Source Catalogs}
\label{sec:bandmerging}

The IRAC map (G09) covers an area comparable to that covered by MIPS
(see Fig.~\ref{fig:where}).  G09 discusses the IRAC source
extractions, and bandmerging with data from the 2 Micron All-Sky
Survey (2MASS, at 1.25, 1.65, and 2.17 \mum; Skrutskie \etal\ 2006). 
In summary, the IRAC and 2MASS point sources were matched by position
with a matching radius of 2$\arcsec$.   The zero points we used for
conversion between magnitudes and flux densities are 1594 Jy, 1024
Jy, and 666.7 Jy for $JHK_s$, respectively; for IRAC's four
channels in order, they are, 280.9, 179.7, 115.0, and 64.13 Jy.

G09 also describes matching to our $BVI_c$ optical catalog, which is
based on CCD imaging for the $\sim2^{\circ}\times2^{\circ}$ region in
the heart of the NAN complex.  There is extracted photometry for about
19,000 stars.   This $BVI_c$ photometry is included in the spectral
energy distributions (SEDs) and discussion here.  The zero points we
used were 4000.87, 3597.28, and 2432.84 Jy for $B$, $V$, and $I$,
respectively.

To match the MIPS data to the IRAC data, we have first bandmerged the
70 \mum\ detections to the 24 \mum\ detections using a 10$\arcsec$
merging tolerance.  By individual inspection, all the 70 \mum\ sources
have 24 \mum\ counterparts.  We assigned the MOPEX-derived 24 \mum\
coordinates to each of the sources.  Then, we merged the 24+70 \mum\
catalog to the IRAC+2MASS catalog using a tolerance of 2$\arcsec$, a
value determined via histograms of nearest multi-band matches and
experience with other star-forming regions (see, e.g., Rebull \etal\
2010). (We note that the absolute pointing accuracy within a single 24
\mum\ observation is $\sim$1.4$\arcsec$, and the internal positional
accuracy within a 24 \mum\ observation can be much smaller than that,
but we have empirically found that, once multiple observations are
combined and compared to other bands, 2$\arcsec$ is the best value to
use.)

Since G09, we have also included data from the Initial Data Release
from the INT Photometric H$\alpha$ Survey of the Northern Galactic
Plane (IPHAS). The IPHAS survey was carried out in narrow-band
H$\alpha$ and Sloan $r^{\prime}$ and $i^{\prime}$ filters, down to red
magnitudes fainter than 20 (see Drew \etal\ 2005 for a general survey
description).   For the purposes of adding these points to our SEDs,
we took the effective wavelengths for H$\alpha$, $r^{\prime}$, and
$i^{\prime}$ to be 0.656, 0.624, and 0.774 \mum, and the zero points
to be 2974.4, 3173.3, and 2515.7 Jy, respectively.

There are IRAS data over parts of out map (though it is in one of the
missing ``wedges''), and both Midcourse Space Experiment (MSX) and
AKARI data over the entire region. We considered including these data
for analysis of YSO candidates, but the source density is high, and
these prior missions have low enough resolution and are shallow, so we
chose not to include these data in, e.g., SEDs. However, we did
compare our results to those from these missions; see
Appendix~\ref{sec:irasmsx}.

We also included literature data from smaller (e.g., not all-sky or
Galactic Plane) surveys; see next section.

\subsection{Literature Ancillary Data}
\label{sec:ancildata}

There are several surveys of the NAN in the literature, and we
compare our catalog to the literature to (a) include additional
optical and near-IR data where possible, (b) identify known YSOs, and (c)
identify known non-YSO (contaminant) objects.  We then used the
population of known YSOs and non-YSOs to inform our color-based
selection of new YSOs.  We now enumerate the literature surveys we
included on a case-by-case basis. Where necessary (usually the very
old surveys), we used original published finding charts and 2MASS to
obtain updated coordinates.  Some objects cannot be updated in this
fashion, as there are multiple close-by objects of comparable
brightness.  In a few cases, we believe that certain sources from the
literature do not really exist, as detailed below.  (Note that prior
identifications appear in Tables~\ref{tab:allysos} and
\ref{tab:missingysos}, discussed below.)

As part of a large Mount Wilson survey for Be stars, Merrill and
Burwell (1949, 1950) identified four Be stars in this region based on
bright H$\alpha$ lines as seen on objective-prism photographs. Three
of the four stars are in our maps, and all three are very bright at
Spitzer bands.  Just one (MWC 1032=HD 198931) is not saturated at all
four IRAC bands.   

Among the first to specifically and systematically study the NAN
complex itself, Herbig (1958) surveyed 6 square degrees and identified
68 stars with H$\alpha$ in emission, which he took to be indicative of
youth.  Many of these objects subsequently have become famous objects,
including well-studied Herbig AeBe (HAeBe) and FU Orionis stars.
Herbig's paper contains several prescient observations: (a) that the
part of the complex identified as NGC 7000 is just a segment of a much
larger reflection and emission nebula, (b) the exciting star(s) is
(are) difficult to identify, (c) the Gulf of Mexico region may contain
the exciting star of the \ion{H}{2} region, and  contains several
young and very red stars.  With respect to (a), the entire complex
with the nebulae is now usually regarded as part of the same molecular
cloud (see, e.g., Feldt \& Wendker 1993, Feldt 1993).  For (b), Herbig
(1958) considers and rejects several stars as the exciting source of
the nebula.  He concludes, agreeing with Osterbrock (1957), that the
source of excitation is likely behind the Gulf of Mexico.  In regards
to (c), 5 of the stars in the Gulf of Mexico region were bright enough
for Herbig (1958) to identify, and he suggested that these objects may
be ``deep in, or behind, the dark cloud.''  Additionally, as described
above, Herbig (1958) identified a cluster of young objects near IC
5070, leading him to conclude that this region was the most active
region of star formation in the complex.  We have used 2MASS and the
finding chart in Herbig (1958) to obtain updated coordinates for these
objects, and they are listed in our Tables~\ref{tab:allysos} and
\ref{tab:missingysos} as ``IC 5070 cluster {\em N}."  Other objects
identified in that paper usually appear in Tables~\ref{tab:allysos}
and \ref{tab:missingysos} as ``LkH$\alpha$" numbers, as per the
convention.  We note here that LkH$\alpha$ 190 may be extended at
Spitzer bands.


The next large, systematic survey of the NAN complex is Welin (1973),
who, like Herbig (1958), looked for stars bright in H$\alpha$.   Welin
(1973) surveyed a larger region than Herbig (1958), finding 141
objects down to $V\sim$16, just 35 of which were previously known. 
Spectral types for several of these objects were also reported.  These
objects appear in Table~\ref{tab:allysos} and \ref{tab:missingysos} as
``UHa'' numbers when no more common identification is known.

In the ground-breaking paper by Cohen \& Kuhi (1979), among
observations for many other clusters, some observations are reported
for objects in the NAN complex.  They identify two new infrared
objects (IRS 1 and IRS 2) in the NAN complex, along with two loose
groupings of stars bright in the infrared: a group near LkH$\alpha$
188 (in the northern Gulf of Mexico region), and
IRS 3, 4, 5, and 6 (near LkH$\alpha$ 185).  These observations were
difficult, obtained with a single-pixel detector drift-scanning across
the region.  By comparison to the Palomar Observatory Sky Survey
(POSS), 2MASS, and our Spitzer images,
we can recover the stars identified as LkH$\alpha$ 188/G1-5 (G2 and G5 are
multiples), though the stars are close enough together that extracting
photometry for each object individually using our current pipeline is
not always possible due to resolution concerns at 8 \mum\ and longer. 
Of the remaining objects identified by Cohen \& Kuhi (1979), we cannot
recover IRS 1-6.  While of course it is possible that YSOs can vary
substantially in the IR, our observations are significantly more
sensitive than these early observations, and a dimming of many orders
of magnitude would be possible before we would not recover them.  We
conclude that these objects do not really exist.  The objects that do
exist appear in Table~\ref{tab:allysos} (and
Table~\ref{tab:missingysos}) using the same notation as Cohen \& Kuhi
(1979).   

In the process of obtaining CO observations of an expanding molecular
shell in this region, Bally \& Scoville (1980) reported on 11 bright
infrared sources.  Using 2MASS and our data, we obtained updated
positions for these objects and included them in our database,
comparing them to subsequent literature as well.  We note that these
objects are all very bright in IRAC, often saturated.  Objects 1, 7,
and 8 have been identified by Comer\'on \& Pasqali (2005; see below)
as asymptotic giant branch (AGB) stars, and thus are not YSO
candidates.  Bally and Scoville's object 2 is in the neck of the
Pelican, and resolves with MIPS into bright nebulosity
and multiple IR sources; these are likely young objects. Object 5 is a
binary (one component of which is identified as an F3 by Strai\v{z}ys
\etal\ 1993). 
Finally, object 9 appears to be the same as object 1. 

Ogura \etal\ (2002) reported on 32 stars bright in H$\alpha$ near
bright-rimmed cloud (BRC) 31, which is near the Pelican Nebula.  Some
of these objects (7, 8, 10, 15, 18, 19, 20, 22, 28) are in regions
that are very bright in the Spitzer bands, and therefore may not be
recoverable at some or all Spitzer bands. Some (9, 19) are not
recovered with Spitzer. One (25) may be linked to more than one IRAC
source.

Villanova \etal\ (2004) studied NGC 6997, which is not at the distance
of the NAN complex, but is covered by our map; see \S\ref{sec:ngc6997}
below.   (Note that this cluster has been commonly confused with NGC
6996; see Laugalys \etal\ 2006b and Corwin 2004.) Villanova \etal\
published $UBVRI$ photometry for 2700 stars in this region, which we
have matched to objects in our catalog as optical supporting data, in
addition to whether or not the objects are members of NGC 6997. The
Villanova \etal\ (2004) optical survey reaches fainter objects than
our survey does; $\sim$1200 objects from this survey do not have
counterparts in our survey.

Comer\'on \& Pasquali (2005), in their quest to find the exciting star
of the complex, studied 19 objects, six of which they determined to be
AGB stars, and five more of which they determined to be carbon stars;
the remaining objects were candidates for the exciting star, e.g.,
young high-mass stars.  Both the young high-mass stars and the
objects confirmed not to be young high-mass stars are useful to
include in our catalog. We note here that their Object 14 appears to
resolve into at least 3 Spitzer sources. Comer\'on \& Pasqali (2005)
identify Bally and Scoville's object 10 as the exciting source for the
nebula.  

We included in our database the spectral types for several NAN complex
objects that were reported as part of large multi-cluster studies;
Shevchenko \etal\ (1993) monitored HAeBes in many clusters, and
Terranegra \etal\ (1994) obtained Str\"omgren photometry of
Orion-population stars.  Here, and in general for this database,
spectral types of objects obtained via spectroscopy were allowed to
supercede spectral type estimates obtained via photometry.

The group at Vilnius Observatory in Lithuania has embarked on a
multi-decade campaign to obtain multiband photometry (from which they
derive estimates of reddening and spectral types) for thousands of
stars in star-forming regions, including hundreds of stars in the NAN
complex.  The {\em Vilnius} bands of observation are {\em U} (0.345
\mum), {\em P} (0.374 \mum), {\em X} (0.405 \mum), {\em Y} (0.466
\mum), {\em Z} (0.516 \mum), {\em V} (0.544 \mum), and {\em S} (0.656
\mum).   Photometrically-derived spectral types (and CCD photometry
where available) from all of these papers were included in our
collection of known objects: Strai\v{z}ys, Mei\v{s}stas, \&
Vansevi\v{c}ius (1989), Zdanavi\v{c}ius \& Strai\v{z}ys (1990),
Strai\v{z}ys \etal\ (1993), Strai\v{z}ys, Corbally, \& Laugalys
(1999), Laugalys \& Strai\v{z}ys (2002), Laugalys \etal\ (2006a,b),
Laugalys \etal\ (2007), Strai\v{z}ys \& Laugalys (2008a,b),
Strai\v{z}ys, Corbally, \& Laugalys (2008), Corbally, Strai\v{z}ys,
and Laugalys (2009), and Strai\v{z}ys \& Laugalys (2009).  Note that,
in addition to photometry and spectral types, Laugalys \etal\ (2006a)
and Corbally \etal\ (2009) specifically identified several YSO
candidates as having strong H$\alpha$ emission. The adopted zero
points, kindly provided by V.\ Strai\v{z}ys (priv.\ comm) are, in the
order {\em UPXYZVS}, 1325, 2112, 4106, 4169, 3949, 3748, and 3141 Jy. 

The IPHAS team, in Witham \etal\ (2008), identified bright H$\alpha$
sources in the first 80\% of their survey, which includes the NAN
region.  The IPHAS team did not attempt to identify the nature of the
sources, simply that they were bright in H$\alpha$. Since YSOs are
expected to be bright in H$\alpha$, the $\sim$80 stars from that
catalog in this region were also matched to our catalog, and the
information that they are bright in H$\alpha$ is folded in to our
assessment of YSO candidacy below. We note that some stars with
H$\alpha$ in emission could be active stars or classical Be stars with
gaseous disks, not necessarily young stars. 

Finally, G09 identified 1657 YSO candidates based on IRAC data. Since
then, we have continued to improve our catalog quality (most notably
identifying some objects incorrectly marked as saturated at 8 \mum),
and, using the same algorithm as described in G09, we can now identify
1750 YSO candidates based just on the IRAC NAN data.  Objects
identified using the G09 methodology but not rediscovered using the
MIPS-based search here are listed explicitly in
Appendix~\ref{sec:iraconly}.

In summary, we merged our Spitzer-based catalog (from
Sec.~\ref{sec:bandmerging}) to these existing data from the
literature.  There is literature information for about 3600 stars
scattered throughout the map, including photometry and spectral types
or other classification (such as if they were oxygen- or carbon-rich
AGB stars, or red clump giant stars), or even just prior
identifications (such as `GSC N' or `TYC N', listed in the tables
below).   Thus, our final multi-wavelength catalog includes MIPS,
IRAC, 2MASS, IPHAS, and other optical data, both from the literature
and that which we ourselves have obtained.  After including all of
this ancillary data, there are spectral types for more than 1200
objects in our catalog, ranging from O5 to M8, with $\sim$400
additional generic AGB, carbon star, and red clump giant
identifications.  These types are largely photometric (as opposed to
spectroscopically obtained), and largely for types K and earlier.  

Just 142 previously-identified objects of any sort are missing
entirely from our Spitzer-based catalog (aside from those faint ones
from Villanova \etal\ 2004); the overwhelming majority of those are
saturated at all Spitzer bands (and thus do not appear in our catalog)
or resolve into multiple pieces when viewed with Spitzer (and thus are
not formally matched to a single Spitzer source).


Of the previously-identified YSOs in the literature (identified in the
tables below), it is interesting to consider what fraction are present
in our database.  About 200 objects in our catalog are identified
specifically as previously-known YSOs, though this is likely to
consist of a mixture of true members and interlopers.  Just $\sim$50\%
of them have measured flux densities in any MIPS band, and 95\% of
them are measured at any IRAC band.  Most of the ``missing'' ones are
saturated or in very crowded regions in the Spitzer images, which is
consistent with the previous surveys being shallower and/or lower
spatial resolution.   (NB: Table~\ref{tab:ysostats} in the discussion
below includes statistics on previously-known objects recovered or not
using G09 and the primarily MIPS-based method presented here.)

\section{Ensemble Statistics, Source Counts, and Contamination}
\label{sec:sourcecounts}

\subsection{Catalog statistics}

\begin{deluxetable}{lrcl}
\tablecaption{Statistics of MIPS point source detections
\label{tab:statistics}}
\tablewidth{0pt}
\tablehead{
\colhead{item} & \colhead{number} & \colhead{fraction of}
& \colhead{notes}\\
& & \colhead{24 \mum\ sample}& }
\startdata
24 \mum\ &                      4334     & 1.00 & 3694 in region covered by all 4 IRAC bands\\ 
70 \mum\ &                       97      & 0.02 \\ 
24 \mum\ \& 70 \mum\ &            95     & 0.02 \\ 
24 \mum\ \& ANY IRAC band       &  3243  & 0.75 & 3100 in region
covered by all 4 IRAC bands (84\%)\\ 
70 \mum\ \& ANY IRAC band       &  85    & 0.02  & 89\% of 70 \mum\ sample\\ 
24 \mum\ \& ALL IRAC bands       &  2582 & 0.60 & 2577 in region
covered by all 4 IRAC bands (70\%) \\ 
70 \mum\ \& ALL IRAC bands       &  59   & 0.01 & 61\% of 70 \mum\ sample \\ 
24 \mum\ \& ANY 2MASS band      &  2972  & 0.69 \\
24 \mum\ \& ANY IPHAS band      &  1824  & 0.42 \\
24 \mum\ \& $V$ or $I_c$ or $r^{\prime}$ or $i^{\prime}$ &  1931   &0.45  \\ 
\enddata
\end{deluxetable}

Table~\ref{tab:statistics} presents MIPS-focused statistics on the
ensemble catalog spanning 0.35 \mum\ through 70 \mum.   About 75\% of
all the MIPS-24 sources have an IRAC counterpart.  However, the IRAC
and MIPS maps do not cover exactly the same area.  Considering just
the region of the MIPS map where there are also IRAC data from all 4
bands, $\sim$85\% of the 24 \mum\ sources have a measured counterpart
at at least one IRAC band (where the 15\% without an IRAC counterpart
are overwhelmingly fainter than $\sim$8th mag, and thus clearly on the
very faint tail of the distribution; see Fig.~\ref{fig:sourcecounts}
below); 60\% of the 24 \mum\ sources have a measured IRAC flux density
at all four IRAC bands.   There are near-IR \jhk\ data covering this
entire region from 2MASS, but 2MASS is relatively shallow; 69\% of the
MIPS-24 sources have a 2MASS counterpart. (We note that this region has
also been covered by the UKIRT Infrared Deep Sky Survey [UKIDSS] to
much deeper limits than 2MASS; we will include these data in a
forthcoming paper.) The optical data are highly affected by reddening;
45\% of all the MIPS-24 sources have an optical counterpart of any
sort (from our data, IPHAS, or the Vilnius photometry).

\subsection{Source Counts and Contamination}

\begin{figure*}[tbp]
\epsscale{1}
\plotone{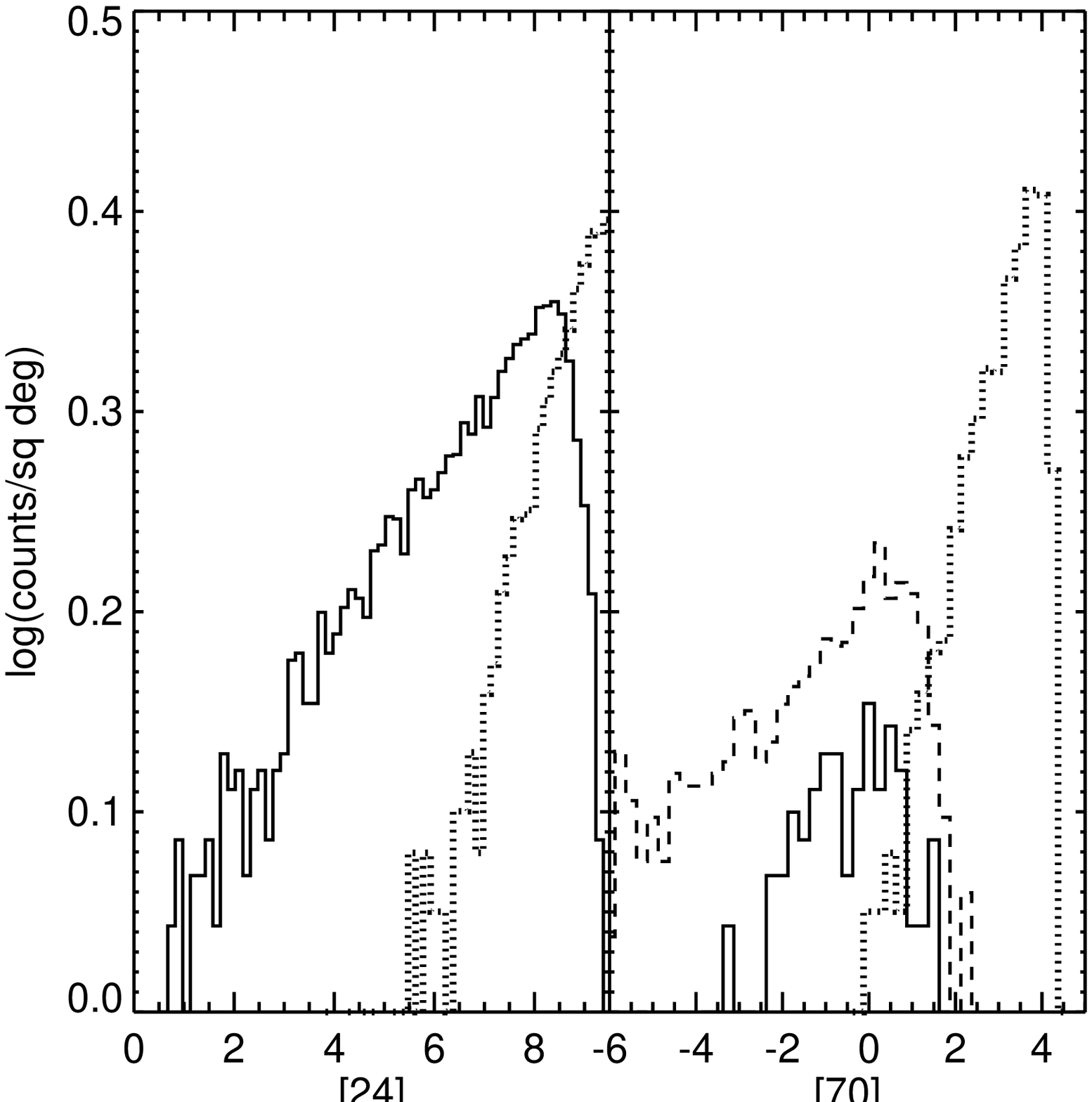}
\caption{Differential number counts at 24 and 70 \mum\ in the NAN
complex.  Extragalactic background source counts from the 6.1 square
degree SWIRE ELAIS N1 field are shown (scaled) for comparison (thick
short-dashed line). For 70 \mum, source counts from the $\sim$8 square
degree Vul OB1 field from MIPSGAL are also shown (scaled) for
comparison (long-dashed line). The sources we see in the NAN are more
likely to be in our Galaxy than extragalactic sources.  }
\label{fig:sourcecounts}
\end{figure*}

MIPS sources in general are a combination of cloud members,
foreground/background stars in the Galaxy, and the extragalactic
background. Since the NAN complex is essentially in the Galactic plane
($b\sim -0.6$) and is viewed along a spiral arm, we expect to see
mostly Galactic sources, with few extragalactic objects, and we expect
an excess over the Galactic background attributable to cloud members. 
Figure~\ref{fig:sourcecounts} shows the observed NAN 24 and 70 \mum\
differential source counts in comparison to observed source counts
from the 6.1 square degree Spitzer Wide-area Infrared Extragalactic
Survey (SWIRE; Lonsdale \etal\ 2003) ELAIS (European Large Area ISO
Survey) N1 extragalactic field\footnote{VizieR Online Data Catalog,
II/255 (J. Surace et al., 2004)} (the Cores-to-Disks [c2d; Evans
\etal\ 2003, 2009] reduction is used here, as in Rebull \etal\ 2007,
and the distribution is scaled for relative areas but not additionally
reddened). The sources that we see in the NAN do not resemble the
source count distribution for extragalactic sources.  For 70 \mum,
scaled source counts from the $\sim$8 square degree Vulpecula OB
association (Vul OB1; Billot \etal\ 2010) from MIPSGAL (Carey \etal\
2009) are also shown for comparison. Like the NAN, Vul OB1 is in the
Galactic plane ($l\sim$60.5, $b\sim$0), the sky background levels are
comparable to those in the NAN, and the Spitzer observation covers an
appreciable area such that there are enough sources to effect a
comparison.  The source counts in the NAN much more closely resemble
that from a star forming region like Vul OB1 than they do an
extragalactic region like SWIRE.  While the shapes of the differential
source count distribution are similar between the NAN and Vul OB1,
there are more sources overall, and more bright sources, in Vul OB1.
In both cases, we have comparable data and data reduction strategies,
and we are looking along a spiral arm. In the case of Vul OB1, the
star-forming region is 2.3 kpc away, which is considerably further
away than the NAN.  As such, the Vul OB1 region encompasses more
foreground volume, and many more foreground sources are included in
the sample.  (More evidence for this can be found in the fact that
most of the NAN 70 \mum\ sources are in the Gulf of Mexico region, as
opposed to more evenly distributed over the field.) 

In order to begin to place constraints on whether we see a higher
count rate of MIPS sources over the generic Galactic background (e.g.,
cloud members), we can compare the source counts within our own maps.
The portion of the map to the northeast can be seen to be very likely
to be ``off-cloud'' by inspection of the cloud morphology at all
available bands (optical through 160 \mum). The general sky brightness is
lower at all bands in this region; see
Figures~\ref{fig:24mosaic}-\ref{fig:3color}. There are just five 70
\mum\ point sources in this very roughly 1 square degree region;
extrapolating this contamination rate to the rest of the map implies
that $\sim$35 of the NAN 70 \mum\ sources may be generic Galactic
background, e.g., contamination. At 24 \mum, in the same region, there
are $\sim$400 sources, implying that as many as $\sim$2500 of the NAN
24 \mum\ sources may be Galactic contamination, or roughly half of all
of the detected point sources. However, this is likely to be a strong
function both of source brightness and location in the map, and
moreover sources that have infrared excesses are more likely to be
cloud members than background, as discussed below. 

Galactic stars contribute substantially to the MIPS source counts in
NAN.  This is different than many of the other nearby star-forming
regions studied with Spitzer (see, \eg, the series of papers from the
c2d Legacy team, such as Rebull \etal\ 2007 for MIPS observations in
Perseus; also see Rebull \etal\ 2010 for Spitzer observations in
Taurus) which are out of the Galactic plane.  The contaminants of
greatest concern to us are the ones that appear to have YSO-like
colors. These include some AGB stars (oxygen- and carbon-rich);
although intrinsically very bright, such that they would be saturated
even at large distances, once they are reddened by the NAN cloud, they
can appear fainter and have thus have colors and overall brightnesses
suggestive of YSOs (see, e.g., Robitaille \etal\ 2008). 

Some AGB (including specifically carbon stars) are already known in
the direction of the NAN; see \S\ref{sec:ancildata} above and
Figure~\ref{fig:3324_4panel} below.  Although this is far from a
complete sample, we can examine the number counts of these objects in
our map in an effort to quantify how many AGB contaminants we might
have.  Twenty-five of these objects are seen in 24 \mum, and two are
seen at 70 \mum\ as well. Their colors suggest that they could indeed
be confused with bright YSO candidates.  All of these objects are
relatively bright, e.g., have [3.6] between 6.6 and 10.7 mag, but this
is likely a bias in that the previous studies tended to focus on the
brighter objects. 

We can also attempt to constrain the AGB contamination by comparison
to other regions studied with Spitzer. Harvey \etal\ (2006) argue that
any AGB star in our Galaxy is likely to be saturated in most IRAC
observations (except for cases of very large \av).  Rebull \etal\
(2007), using data from Blum \etal\ (2006) obtained in the Large
Magellanic Cloud (LMC) as an aid to quantifying the AGB contamination
in Perseus, find that a typical AGB star (in the direction of Perseus)
would appear so bright as to be saturated in the 2MASS survey, though
not MIPS.  Harvey \etal\ (2007b) argue that, in their 0.58 square
degree field in Serpens (at $b\sim5\deg$), they expect up to 6 AGB
stars and estimate that they actually find 3.  Scaling from those
values to our $\sim$6 square degree field, we might expect $\sim$30-60
AGB contaminants. However, in spectroscopic follow-up of half of the
the Spitzer-selected YSO candidates from Serpens, Oliveira \etal\
(2009) found a much higher AGB contaminant rate among the
Spitzer-selected YSO candidates -- 25\%. Unlike Perseus, where we are
looking out of the Galaxy and so there is an `edge' to the
distribution of AGBs, both the NAN and Serpens have sightlines in the
Galactic plane, where extinction cuts off the AGBs before the Galactic
population really falls off. Extinction makes the AGBs look even more
like YSO candidates, because it can make an object selected from a
single color-magnitude diagram appear to have an IR excess, and more
generally because it makes the AGBs look fainter, comparable to YSOs
in the cloud. Thus, a high contamination rate among the YSO candidates
from AGB stars is expected, probably anywhere from the $\sim$5\%
estimated originally by a scaled Harvey \etal\ (2007b) to the
$\sim$25\% based on the Oliveira \etal\ (2009) spectroscopic follow-up
of half the Spitzer-selected Serpens YSO candidates. 

While of course any individual YSO candidate we find will require
spectroscopic followup to confirm its YSO status, we conclude that
even in the worst contamination case, the majority of objects selected
as having red MIPS colors that we report here as YSO candidates are
indeed likely to be young stars associated with the NAN cloud complex.
Statistically, other properties (e.g., position in multiple
color-magnitude diagrams, clustering) make it less likely that a given
object is a contaminant, and more likely to be a YSO, as we discuss in
the next section.

\section{Selection of YSO candidates}
\label{sec:pickysos}

There is no single Spitzer color selection criterion (or set of
criteria) that is 100\% reliable in separating members from non-member
contaminants.  Many have been considered in the literature (e.g.,
Allen \etal\ 2004, Rebull \etal\ 2007, Harvey \etal\ 2007, Gutermuth
\etal\ 2008, 2009, Rebull \etal\ 2010). Some make use of just MIPS
bands, some make use of just IRAC bands, most use a series of color
criteria, and where possible, they make use of (sometimes substantial)
ancillary data.  In our case of the NAN, because it is in the Galactic
plane and along a spiral arm, as discussed above, the contaminants
having YSO-like colors are largely expected to be AGB stars and not
galaxies; we can use that information to our advantage.  We have
shorter-wavelength ancillary data for many of the objects in our
catalog (see \S\ref{sec:ancildata}) that can be used to constrain
reddening due to extinction and separate out things resembling
reddened photospheres. It is still most effective to separate the
members from the diskless stars and galaxies (if not the less common
AGBs) by using infrared colors as a discriminant, so we are still
primarily Spitzer-driven in our source selection.  Our inventory of
the no or very low IR excess objects (Class III or its high-mass
equivalent) is incomplete since we are using red colors (interpreted
as an IR excess due to a circumstellar disk) to select members. 

We imposed an IRAC-only YSO selection in G09.  Now, we include
MIPS data, investigating candidates found via a MIPS-based selection
and comparing this to the IRAC-only YSO selection from G09.  In
order to construct the best possible list of YSO candidates, we start
with a simple color selection, then incorporate information from the
ancillary data we have amassed.  A complete list of the YSO candidates
selected here is included in the text below; objects selected using
the IRAC-only YSO selection from G09 but not recovered here are
reported separately in the Appendix.

\subsection{The $[3.6]$ vs.\ $[3.6]-[24]$ Diagram}
\label{sec:3324}

Of the MIPS bands, 24 \mum\ data are available for the largest number
of sources. In order to identify sources with excesses at 24 \mum, we
compare 24 \mum\ flux densities to those from a shorter wavelength
band. 

In recent papers such as Rebull \etal\ (2007) or Rebull \etal\
(2010),  the $K_s$ vs.\ $K_s-[24]$ color-magnitude diagram (CMD) was
used to find YSO candidates.  In cases of nearby star forming regions,
or regions where there are no IRAC data, this can be effective.
However, there are some good reasons to use [3.6] vs.\ [3.6]$-$[24]
rather than $K_s$ vs.\ $K_s-[24]$.  There is an intrinsic spread in
$K_s-[24]$ photospheric colors that is not present in [3.6]$-$[24]
because late type stars are not colorless at $K_s-[24]$ (Gautier
\etal\ 2007).  The effects of reddening are stronger at \ks\ than at
3.6 \mum. And, if 2MASS is the only source of \ks, even short 3.6
\mum\ integrations can reach fainter sources than 2MASS does. In the
specific case of the NAN complex, the area covered by MIPS is
well-matched to that by IRAC, we have substantial reddening, and 2MASS
is shallow with respect to the brightnesses expected for late type
young stars at the distance of the NAN (and with respect to the depth
of our IRAC data; see Table~\ref{tab:statistics}).   For the above
reasons, we use [3.6] vs.\ [3.6]$-$[24] as our primary mechanism for
selecting YSO candidates.  

\begin{figure*}[tbp]
\epsscale{.9}
\plotone{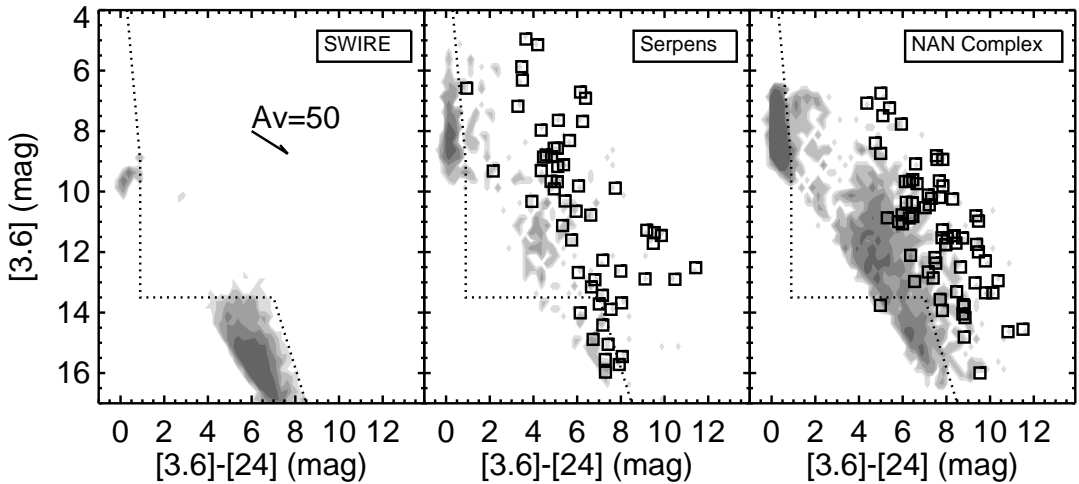}
\caption{$[3.6]$ vs.\ $[3.6]-[24]$ contour plots for objects in SWIRE
(left; $l$=85$\arcdeg$, $b$=+45$\arcdeg$, 6.1 sq.\ deg.), Serpens
(middle; $l$=32$\arcdeg$, $b$=+5$\arcdeg$, 0.85 sq.\ deg.), and the
entire NAN complex (right; $l$=85$\arcdeg$, $b$=$-$0.6$\arcdeg$;
$\sim$7 sq.\ deg.).   Objects in SWIRE are expected to be mostly
galaxies (objects with $[3.6]\gtrsim$14) or stellar photospheres
(objects with $[3.6]-[24]\lesssim$1).   A box denotes that the
underlying object was also detected in 70 \mum.  Objects that are
the most obvious candidate young objects have colors unlike those
objects found in SWIRE, e.g., $[3.6]\lesssim$14 and
$[3.6]-[24]\gtrsim$1, or very red colors if $[3.6]\gtrsim14$. The
dotted line denotes the dividing line between the region clearly
occupied by SWIRE-type contaminants (galaxies and diskless stars) and
YSOs. }
\label{fig:3324_3panel}
\end{figure*}

\begin{figure*}[tbp]
\epsscale{.9}
\plotone{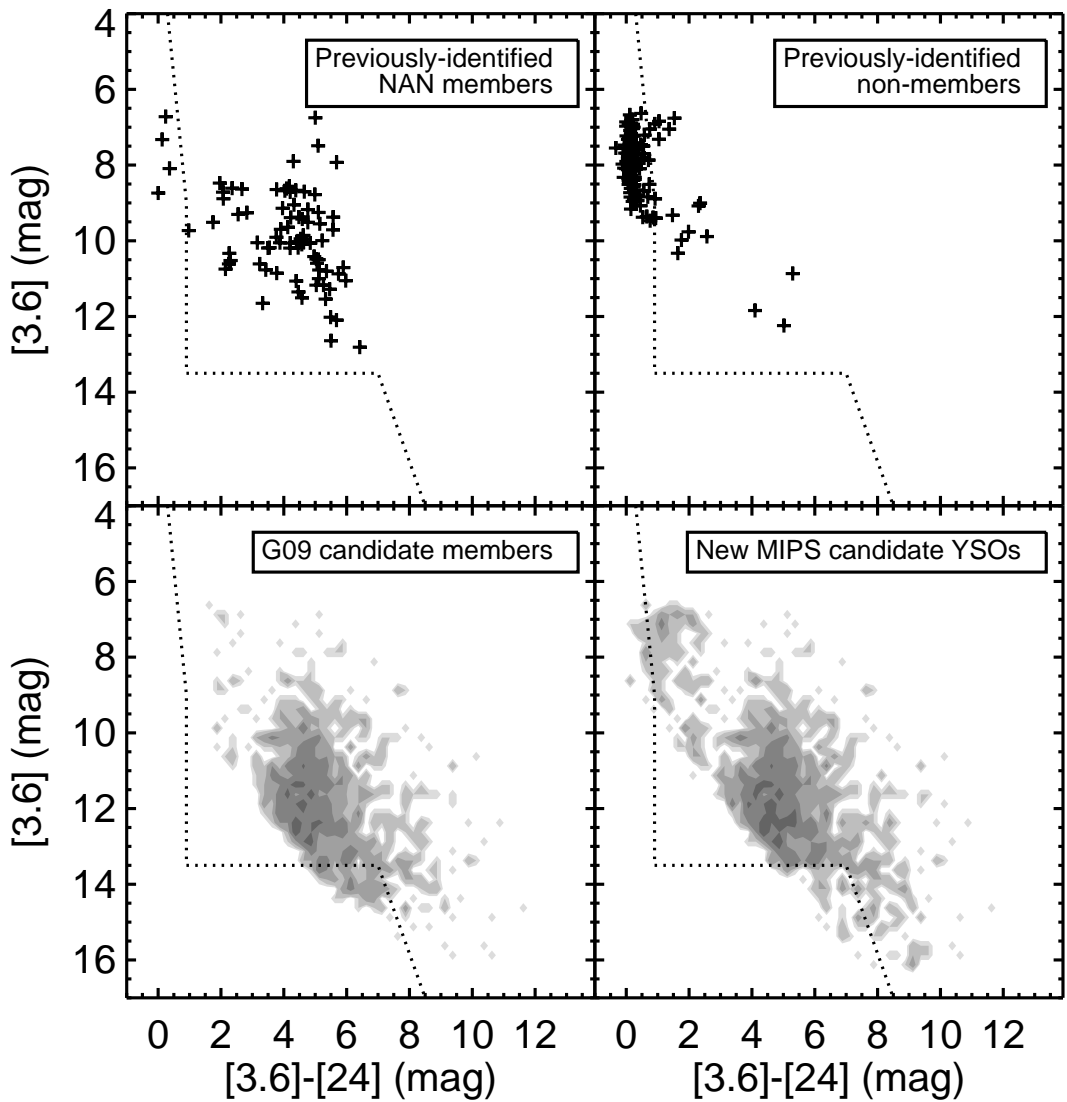}
\caption{$[3.6]$ vs.\ $[3.6]-[24]$ plots for subsets of the NAN
sample; see Figure~\ref{fig:3324_3panel} for comparison. As before,
dotted line denotes the dividing line between the region clearly
occupied by SWIRE-type contaminants (galaxies and diskless stars) and
YSOs. Here the panels are previously-identified YSOs, previously
identified contaminant objects, YSO candidates from G09, and the YSO
candidates identified in this work, some of which rediscover those
from G09 (see \S\ref{sec:allysos}).  The distributions of YSO
candidates look different than the previously-identified non-members
and more like the previously-identified members.}
\label{fig:3324_4panel}
\end{figure*}

\begin{figure*}[tbp]
\epsscale{.9}
\plotone{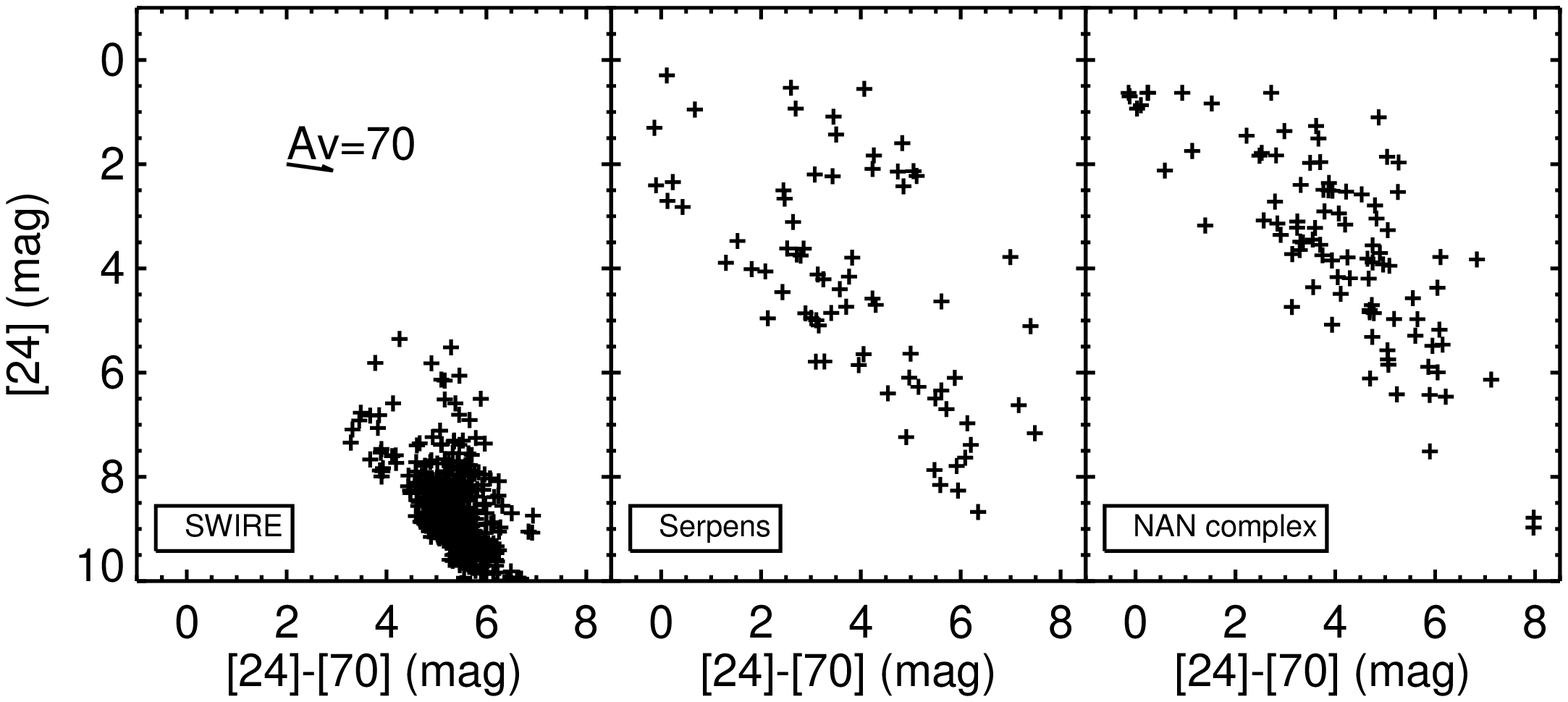}
\caption{$[24]$ vs.\ $[24]-[70]$ plots for objects in SWIRE (left),
Serpens (middle), and the entire NAN complex (right). Again, the
distribution of objects in the NAN more closely resembles that from
Serpens than SWIRE, e.g., are more likely to be young stars than
galaxies.}
\label{fig:2470}
\end{figure*}

Figure~\ref{fig:3324_3panel} shows the [3.6] vs.\ [3.6]$-$[24] plot
for the NAN sample in comparison to the 6.1 sq.\ deg.\ SWIRE (Lonsdale
\etal\ 2003) ELAIS N-1  and 0.85 sq.\ deg.\ c2d Serpens (Harvey \etal\
2007a) samples.  Ordinary stellar photospheres (likely foreground or
background stars) have $[3.6]-[24]\sim$0, and the galaxies as seen in
SWIRE make up the large, elongated source concentration near
[3.6]$-$[24]$\sim$6, [3.6]$\sim$15.  Objects not in this region, e.g.,
the brighter and/or redder objects (those to the right of the dotted
line in Figure~\ref{fig:3324_3panel}), are less likely to be part of
the Galactic or extragalactic backgrounds, and more likely to be YSOs
with a 24 \mum\ excess.  Both the Serpens and NAN clouds have
substantial populations of objects meeting these criteria.  The
distribution of objects found in the NAN is much more similar to that
in Serpens than in ELAIS N-1. 

Reddening can move higher mass objects from the photospheric locus
into the region where lower mass YSOs are found.  The reddening is far
from uniform, with regions of \av\ up to $\sim$30 mag, according to
the Cambr\'esy \etal\ (2002) extinction map.  The extinction is
predominantly generated by the clouds in the NAN.  As seen in
Figure~\ref{fig:3324_3panel}, even \av$\sim$50 does not move objects
far enough from the photospheric locus into the region where most of
the objects are found in the ``red'' region of the diagram (to the
right of the dotted line). However, the objects that are more
moderately red, those immediately to the right of the dotted line,
could appear there because of reddening, as opposed to due to an
excess due to a circumstellar disk. We can attempt to remove such
simply reddened photospheres by incorporating the ancillary data we
have collected; see below.

Given the distance to the NAN, and given the background counts
analysis above, there are likely to be many legitimate NAN cloud
members overlapping the region populated by the SWIRE galaxies, but it
is difficult to distinguish the populations given the data we have.
For this work, we limit ourselves to the more obvious candidates,
e.g., primarily the objects brighter and/or redder than the dotted
line in Figure~\ref{fig:3324_3panel}, with the caveat that the
selection is further modified, here and below.  The first modification
to this selection is that we look at the significance of the excess. 
If the excess is seen only at 24 \mum, and the measured flux density
is less than 5$\sigma$ above the expected photosphere as extrapolated
from a Rayleigh-Jeans slope tied to 3.6 \mum, it is less likely to be
a real excess source.  Conversely, for the objects closest to the
photosphere locus (e.g., [3.6]$<$10), we accept as candidates those
objects where the 24 \mum\ excess is more than 8-10$\sigma$ above the
extrapolated photosphere, depending on whether the excess is seen at
any other bands (e.g., if there is a small excess at 24 \mum\ and a
clear excess at IRAC bands, this is more believably an object with a
true IR excess than if there is no excess at IRAC bands, just a small
excess at 24 \mum, and no detection at 70 \mum).   This has the effect
of omitting some sources that technically meet the color criteria, and
including some sources that are just blueward of the dotted line in
Figure~\ref{fig:3324_3panel}.  There are $\sim$1200 sources identified
from 3.6 and 24 \mum\ data in this fashion.

Figure~\ref{fig:3324_4panel} compares the distribution of various
subsets of the NAN data in the [3.6] vs.\ [3.6]$-$[24] parameter
space.  Based on our collection of literature data above
(\S\ref{sec:ancildata}), we have a pre-assembled set of NAN members
and non-members, which are shown in Figure~\ref{fig:3324_4panel}. By
and large, these objects fall where we expect them to based on
Figure~\ref{fig:3324_3panel} -- most of the NAN members appear as red
in this diagram, which we interpret to mean that they have infrared
excesses, and most of the non-members do not. However, it is notable
that there are outliers in both cases -- some non-members appear to
have excesses (largely the AGB contaminants noted above), and there
are some members without excesses (which were most likely detected
using other mechanisms besides IR colors, such as H$\alpha$, although
we note that intrinsic stellar variability could also be a
contributing factor).  Figure~\ref{fig:3324_4panel} also shows the
distribution of the YSO candidates selected in G09 as well as the
distribution of YSO candidates from our final selection below
(\S\ref{sec:allysos}).  Note that the candidates selected below often
recover sources from G09; see Table~\ref{tab:ysostats} below and
associated discussion.  In both cases, the distribution of YSO
candidate sources is similar, with the highest density of sources
found near [3.6]$-$[24]$\sim$5, [3.6]$\sim$12. 

Figure~\ref{fig:2470} shows [24] vs.\ [24]$-$[70] for the same three
samples as Figure~\ref{fig:3324_3panel}, namely SWIRE ELAIS N-1, c2d
Serpens, and the NAN sample. The distribution of sources found in the
NAN again is a much closer match to that from Serpens than from SWIRE.
Few of the objects seen at 70 \mum\ in the NAN are likely galaxies;
several are very red.

\subsection{Incorporating ancillary data}
\label{sec:anciltweaks}

As discussed above (\S\ref{sec:ancildata}), we have included in our
database a wide variety of ancillary data. These data are not always
well-matched to our Spitzer data in depth or area. But they can
contribute to the assessment of any one object as to its likelihood of
being a YSO.

Rebull \etal\ (2010), working in Taurus, had a similar problem to
overcome in that there were $\sim$44 square degrees of Spitzer data,
considerable foreground and background contamination, but also
considerable ancillary data.  The approach taken in that study was to
incorporate, in addition to the easily quantifiable Spitzer magnitude
and color criteria, qualitative judgments based on what data were
available.  These included but were not limited to: relative brightness
at all available bands; amplitude of excess; shape of SED; apparent
(projected) proximity to other previously identified cluster members
or non-members; previous identifications; and star counts.  

For the NAN data, though the complete catalog is comparable in size to
that from Taurus, we have many more objects with YSO-like colors to
consider, and different ancillary data.  However, our fundamental
approach is similar to that in Taurus.  We started with the objects
selected in the [3.6] vs.\ [3.6]$-$[24] CMD, and we folded in several
considerations to assess the liklihood of youth and therefore
membership in the NAN. In each case, the assessment can be described
as an educated statistical guess (e.g., an object selected in the
[3.6] vs.\ [3.6]$-$[24] CMD is statistically more likely than an
object not selected in that diagram to be a YSO), but 
that does not ensure that all YSOs will be selected, nor that all
objects that are selected are YSOs. Objects that had no available data
for any particular one of these items were unaffected by the lack of
such data.  (NB: The description here includes forward-references to
columns in Table~\ref{tab:allysos2}, which tabulates whether a given
YSO candidate met the listed criterion.)
\begin{itemize} 
\item If the object appears to the right of the dotted line in
Fig.~\ref{fig:3324_3panel}, e.g., if it is in the space occupied by
most YSO candidates in the [3.6] vs.\ [3.6]$-$[24] CMD, then it is a
likely YSO. (column 5) 
\item The fainter the object is in [3.6], the more likely that it is a
galaxy even if it is red. (column 6) 
\item If the object is also selected as bright and red (e.g., a likely
YSO) in the \ks\ vs.\ \ks$-[24]$ CMD, then it is more likely to be a
YSO, but if those data (\ks\ and [24]) are available for the target
and it is not in the bright and red regime, then it is less likely to
be a YSO. (column 7) 
\item If an object is detected at 70 \mum, it is more likely to be a
YSO, because the 70 \mum\ survey is effectively less sensitive than
the 24 \mum\ survey (e.g., reaches only the brighter excess objects in
the NAN -- see Fig.~\ref{fig:3324_3panel} and  Fig.~\ref{fig:2470}). 
(column 8) 
\item If the object was found via the IRAC-based method used in G09,
then it is more likely to be a YSO; if it could have been found using
that method (e.g., observations exist at all IRAC bands) but was
rejected, then it is less likely to be a YSO. (column 9) 
\item Because the complete map we have extends past the limits of the
bright nebulosity and presumably past the edge of the molecular cloud,
if the object was closer to the middle of the nebula, e.g., within 
$\sim 1 \arcdeg$ radius of the approximate center  ($\alpha, \delta$=
313.752$\arcdeg$, 44.231$\arcdeg$ or 20:55:00.5, +44:13:52), then it
is more likely to be a YSO than if it is further away. (column 10)
\item Because YSOs are more likely to be clustered, if the candidate
was in a cluster (see discussion in \S\ref{sec:clusterdefinition}
below), it is more likely to be a YSO than if it is not in a cluster.
(column 11) 
\item If an object appeared in the literature as a
previously-identified YSO using any wavelength, it is substantially
more likely to be a legitimate YSO. Similarly, if it was previously
identified as a likely contaminant, it is substantially less likely to
be a legitimate YSO. (column 12) 
\item In the \ic\ vs.\ $V-$\ic\ CMD, if the object appears
substantially above the zero-age main sequence (ZAMS), close to where
most of the previously-identified YSOs are found, then it is more
likely to be a YSO; if the object appears below this (or below the
ZAMS itself) it is less likely to be a YSO, at least one belonging to
the NAN, unless dominated by scattered light at optical wavelengths.
(column 13) 
\item Similarly, in the IPHAS $r^{\prime}$ vs.\
$r^{\prime}-i^{\prime}$ CMD, if the object appears substantially above
the ZAMS (assuming no reddening), near where most of the
previously-identified YSOs are found, then it is more likely to be a
YSO; if it appears below this (or below the ZAMS itself), it is less
likely to be a YSO. (column 14) 
\item If the object was already identified as a bright H$\alpha$
source in IPHAS (Witham \etal\ 2008; column 15), or appears in the
$r^{\prime}-H\alpha$ vs.\ $r^{\prime}-i^{\prime}$  diagram as
substantially abvove the unreddened main sequence locus (column 16),
it is more likely to be a YSO; if the object is below this, it is less
likely to be a YSO. 
\end{itemize}

Each object was individually initially assessed resulting in a
distribution of confidence levels, which we arbitrarily bin into three
groups -- $\sim$700 in grade A, $\sim$500 in grade B, and $\sim$200 in
grade C.  However, there is still more information that can be
included.

\subsection{Further considerations}
\label{sec:downgrade}

Even with all of the Spitzer+ancillary data above, there are still
sources left in the pool of possible candidates that are unlikely to
be true YSOs. The SED for each YSO was examined by hand, along with
each object's status for each item from the prior semi-automatic step.
Objects that were too bright (likely foreground and/or AGB stars) or
were clearly reddened photospheres (likely background giants) or
had SEDs resembling known extragalactic objects or had very low
significance excesses or very sparse data (e.g., not many measurements
beyond 3.6 and 24 \mum) were removed from the list ($\sim$300
objects) or downgraded in confidence level ($\sim$100 objects).  (See,
e.g., Rebull \etal\ 2010 for discussion of similar extragalatic and
very bright sources.)

\label{sec:upgrade}

Similarly, there are sources that need to be added to the list of
candidates. The color cut in the [3.6] vs.\ [3.6]$-$[24] diagram above
summarily rejected sources in the region of this diagram occupied by
SWIRE galaxies, but it is known that YSOs can appear in this portion
of the diagram.  As discussed above (\S\ref{sec:sourcecounts}), the
extragalactic contamination is likely to be low in the NAN region, and
moreover, where there is very high \av, such as in the Gulf of Mexico
cluster (see \S\ref{sec:gulf} below), it is unlikely that sources of
such red color are extragalactic contaminants. Similarly, there are
sources where there is a 24 \mum\ measurement, but the 3.6
\mum\ measurement is missing due to source confusion or reddening or
other similar issues; these objects could not be selected in the [3.6]
vs.\ [3.6]$-$[24] diagram but are unlikely to be
background galaxies. There are several ($\sim$60) sources like this in
the NAN; we have added these sources as grade B or C confidence YSO
candidates and indicated them as such in the Tables below.  

\subsection{Complete list of MIPS-selected YSO candidates}
\label{sec:allysos}

The complete list of 1286 MIPS-selected YSO candidates appears in
Table~\ref{tab:allysos}, along with Spitzer measurements and, if
available, prior identifications and spectral types. 
Table~\ref{tab:allysos2} repeats the entire list, along with their
properties as derived here, including the assigned cluster membership,
if applicable, the final YSO quality flag (grades A/B/C), the YSO
class (see \S\ref{sec:ysoclass}), and which of the above YSO criteria
a given object met. In each case (for each individual criterion),
``yes'' means the object met that given criterion for YSO selection,
``no'' means it did not, and ``no data'' (``\nodata'') means that it
was unable to be assessed for that criterion.  For example, column 5
(abbreviated ``sel 3324'') indicates whether each object was selected
as a YSO candidate in the [3.6] vs.\ [3.6]$-$[24] diagram (``yes''),
or it was not (``no''), or, it was missing a 3.6 or 24 micron
measurement, and thus could not be assessed in the [3.6] vs.\
[3.6]$-$[24] diagram (``\nodata'').  The column definitions are listed
in \S\ref{sec:anciltweaks} above.

\begin{figure*}[tbp]
\epsscale{.9}
\plotone{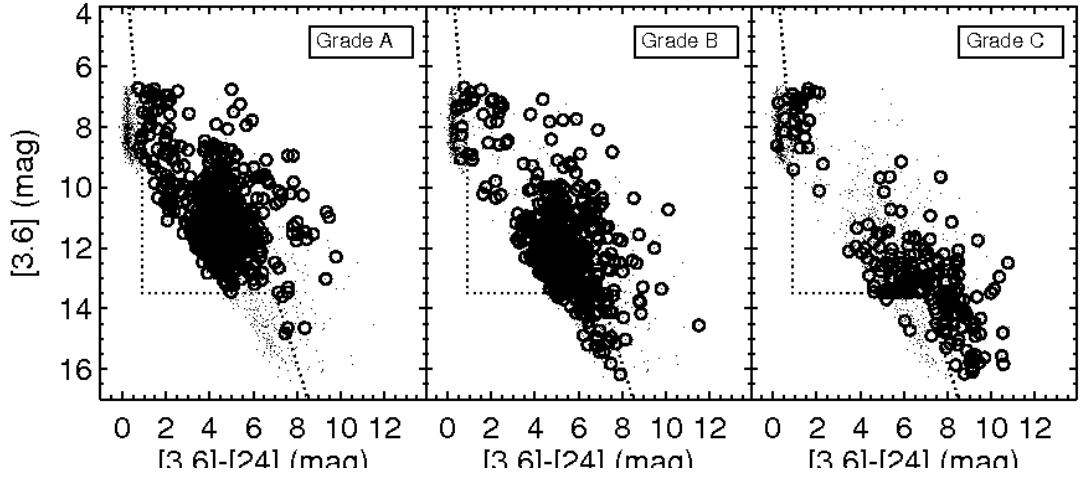}
\caption{$[3.6]$ vs.\ $[3.6]-[24]$ plot, with final grade A, B, and C
YSO candidates indicated. There are 511 grade A,
523 grade B, and 252 grade C sources.  The grade is listed in
Table~\ref{tab:allysos2} for each object.}
\label{fig:3324_final}
\end{figure*}

As in Taurus, based on the criteria above
(\S\ref{sec:anciltweaks}-\S\ref{sec:upgrade}), we have assigned grades
to each source to reflect the confidence in its identification as a
YSO. There are in the end 511 grade A, 523 grade B, and 252 grade C
(with a subset of 23 of those being marked ``C-'').  The highest
confidence YSO candidates generally have at least $\sim$5 or 6 of the
criteria listed above in their favor (yeses in
Table~\ref{tab:allysos2}), and the lowest confidence have $\sim$3. The
cumulative YSO candidate list is plotted in
Figure~\ref{fig:3324_final}, separated by grade. The faintest ones are
overall more likely to be contaminants, the clump of bright grade C
candidates are some of the ones that were manually downgraded (in
\S\ref{sec:downgrade}) due to less significant excesses, and the ones
in the region occupied by SWIRE galaxies are the ones manually
promoted in \S\ref{sec:upgrade}.  For the analysis in the
rest of this paper, all of the candidates are considered as one large
set.

Table~\ref{tab:ysostats} summarizes a variety of statistics on the YSO
sample selected here, that selected by G09 methodology, and the
initial sample of YSOs from the literature.   Note that, because our
MIPS survey is shallower than our IRAC survey for relatively blue
objects, $\sim$40\% of the objects selected using the G09 IRAC
methodology do not have MIPS counterparts! The objects selected by
G09/IRAC but not recovered by our MIPS-based selection are listed in
Appendix~\ref{sec:iraconly}.  The objects found by our MIPS-based
selection but not by G09/IRAC, in most cases, are those with small or
no excesses at IRAC bands but clear excesses at 24 microns (and
sometimes longer wavelengths as well). Previously-identified YSOs from
the literature that were not recovered (via IRAC or MIPS selection) do
not have an infrared excess that we could measure, and they appear in
Appendix~\ref{sec:iraconly} as well.  These objects may or may not be
legitimate NAN members; we cannot determine this from our data.

In certain contexts below, the objects that were IRAC-selected
but not recovered by MIPS, and those previously-identified YSOs not
recovered by either IRAC or MIPS, may be included in the discussion
with the MIPS-selected objects.  These instances are all clearly noted.

\include{monsterheaders}


\begin{deluxetable}{rrl}
\tablecaption{Statistics of YSO candidate sample\label{tab:ysostats}}
\tablewidth{0pt}
\tabletypesize{\small}
\tablehead{
\colhead{item} & \colhead{number} & \colhead{notes}}
\startdata
number previously-known YSOs in database (hereafter ``prev.\ known YSOs'') &  201 \\ 
number prev.\ known YSOs with any IRAC measurement                 &  197 \\ 
number prev.\ known YSOs with IRAC measurements in all 4 bands     &  151 \\ 
number prev.\ known YSOs with [24] measurement                      &  108 & 54\% of prev.\ known YSOs\\ 
number YSO candidates identified by G09 method (hereafter ``G09 YSOs'')    & 1750 & \\ 
number G09 YSOs with [24] measurement                               & 1012 & 58\% of G09 YSOs\\ 
number YSO candidates identified here                              & 1286 \\ 
number prev.\ known YSOs recovered as YSOs by G09                  &   77 \\ 
number prev.\ known YSOs recovered as YSOs here                    &   80 \\ 
number G09 YSOs recovered as YSOs here                             &  954 & 94\% of G09 YSOs with M24\\ 
number prev.\ known YSOs recovered as YSOs by G09 and here         &   70 \\ 
number prev.\ known YSOs not recovered as YSOs by G09 or here      &  114 & possible non-members?\\ 
number G09 YSOs not recovered as YSOs here                         &  796 \\ 
number G09 YSOs with [24] measurements not recovered as YSOs here   &   58 \\ 
\enddata
\end{deluxetable}


\section{Properties of the Ensemble of YSO Candidates}
\label{sec:ensemble}

\subsection{Optical \& Near-IR Properties}

We have $V$\ic\ data for 14\% and $JHK_s$ for $\sim$80\% of our YSO
candidates.   The $V$\ic\ objects tend to be the brighter and/or less
embedded of the YSO candidate sample.  

\begin{figure*}[tbp]
\epsscale{.9}
\plotone{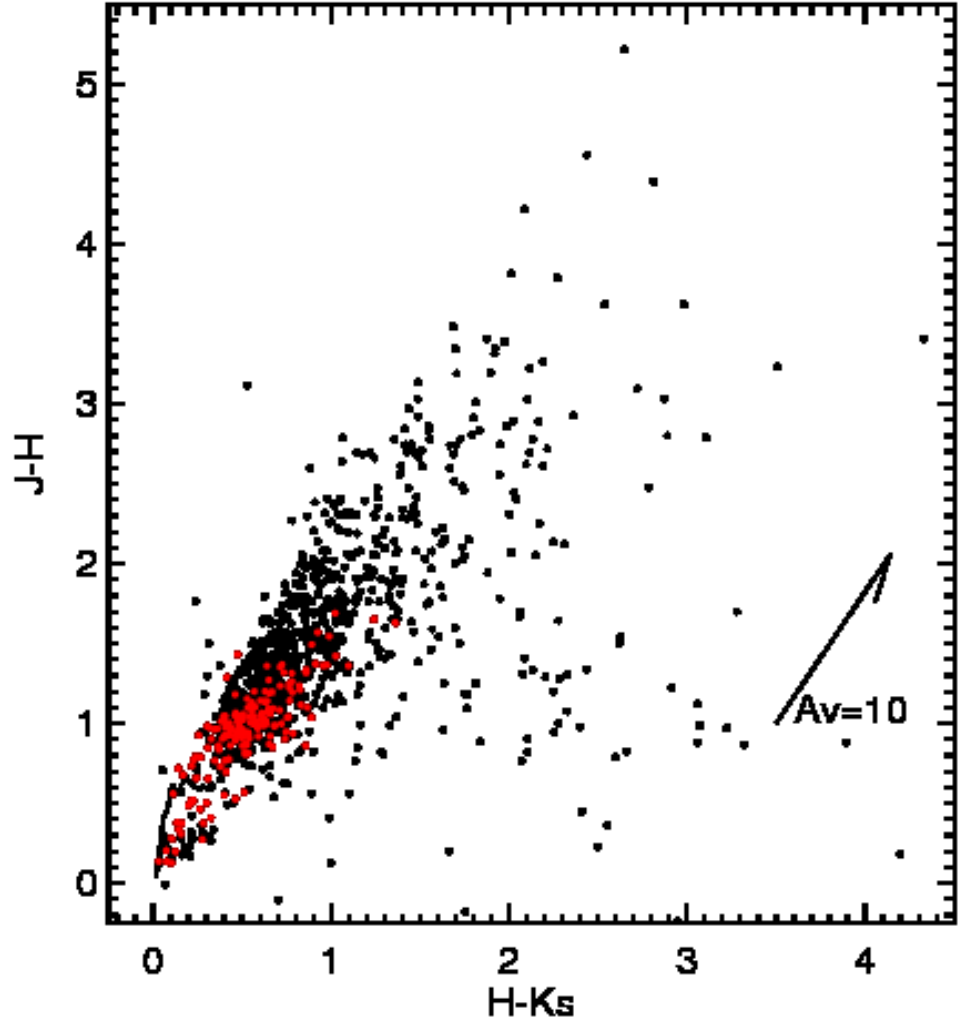}
\caption{$J-H$ vs.\ $H-$\ks\ for the 80\% of our YSO candidates with
such data available. Red points indicate that the objects are also
detected at $V$\ic; these are generally the less-embedded
objects. A reddening vector is indicated, as is the main sequence
locus (line in lower left).}
\label{fig:jhk}
\end{figure*}

Figure~\ref{fig:jhk} shows the $J-H$ vs.\ $H-$\ks\ color-color diagram
for the candidates.  The subset of objects with $V$\ic\ data are
indicated. As suggested by earlier figures, there is generally high
\av\ in the direction of the NAN. \av$\sim$10 to 20 are common. 
Near-infrared evidence for the disks selected by the Spitzer color
criteria are apparent.

\subsection{YSO Classes}
\label{sec:ysoclass}

\begin{figure*}[tbp]
\epsscale{1}
\plotone{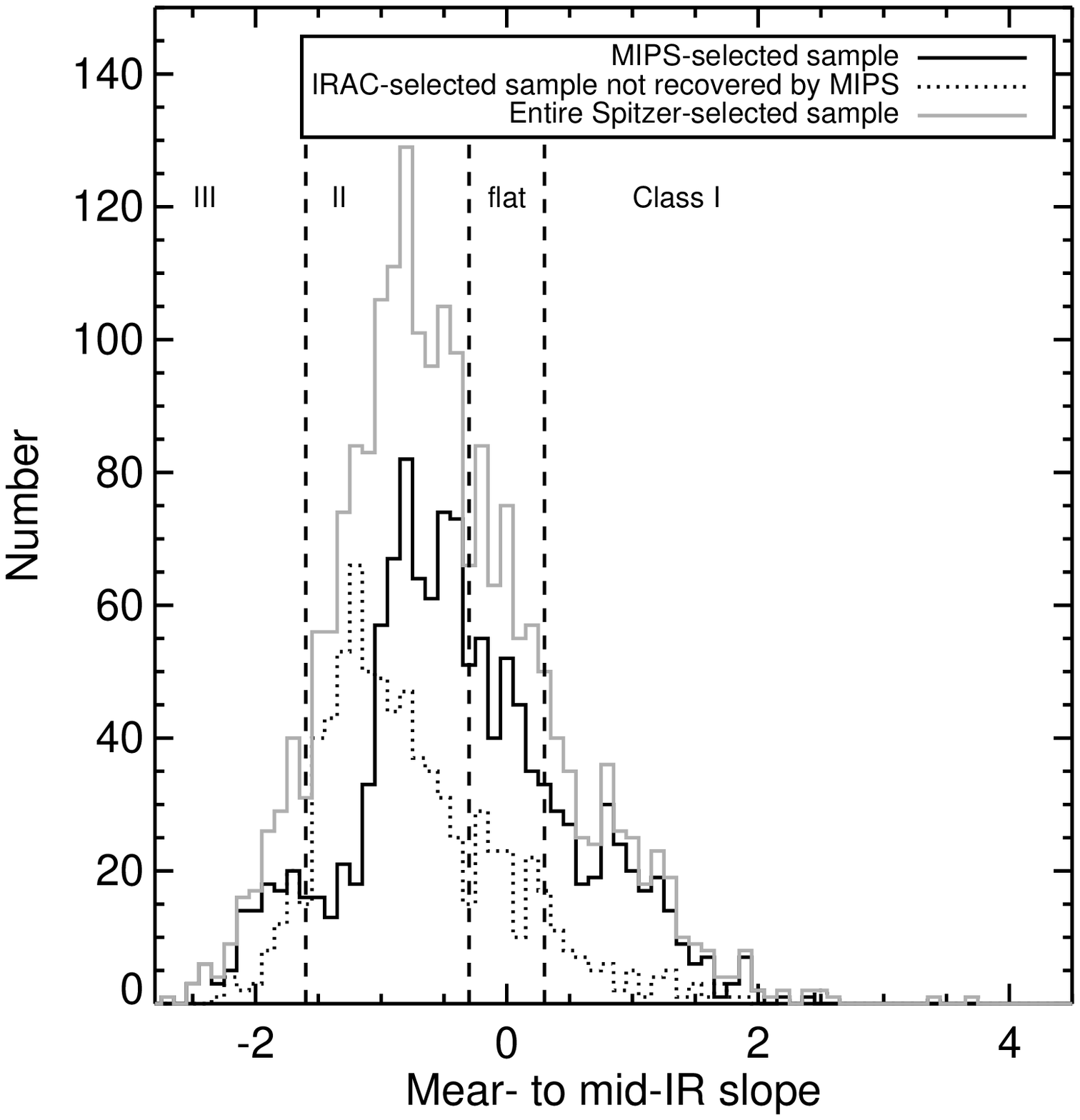}
\caption{Histogram of SED slopes, calculated as described in the text,
for the MIPS-selected sample (solid line) and the IRAC-selected sample
not recovered by MIPS (dotted lines). The entire Spitzer selected
sample (MIPS and IRAC selected) is the grey solid line, e.g., the sum
of the other two histograms. Dashed lines denote the
adopted divisions between Class I, flat, Class II and Class III.  Note that
since we are identifying YSOs by their IR excess, our Class III sample
is defined to be incomplete.  The bulk of our Spitzer-selected sample
of YSO candidates have SED slopes consistent with those of Class II
YSOs.}
\label{fig:slopehisto}
\end{figure*}

In the spirit of Wilking \etal\ (2001), we define the near- to mid-IR
slope of the SED, $\alpha = d \log \lambda F_{\lambda}/d \log 
\lambda$,  where  $\alpha > 0.3$ for a Class I, 0.3 to $-$0.3 for a
flat-spectrum  source, $-$0.3 to $-$1.6 for a Class II, and $<-$1.6
for a Class III.  For each of the YSO candidate objects in our sample,
we performed a simple ordinary least squares linear fit to all
available photometry (just detections, not including upper or lower
limits) between 2 and 24 $\mu$m, inclusive.  Note that errors on the
infrared points are so small as to not affect the fitted SED slope. 
The precise definition of $\alpha$ can vary, resulting in different
classifications for certain objects. Classification via this method is
provided specifically to enable comparison within this paper via
internally consistent means. Note that the formal classification puts
no lower limit on the colors of Class III objects (thereby including
those with SEDs resembling bare stellar photospheres, and allowing for
other criteria to define youth).  By searching for IR excesses, we are
incomplete in our sample of Class III objects.

Figure~\ref{fig:slopehisto} shows a histogram of the SED slopes,
calculated as described, for the MIPS-selected sample, the
IRAC-selected sample not recovered by the MIPS-based search, and  the
complete Spitzer-selected sample (IRAC-selected plus MIPS-selected).
The objects not recovered by MIPS in general have more negative
slopes, because the objects that are not detected by MIPS are the ones
that are the faintest in the MIPS bands, e.g., the ones whose SEDs are
falling through the IRAC bands.  The IRAC-selected sources with
steeply rising SEDs (suggesting that they should have been detected at
24 \mum) that are not recovered by our MIPS selection are either off
the edge of the MIPS map or saturated at MIPS bands.  

We repeated the spectral index analysis with slopes fit to just
available points between 2 and 8 microns (e.g., ignoring any points at
24 \mum\ and longer). The peaks of the histograms of the MIPS- and
IRAC-selected samples move closer together, but the MIPS-selected peak
is still to the right of the IRAC-selected peak, and the relative
fractions of sources in the Class bins do not change.

The MIPS-selected sample is nearly half Class IIs (48\%).  Including
the IRAC-selected sample as well, the fraction of Class IIs among the
selected IR-excess objects rises to 56\%. For either sample, there are
about 8\% Class IIIs, and the remaining objects are evenly split
between Class Is and flat objects (with an additional 2\% where there
are too few points to calculate a slope). The relative numbers are
quite comparable to the fraction of such objects found in Serpens by
Harvey \etal\ (2007b) or Perseus (Rebull \etal\ 2007), suggesting that
these regions are all comparable in the status of the circumstellar
material and hence possibly their age.

\subsection{Projected Location of the YSO Candidates}

\begin{figure*}[tbp]
\epsscale{1}
\plotone{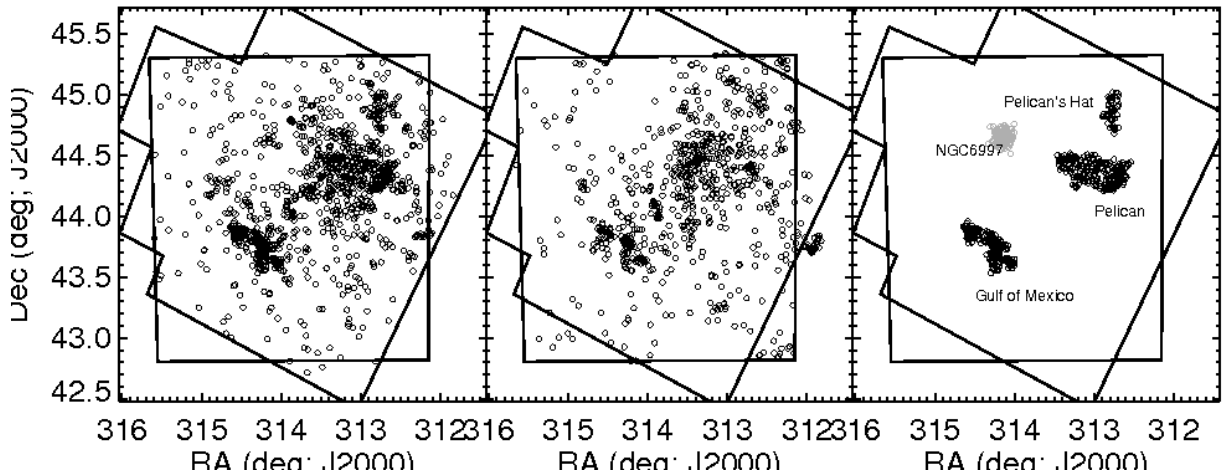}
\caption{Projected locations of (left) entire set of MIPS-selected YSO
candidates, (middle) set of IRAC-selected YSOs not recovered by the
MIPS-based selection, and (right) clusters as discovered in the
MIPS-based distribution discussed further below. In all
three panels, the polygons from Fig.~\ref{fig:where} are reproduced,
indicating the approximate IRAC and MIPS coverage; note also that in
all three panels, only hollow circles are used as symbols -- black
portions come from overlapping symbols. In the right-most panel, NGC
6997 is represented in grey because it is actually a background (or
possibly foreground) object, it is defined in the literature (as
opposed to defined on the basis of YSO candidates here), and primarily
consists of non-IR excess sources. It is included here for reference;
see discussion in  \S\ref{sec:ngc6997}.}
\label{fig:whereyso}
\end{figure*}

Figure~\ref{fig:whereyso} shows the projected locations of all of the
YSO candidate sources selected using our MIPS-based selection. 
Several clusters of objects can be seen by eye here; we will consider
clustering formally in the next section. There is also a distributed
population.  The fact that there is a substantial distributed
population here is consistent with results from many other Spitzer
studies of star-forming regions, where substantial numbers of YSO
candidates are found far from the canonical centers of star formation.

In Figure~\ref{fig:24mosaic}, we saw large spatial variation in the 24
\mum\ surface brightness due to the ISM, which renders the limiting
flux density a function of position in the image.  Similar structures
can be seen at the IRAC bands (see G09).  It is difficult to assess if
this has affected the apparent distribution of the selected YSO
candidates.  The distribution of YSOs is not uniform, and it is
expected based on physical arguments that the distribution of YSOs
will be coupled to the ISM, especially at these young ages.  We
examined the distribution of YSO candidates (and all 24 \mum\ sources)
by plotting them on top of the 24 \mum\ image. We do not see a very
strong effect of the ISM brightness on our YSO candidate
distribution.  Qualitatively, we note that the bright ``streaks'' in
the center of the map have few YSO candidates, and also fewer 24 \mum\
sources in general; the even brighter ISM in the region of the Pelican
cluster hosts many YSO candidates.  We conclude that the ISM surface
brightness is not a dominant effect in the structure seen in the YSO
candidate distribution.  However, many of the YSO candidates are
projected against the regions of highest extinction (which correlates
well with the MIPS surface brightness), with concentrations near the
Gulf of Mexico and the greater Pelican region,

As discussed above, there are a significant number of IRAC-selected
YSO candidates that are not recovered by our MIPS-based selection, and
these appear in the second panel of Figure~\ref{fig:whereyso}. 
Clustering similar to the left panel can be seen by eye, even in this
residual IRAC-only selection.

\section{Clusters of YSO Candidates Within the NAN Complex}
\label{sec:clusters}

\subsection{Defining the clusters}
\label{sec:clusterdefinition}

G09 discussed in some detail a formal definition of clustering, using
a Kernel method to estimate the density of YSO candidates (NB: not the
density of sources in the entire catalog) as a function of
$\alpha,\delta$. Using the same formalism, we computed a YSO density
map of our final set of MIPS-selected YSO candidates.  There are three
main clusters that can be identified out of this sample.  These three
clusters appear in the rightmost panel of Figure~\ref{fig:whereyso}. 
(A background cluster is also indicated in that figure for
reference; see discussion in \S\ref{sec:ngc6997}.)

\begin{figure*}[tbp]
\epsscale{.9}
\plotone{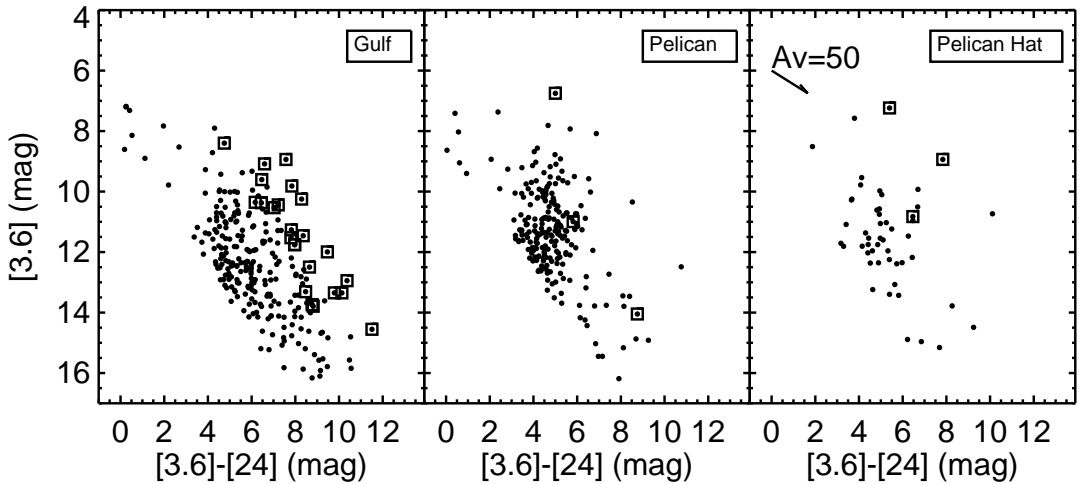}
\caption{[3.6] vs.\ $[3.6]-[24]$ plots for YSO candidates in three
clusters located in the NAN complex; compare to
Fig.~\ref{fig:3324_3panel} or \ref{fig:3324_4panel}.  An additional
box around a point indicates detection at 70 \mum. The Pelican's Hat
cluster is the least populous.  There is a wide range of colors found
in each of these clusters. The distribution of objects in
the Pelican cluster is more clumped than that from the Gulf of Mexico,
perhaps due to the more substantial reddening in the Gulf.}
\label{fig:3324clusters}
\end{figure*}

\begin{figure*}[tbp]
\epsscale{.9}
\plotone{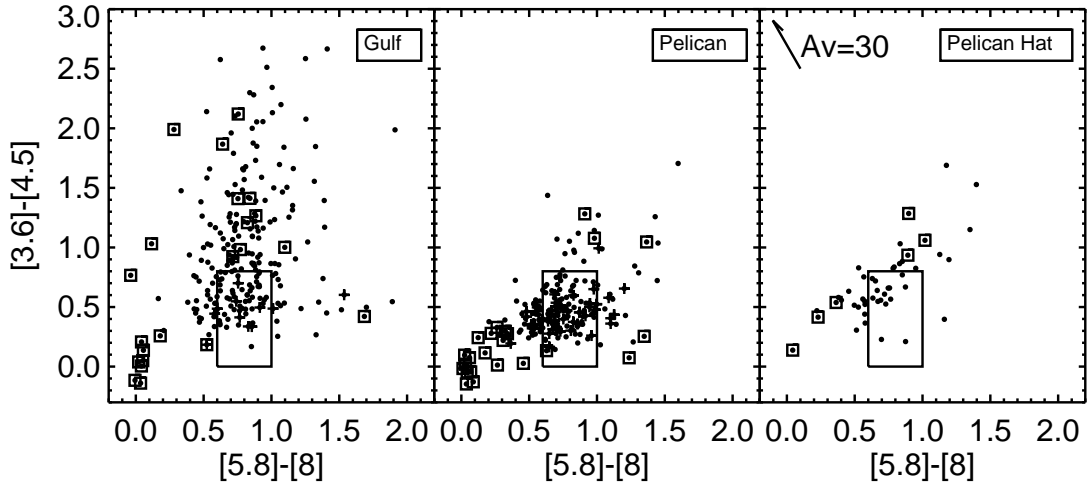}
\caption{IRAC color-color diagram for MIPS-selected YSO candidates in
three clusters located in the NAN complex.  Objects previously
identified as YSOs in the literature and recovered here are indicated
by a $+$ sign; an additional box around a point denotes objects
identified here but not in G09. Note that YSO candidates identified in
G09 but not here do not appear in these plots.  The large box comes
from Allen \etal\ (2004) and indicates the unreddened location of
likely Class II objects.   As in Fig.~\ref{fig:3324clusters}, there is
a wide range of colors of YSO candidates, and the Gulf of Mexico
cluster seems to be the most reddened. }
\label{fig:iracclusters}
\end{figure*}

We name these three clusters based on their position with respect to
the nebular features -- the Gulf of Mexico cluster is located in the
Gulf of Mexico region of the North American Nebula; the Pelican
cluster overlaps with the Pelican's neck, head, and beak; and the
third cluster appears above the Pelican's head, hence ``Pelican's
Hat.'' We discuss the properties of these clusters in turn below.
Cluster membership is indicated for individual objects in
Table~\ref{tab:allysos2}.  Ensemble properties are summarized in
Table~\ref{tab:clusterstatistics}; plots of [3.6] vs.\ [3.6]$-$[24]
are in Figure~\ref{fig:3324clusters} and plots of [3.6]$-$[4.5] vs.\
[5.8]$-$[8] are in Figure~\ref{fig:iracclusters}.  

Because clustering is considered in the selection of YSO candidates
above (\S\ref{sec:anciltweaks}), and because some sources were added
by hand in regions of high YSO density (see \S\ref{sec:upgrade}),
identifying and defining clusters is somewhat of a recursive problem,
with the potential for biased results.   We also ran the cluster
identification algorithm on the initial and least biased sample of
[3.6] vs.\ [3.6]$-$[24] selected objects (\S\ref{sec:3324}).  The
three clusters were unambiguously recovered even in this sample, so we
conclude that they are independent of the details of our YSO selection
algorithm.  (We note that heavily reddened K and M giants
located behind a very dense dust cloud could appear to be clustered
YSOs; see, e.g., Strai\v{z}ys \& Laugalys 2008a.)

In the left and middle panels of Figure~\ref{fig:whereyso}, there are
additional smaller clumpings of sources to which one's eye is drawn,
most notably a clump on the far West of the distribution.  It has a
linear string of YSOs in the MIPS-selected sample and a slightly
offset clump of sources on the edge of the IRAC map from the
IRAC-selected YSOs. Neither of these groupings appear 
in the clustering analysis above.  

There are legitimate YSOs here outside of the clusters, notably the
outbursting Class I object PTF10nvg (Covey \etal\ 2011), which is
South of the Pelican cluster.

Parts of two of the clusters described here (the Gulf of Mexico and
the Pelican) have been previously identified in the literature (based
primarily but not exclusively on H$\alpha$);  our data suggest
expanded spatial definitions (and far more members) for these
clusters.  The Pelican's Hat cluster is an entirely new discovery
based on the MIPS data.  In the Appendix
(\S\ref{sec:appendixclusters}), we discuss additional clusters from
the literature that are sufficiently unremarkable in the MIPS maps
that they are mentioned here solely for completeness.

The SEDs presented below are all in log $\lambda F_{\lambda}$ in cgs
units (erg s$^{-1}$ cm$^{-2}$), against log $\lambda$ in microns. 

\begin{deluxetable}{lrrrr}
\tablecaption{Statistics of MIPS-selected YSO candidate detections
in clusters   
\label{tab:clusterstatistics}}
\tablewidth{0pt}
\tablehead{
\colhead{item}  & \colhead{Gulf of Mexico} &
\colhead{Pelican}  & \colhead{Pelican's Hat}  }
\startdata
24 \mum\ &                         283  &   247  &   51  \\ 
70 \mum\ &                          29  &     3  &    3  &   \\ 
24 \mum\ \& 70 \mum\ &              29  &     3  &    3  &   \\ 
24 \mum\ \& ANY IRAC band       &  269  &   233  &   51  &  \\ 
70 \mum\ \& ANY IRAC band       &   26  &     3  &    3  &   \\ 
24 \mum\ \& ALL IRAC bands      &  240  &   216  &   48  &   \\ 
70 \mum\ \& ALL IRAC bands      &   22  &     2  &    3  &   \\ 
24 \mum\ \& ANY 2MASS band      &  162  &   210  &   43  &  \\ 
24 \mum\ \& ANY IPHAS band      &  36   &   143  &    5  &        \\
24 \mum\ \& $V$ or $I_c$ or $r^{\prime}$ or $i^{\prime}$ &  36 & 148 & 5 & \\ 
approximate area (sq.\ deg.)    &  0.105   & 0.135  & 0.051 &   \\
\hline
class I sources  & 109 (39\%) & 31 (13\%) & 15 (29\%) \\
flat sources     & 91 (32\%)  & 33 (13\%) & 20 (39\%) \\
class II sources & 68 (24\%)  & 157 (64\%) & 16  (31\%) \\
class III sources (incomplete) & 4 (1\%) & 12 (5\%) & 0 (0\%) \\
insufficient data for class & 11 (4\%) & 14 (6\%) & 0 (0\%) \\
\enddata
\end{deluxetable}

\clearpage

\subsection{Gulf of Mexico}
\label{sec:gulf}

\begin{figure*}[tbp]
\epsscale{1.0}
\plotone{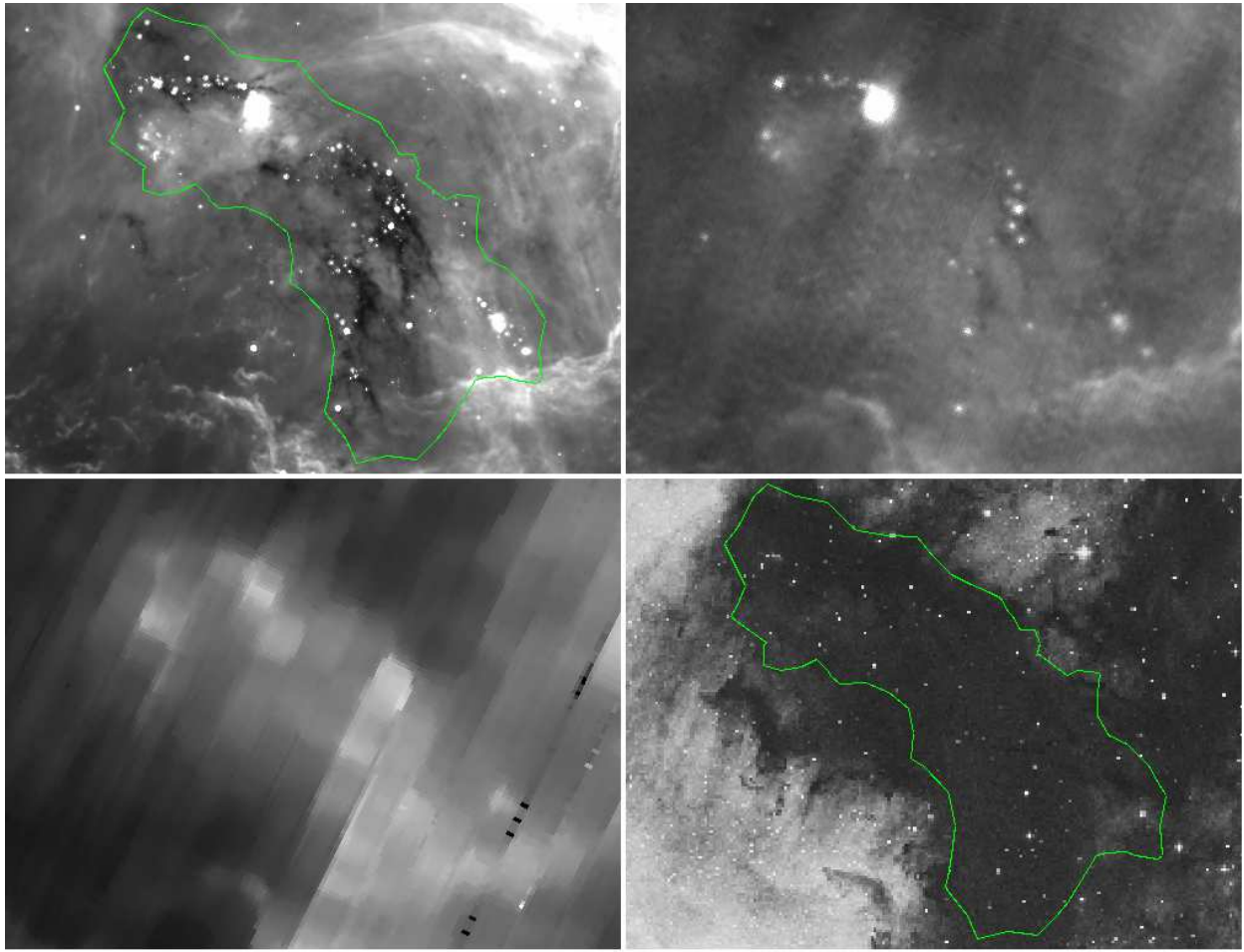}
\caption{Mosaic of Gulf of Mexico at 24 \mum\ (top left), 70 \mum\
(top right), 160 \mum\ (lower left), and, for comparison, POSS-red
(lower right).  The greyscale colors correspond to a
histogram-equalized stretch of surface brightnesses in each case. The
contour (plus YSO candidacy) defines cluster membership, and is
essentially the \av=10 contour from the Cambr\'esy \etal\ (2002)
extinction map. }
\label{fig:gulfimg}
\end{figure*}

The ``Gulf of Mexico'' region is the area of highest extinction in the
NAN complex; the Cambr\'esy \etal\ (2002) extinction map peaks at
\av$\sim$30, and is likely a lower estimate to the true extinction due
to the fact that it was calculated using near-IR data and is lower spatial
resolution than our MIPS data.  Herbig (1958) identified several stars
bright in H$\alpha$ at the northernmost part of the cluster we
describe below. Cambr\'esy \etal\ (2002) find three subclusters within
this cluster using 2MASS star counts (in their notation, 1, 2, and
3a/b); G09 also finds 3 subclusters within this cluster using IRAC.

As can be seen in Figure~\ref{fig:gulfimg}, the image of this region
is striking at 24 \mum\ (and 8 \mum; not shown), with infrared
emission and a filamentary dark cloud.  Parts of
the cloud are dark at 24 \mum\ and even at 70 \mum; this region has
the highest surface density of 70 \mum\ point sources in the complex. 
Counterparts at 160 \mum\ can be seen as well, though no viable
photometry can be extracted from our 160 \mum\ data.  
Many jets are visible at the IRAC
bands, especially at the northern end of the cluster.

The cluster membership is defined as discussed above, and is
represented as a contour in Figure~\ref{fig:gulfimg}; that contour is
very close to the \av=10 contour from the Cambr\'esy \etal\ (2002)
extinction map. Using the G09 definition of candidate (IRAC) YSOs and
clustering metrics, this contour is similar to that for 1500
(IRAC-selected) YSO candidates per square degree.  It is about
30$\arcmin$ across at its widest part, corresponding to about 4.5 pc
at a distance of 520 pc.  There are 283 objects seen at 24 \mum\ that
we define as being part of the cluster (see
Table~\ref{tab:clusterstatistics}), which is $\sim$2500 per square
degree.  Note that some objects not explicitly selected via the color
mechanisms in \S\ref{sec:pickysos} above (such as those objects seen
only at 24 \mum) have been added to the YSO candidate list on the
basis of their projected location in this cluster (and 24 \mum\ flux
density). At 70 \mum, 10\% of the 24 \mum\ sample have counterparts.
Most of the objects, 85\%, are seen at all four IRAC bands, but 5\%
are so embedded that they lack any IRAC counterparts at all.

Figures~\ref{fig:3324clusters} and \ref{fig:iracclusters} show the
[3.6] vs.\ [3.6]$-$[24] and the IRAC color-color diagram for this
cluster in context with the other clusters discussed here.  The Gulf
cluster has by far the most 70 \mum\ detections (see
Table~\ref{tab:clusterstatistics}), and the widest range of colors of
YSO candidates in both these diagrams. While several objects in the
Gulf have extreme IR colors, a major contributing factor to the
dispersion in colors is the high \av.

Our primary YSO candidate selection mechanism above  requires MIPS,
and we defined clusters based on the location of those MIPS-selected
YSO candidates. But, for relatively blue sources, the IRAC survey is
more sensitive than the MIPS survey, and there are most likely
legitimate Gulf cluster members not detected at MIPS bands. There are
$\sim$4100 IRAC-1 and IRAC-2 sources without MIPS counterparts within
the \av=10 contour ($\sim$40,000 per square degree (compared to an
average of $\sim$60,000 per square degree over the whole map).  Just
$\sim$800 of the objects in this region are detected at all four IRAC
bands ($\sim$8000 per square degree). Of those objects, just 92
IRAC-selected YSO candidates (requiring selection in all four IRAC
bands) are unrecovered by our MIPS-based selection, and could be
legitimate cluster members. There are additional likely YSOs that we
list as YSO candidates based on 5.8 or 8 through 70 \mum\ properties,
but they do not have detections at all four IRAC bands.  Such sources
without 3.6 or 4.5 \mum\ detections would not be selected via the G09
mechanism.  The 92 IRAC-selected YSO candidates unrecovered by our
MIPS-based selection is the minimum number of additional cluster
members.  Adding those 92 to the 283 we define (based on MIPS) as the
cluster membership, we derive a YSO census of 375 members and a
surface density of $\sim$3500 per square degree. 


To constrain the maximum number of true cluster members, we can look
at overall source counts.  The number per unit area of 4-band IRAC
detections (without MIPS detections) on the edges of the map, in
regions of much lower \av, can be taken as roughly the number of
background+foreground counts; these are $\sim$11,000 per square
degree. Considering those objects with IRAC-1 and 2 detections (but no
MIPS-24), there are $\sim$70,000 sources per square degree on the map
edges.  It is thus unlikely that {\em all} projected IRAC-only sources
could be cluster members, but the IRAC-selected YSO candidates are far
more likely to be members than any given IRAC-only source.  Additional
observations and analysis are needed to distinguish the true cluster
members from the foreground stars.

Of 375 cluster members, just eleven were identified in the literature
prior to our Spitzer observations as cluster members (with 303 
identified in G09).  All of the objects known prior to Spitzer are in
the northernmost $\sim6\arcmin$ of the cluster, and eight of them are
in the northermost $\sim1.5\arcmin$ of the cluster, so the size of the
cluster is redefined as about 10-100 times larger (in projected area)
than previously realized. 

\begin{figure*}[tbp]
\epsscale{1}
\plotone{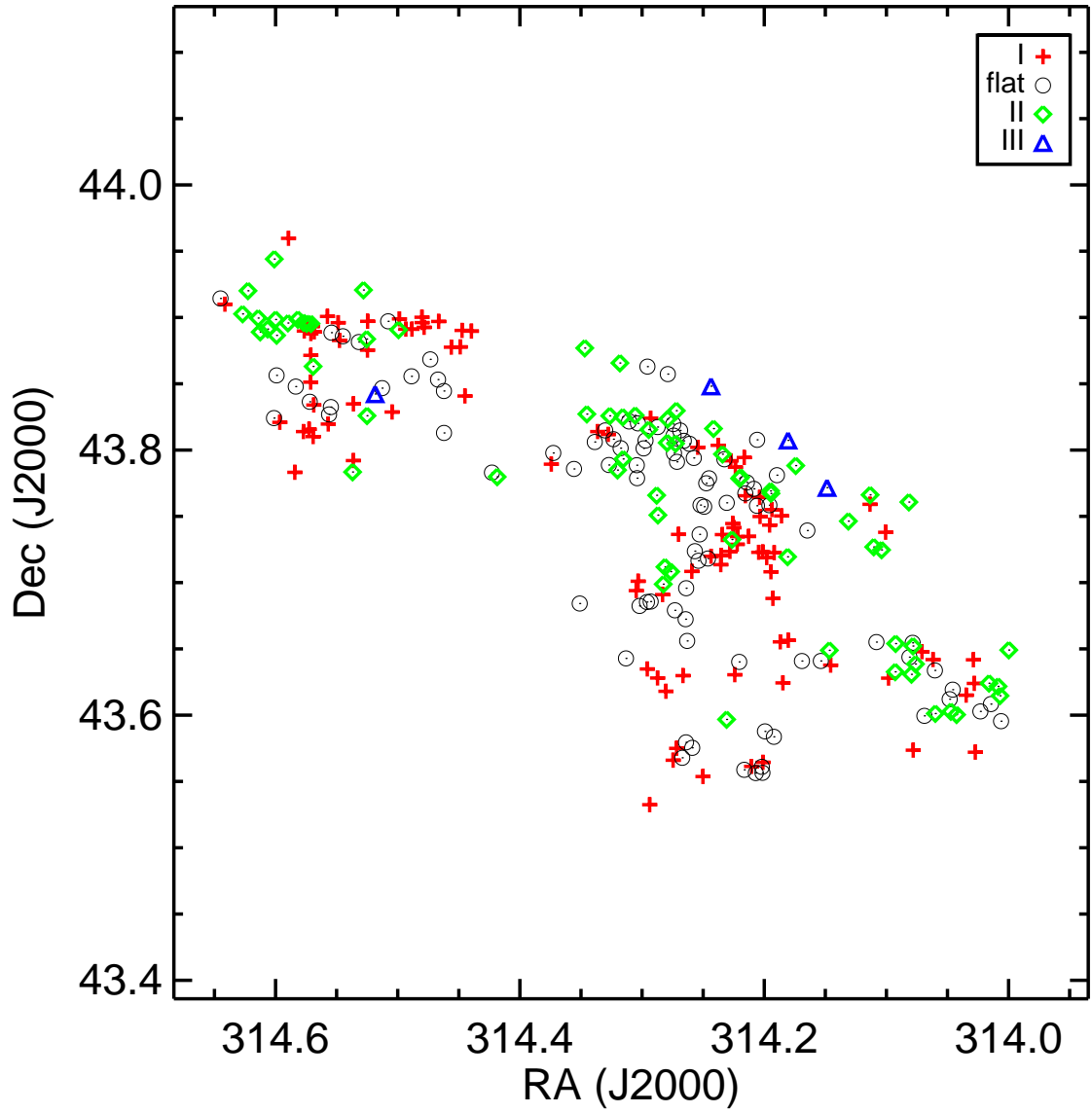}
\caption{
Location of YSOs, color-coded by their SED slope, in the Gulf
of Mexico.  The Class I and flat objects are more clustered than the
Class II and Class III objects. }
\label{fig:gulfspace}
\end{figure*}

Figure~\ref{fig:gulfspace} shows the projected spatial location of the
MIPS-selected objects, coded by their SED slope.  The Class I and flat
objects are more clustered than the Class II and Class III objects.  
Table~\ref{tab:clusterstatistics} lists the relative fractions of
Class I, flat, II, and III MIPS-selected YSO candidates in the
cluster. If the candidates found using IRAC only are included, the
relative fractions are 32, 30, 34, and 2\%, respectively (with 3\% for
which no slope can be calculated); the largest increase is in the
Class II category, consistent with Figure~\ref{fig:slopehisto}.  The
dense embedded cluster found in Serpens South (Gutermuth \etal\ 2008)
bears some morphological similarity to this region.  The mechanism
used by those authors to select YSOs is very different than what we
use, but they find about 60\% Class I+flat objects. In the Gulf, we
find that about 62\% of all the YSO candidates are Class I or flat
(71\% of just the MIPS-selected sample).  Within the NAN complex, the
Gulf has the largest fraction of Class I objects.  The overwhelming
majority of the NAN objects seen in the [24] vs.\ [24]$-$[70] diagram
(Fig.~\ref{fig:2470}) are located in the Gulf. We identify this region
in the NAN complex to have the most embedded objects (presumably the
youngest?), the highest surface density of YSO candidates, and to be
where star formation is actively occurring.

\begin{figure*}[tbp]
\epsscale{0.8}
\plotone{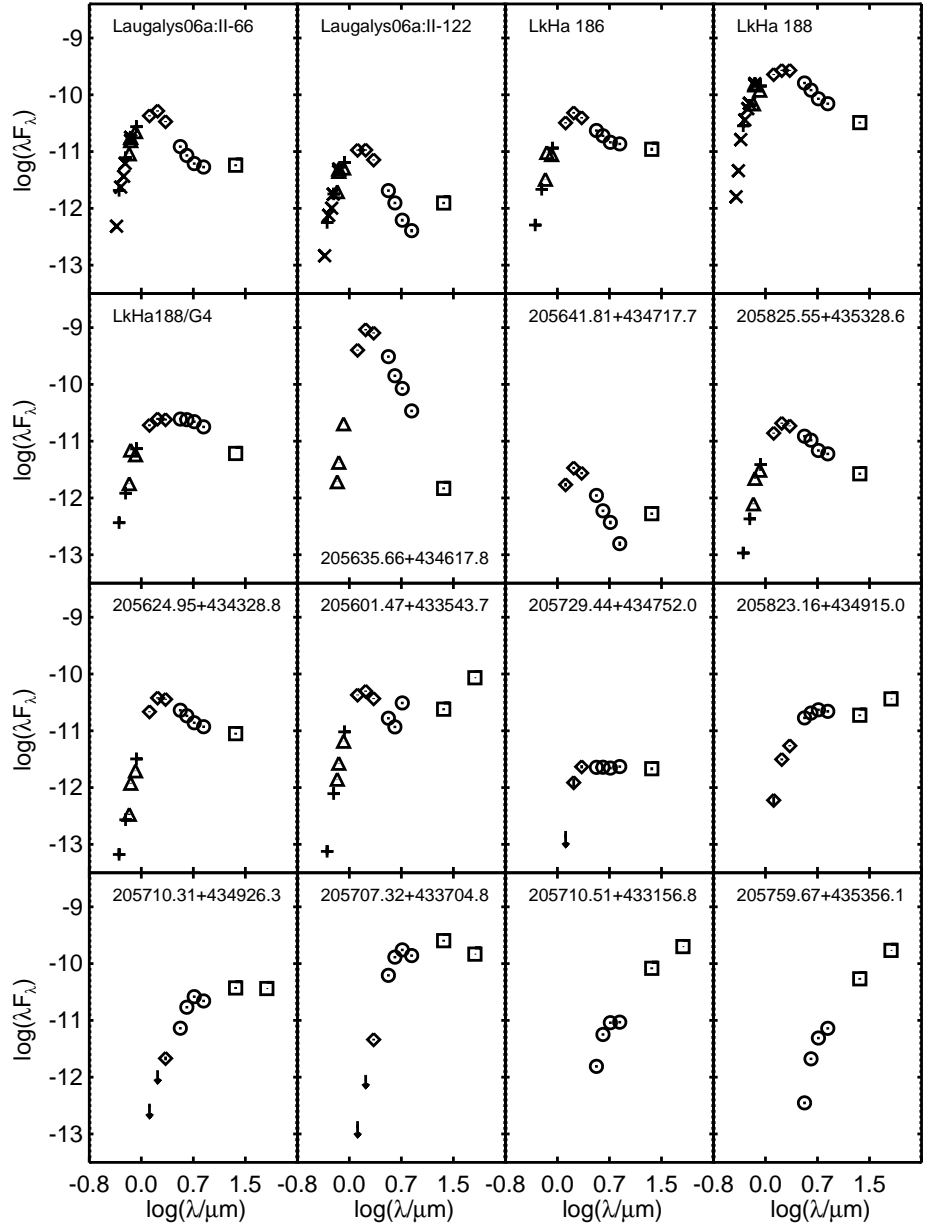}
\caption{
Sample of selected YSO candidate SEDs seen in the Gulf of Mexico
cluster;  $\lambda F_{\lambda}$ is in erg s$^{-1}$ cm$^{-2}$, and
$\lambda$ is in microns.  Previously-identified YSOs are denoted by
their previous name; new objects have IAU-compatible names.  Vertical
error bars are typically smaller than the points. Symbols
denote origin of photometry: $\times$={\em Vilnius} system,
triangles=IPHAS, $+$= Johnson (from the literature), diamonds=2MASS,
circles=IRAC, squares=MIPS; downward-pointing arrows are upper
limits.  Note that there is a wide range of SED types, from
essentially photospheric (e.g., 205635.66+434617.8) to deeply embedded
(e.g., 205759.67+435356.1).}
\label{fig:gulfseds}
\end{figure*}

Figure~\ref{fig:gulfseds} shows a sample of 16 SEDs selected out of
our proposed cluster membership to represent the range of colors and
SED types found in this cluster.  One object in this Figure is
essentially photospheric at all available bands (which may mean that
it is a foreground object superimposed on the dense cluster region and
erroneously included as a cluster member solely based on its location
on the sky), while others in this figure are deeply embedded,
appearing only at IRAC+MIPS, not 2MASS or optical.  Notably, the
object (PTF10qpf) recently identified by Miller \etal\ (2011) as an FU
Ori-like outburst was present in our pre-outburst observations of this
cluster.

\subsection{Pelican Cluster}
\label{sec:pelican}

\begin{figure*}[tbp]
\epsscale{1.0}
\plotone{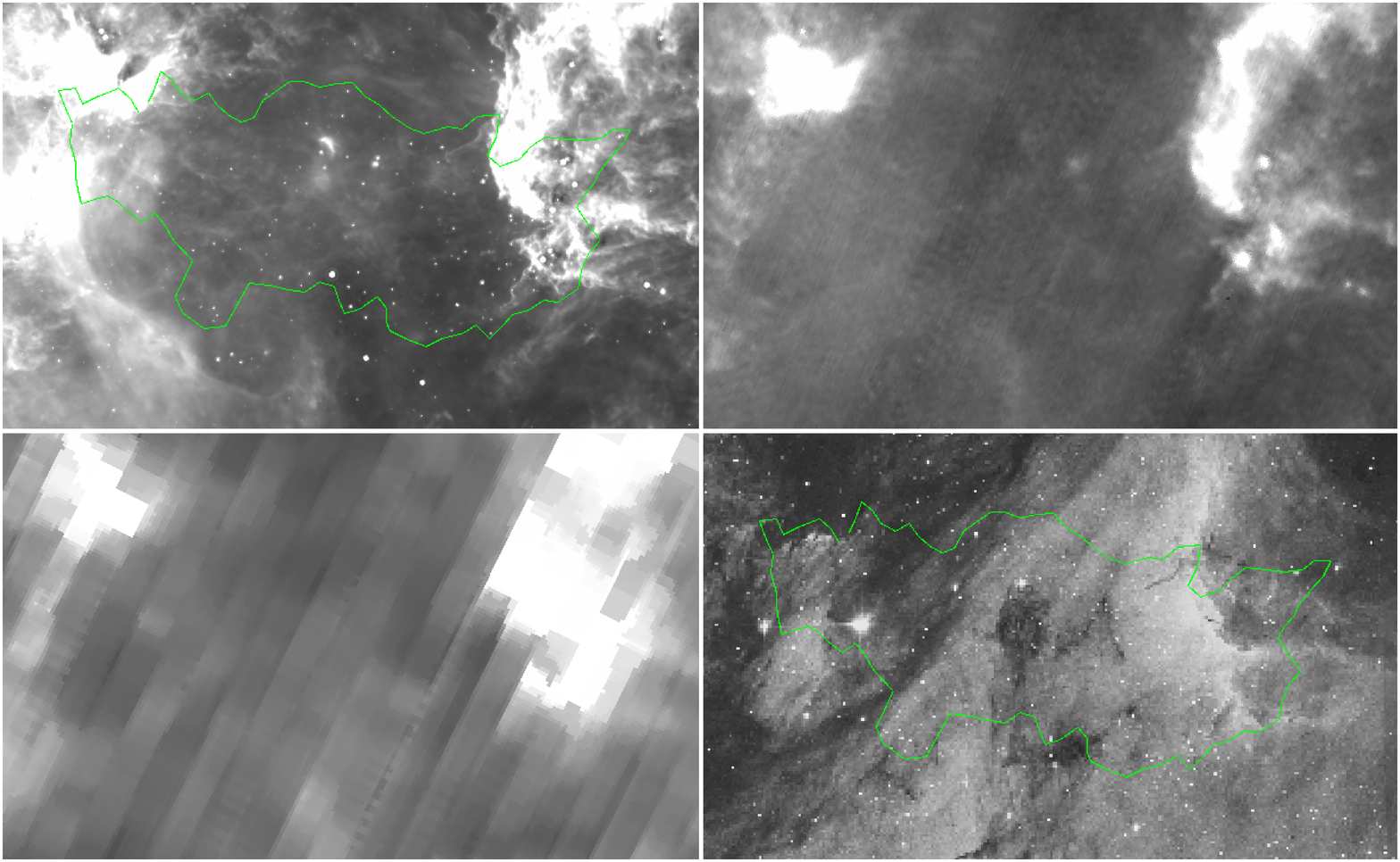}
\caption{Mosaic of the Pelican Cluster at 24 \mum\ (top left), 70
\mum\ (top right), 160 \mum\ (lower left), and, for comparison,
POSS-red (lower right).  The greyscale colors correspond to a
histogram-equalized stretch of surface brightnesses in each case. The
contour (plus YSO candidacy) defines cluster membership.  This contour
encloses two \av$\sim$20 peaks (the brightest regions in the 24 \mum\
map). }
\label{fig:pelimg}
\end{figure*}

\begin{figure*}[tbp]
\epsscale{1.0}
\plotone{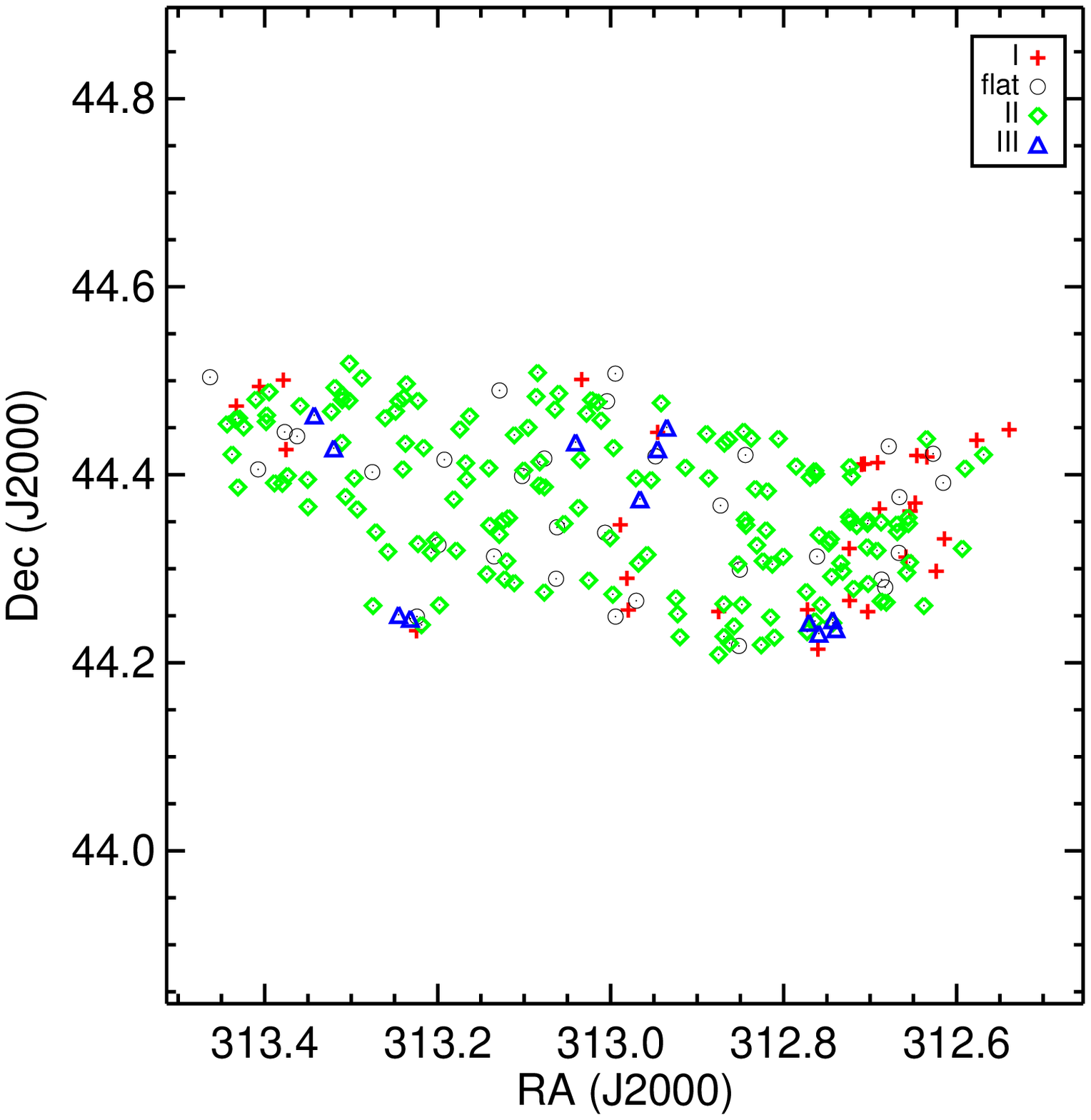}
\caption{Location of YSOs, color-coded by their SED slope, in the
Pelican cluster.  The Class I and flat objects are more clustered and
located closer to the \av\ peaks than the Class II and Class III
objects.}
\label{fig:pelspace}
\end{figure*}

Herbig (1958) identified a cluster of 12 young objects bright in
H$\alpha$ near IC 5070 (the Pelican Nebula).  This cluster was denser
than the other clusters Herbig identified in the NAN complex, leading
him to conclude that this region hosted the most active region of star
formation in the complex.   While all of these objects are seen at
2MASS and IRAC bands, none are detected at 24 \mum, and none have
[3.6]$-$[24] excesses; moreover, none are even identified by G09 as
IRAC-selected YSOs.  There is, however, still a noticable clumping of
Spitzer-identified YSO candidates in this general region; see
Figure~\ref{fig:whereyso}.

Figure~\ref{fig:pelimg} shows the Pelican Cluster region in 24, 70,
and 160 \mum\ as well as a POSS-red image.  There are many point sources
at 24 \mum; some sources can be seen in the 70 \mum\ image, but just
three can be identified using our pipeline -- most are slightly
extended or in regions of extremely bright background.   No point
sources are discernible at 160 \mum.

The contour defining the cluster membership is shown in
Figure~\ref{fig:pelimg} as well.  Interestingly, unlike the Gulf
cluster in \S\ref{sec:gulf} above or in the Pelican's Hat cluster in
\S\ref{sec:pelhat} below, the definition of the cluster as returned by
the formal clustering metrics does not trace an \av\ contour.  Rather,
it links two \av$\sim$20 peaks, seen in the 24 \mum\ image as the
brightest surface brightness regions.  It also overlaps with more than
half the projected area of two clusters from Cambr\'esy \etal\ (2002),
clusters \#5 and 6, though the regions specifically called out by
Cambr\'esy \etal\ often include regions too bright and nebulous for
many MIPS-24 point sources to have been recovered.  Similarly, the G09
clusters 5 and 6 are both located within our MIPS-based cluster
contour.

Table~\ref{tab:clusterstatistics} lists characteristics of this
cluster as we now define it.  Forty-eight of the 247 cluster members
are previously identified as YSOs in the literature, so we have
expanded the likely cluster membership by a factor of $\sim$5.  Herbig
(1958) identified 13 more H$\alpha$ sources in this region, but did
not explictly call them out as being part of the IC 5070 sources he
identified as a cluster; Ogura \etal\ (2002) also found several
H$\alpha$ sources here.  In contrast to the twelve original Herbig
(1958) IC 5070 sources, we detect these Ogura H$\alpha$ sources at 24
\mum.    The fact that nearly 20\% of the cluster membership has
already been identified in the literature as young stars is consistent
with there being overall lower \av\ towards this cluster than the Gulf
of Mexico and Pelican's Hat clusters identified in this paper.  Unlike
the Gulf of Mexico cluster, we did not add manually additional objects
to the cluster inventory on the basis of 24 \mum\ flux density plus
clustering.  Because the average \av\ is lower here than in the Gulf,
the risk of contamination is higher.  However, a high fraction of the
IRAC-selected (but not MIPS-recovered) sources are likely to be
cluster members, as evidenced by the clustering found in G09 (or, for
that matter, Cambr\'esy \etal\ 2002).  

\begin{figure*}[tbp]
\epsscale{0.8}
\plotone{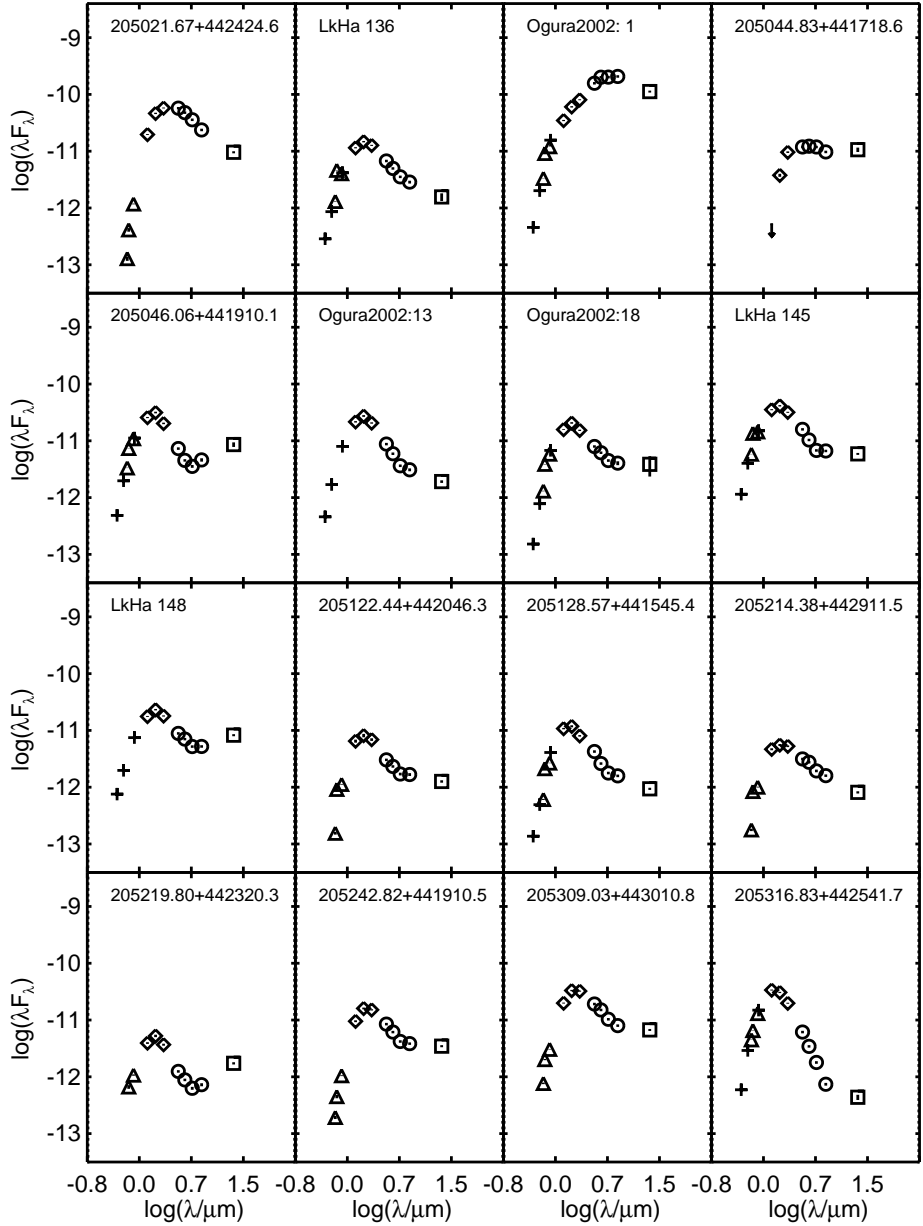}
\caption{
Sample of selected YSO candidate SEDs seen in the Pelican cluster; $\lambda
F_{\lambda}$ is in erg s$^{-1}$ cm$^{-2}$, and $\lambda$ is in
microns. Previously-identified YSOs are denoted by their previous
name; new objects have IAU-compatible names.  Vertical
error bars are typically smaller than the points. Symbols
denote origin of photometry: $\times$={\em Vilnius} system,
triangles=IPHAS, $+$= Johnson (from the literature), diamonds=2MASS,
circles=IRAC, squares=MIPS; downward-pointing arrows are upper
limits.   }
\label{fig:pelseds}
\end{figure*}

Figures~\ref{fig:3324clusters} and \ref{fig:iracclusters} and
Table~\ref{tab:clusterstatistics} show that, although it contains
objects of a wide range of colors, the Pelican cluster is almost
two-thirds Class II objects; in both  Figures~\ref{fig:3324clusters}
and \ref{fig:iracclusters}, there are regions of higher source
density.    Among the three clusters under consideration here, this
one is significantly more tightly clumped in color space or in class
distribution, with a much higher fraction of Class IIs and much lower
fraction of Class I or flat objects.  This is almost certainly related
to the relatively lower \av\ (and thus less smearing in the color
spaces) in the direction of this cluster, but also may be telling us
something about the relative degree of embeddedness and possibly age
of this cluster compared to the other two.  If this cluster is indeed
the oldest of the 3 clusters discussed here, it has had more time to
disperse the surrounding cloud (accounting for the lower \av) and for
the objects to move from their original natal location (accounting for
the more spatially dispersed, less tightly clustered locations).
Figure~\ref{fig:pelspace} shows the (projected) locations of the
cluster members.  Most of the Class I and flat objects are seen on the
East and West edges of the cluster, close to the \av\ peaks.
Figure~\ref{fig:pelseds} shows a sample of SEDs selected out of our
proposed cluster membership to represent the range of colors and SED
types found in this cluster.

\subsection{Pelican's Hat}
\label{sec:pelhat}

\begin{figure*}[tbp]
\epsscale{1.0}
\plotone{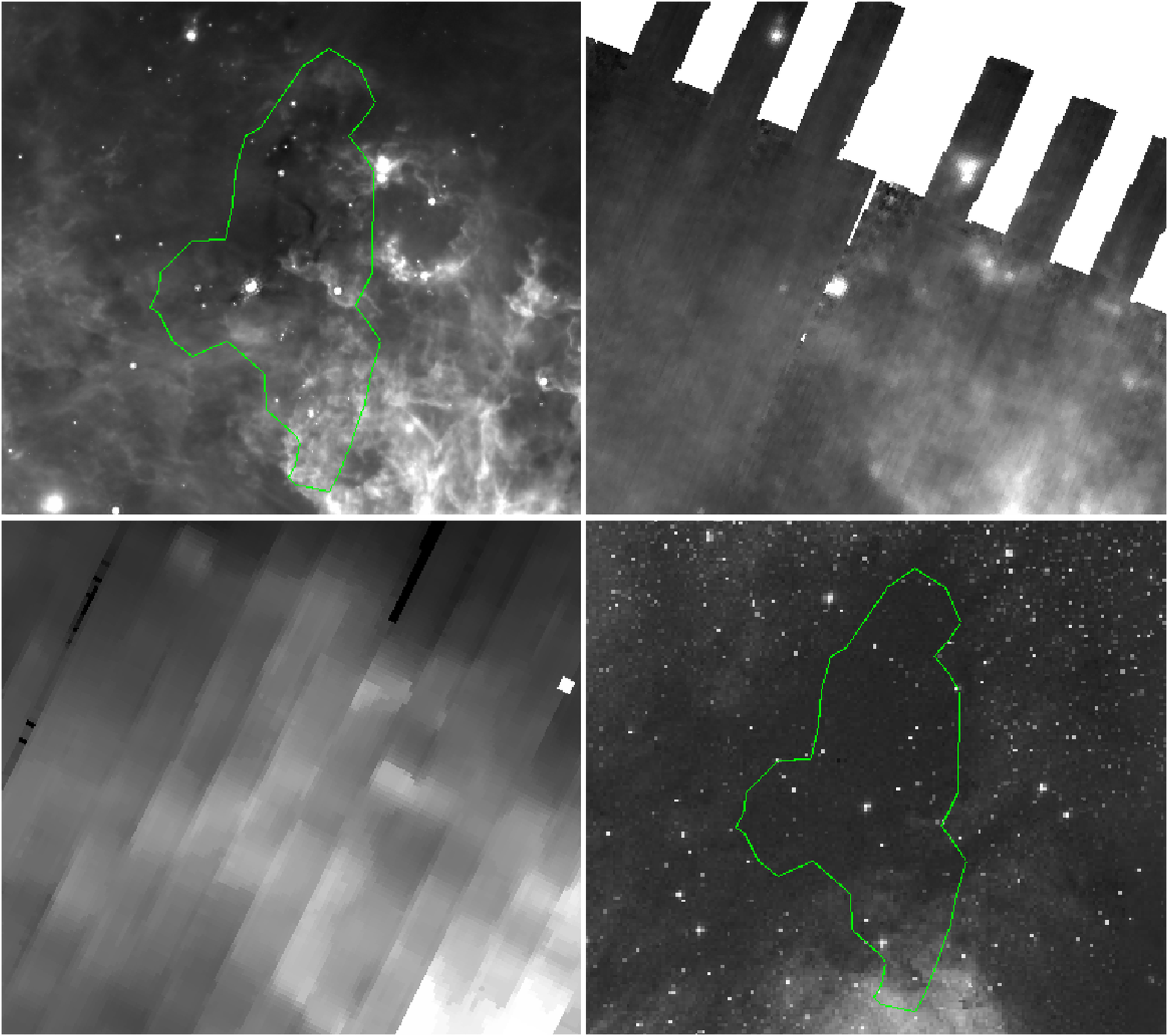}
\caption{Mosaic of the Pelican's Hat Cluster at 24 \mum\ (top left),
70 \mum\ (top right), 160 \mum\ (lower left), and, for comparison,
POSS-red (lower right).  The greyscale colors correspond to a
histogram-equalized stretch of surface brightnesses in each case. The
contour (plus YSO candidacy) defines cluster membership.  This contour is
essentially an \av=8 contour from the Cambr\'esy \etal\ (2002)
extinction map. }
\label{fig:pelhatimg}
\end{figure*}

\begin{figure*}[tbp]
\epsscale{1.0}
\plotone{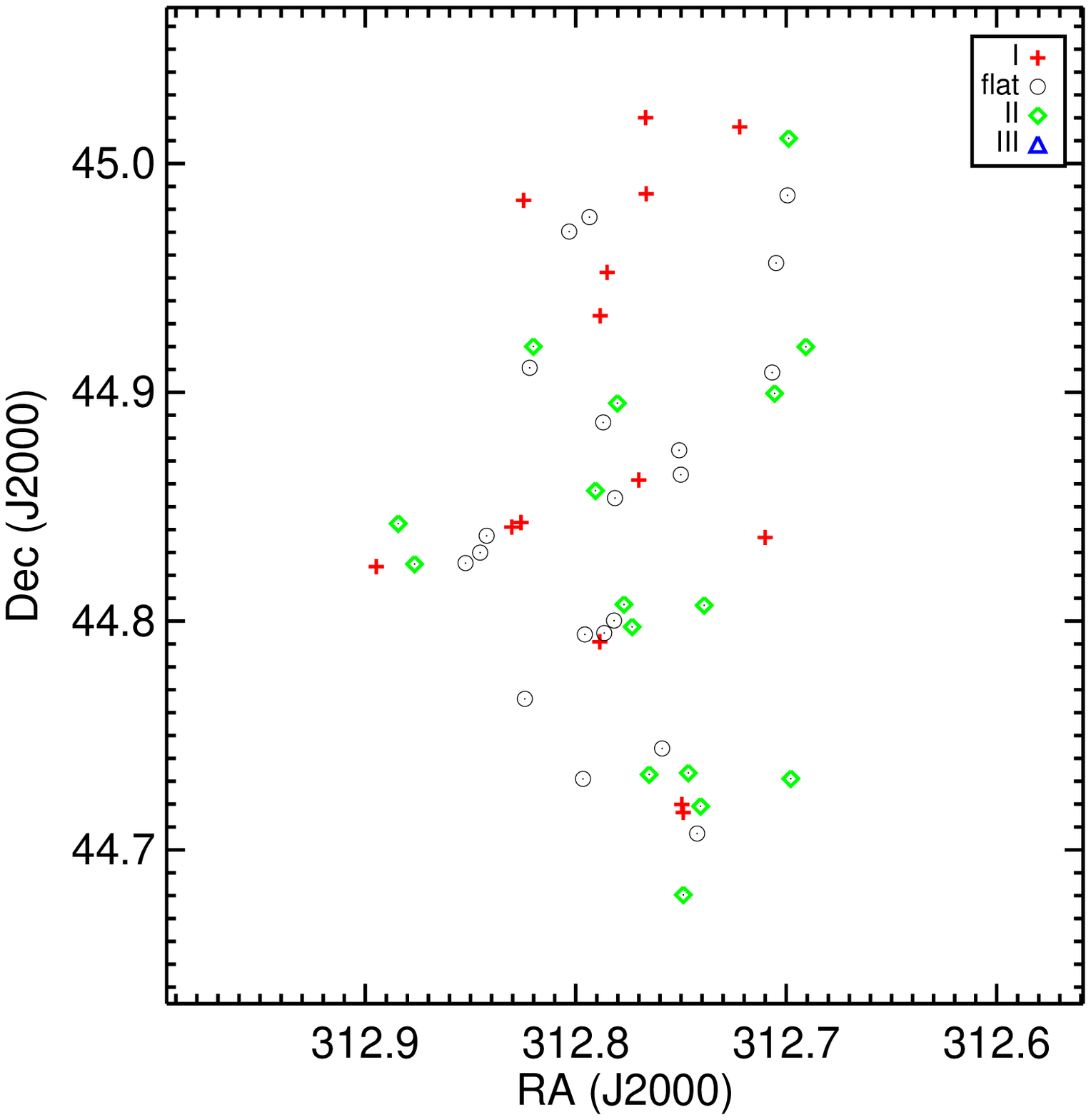}
\caption{Location of YSOs, color-coded by their SED slope, in the
Pelican's Hat cluster. Notation is as in similar earlier figures.  }
\label{fig:pelhatspace}
\end{figure*}

North of the Pelican's head, in roughly the same location as would be
found a hat on its head, is a previously unidentified dark lane at 24
\mum, similar to but much smaller than the Gulf of Mexico; see
Figure~\ref{fig:pelhatimg}. This sinuous lane is about 10$\arcmin$
long ($\sim$1.5 pc at a distance of 520 pc), with two bright 24 \mum\
sources at the south end, and several other nearby objects detected at
24 \mum.  The definition that emerges as part of the cluster formalism
above corresponds very closely to an \av$\sim$8 contour from the
Cambr\'esy \etal\ (2002) extinction map.  

None of the objects found in this region correspond to previously
known YSOs or to any other prior identifications. As seen in
Figure~\ref{fig:pelhatimg}, the region is on the edge of our 70 \mum\
map, and so viable data are obtained only for part of the cluster.
Texture in the nebulosity (including some nearby patches not formally
included in the cluster definition) can be seen. The 160 \mum\ image
has hints of the same structure.

Of the three clusters highlighted in this paper, this is the smallest
in projected area, number of sources, and surface density. There are
51 sources we define to be part of this region. 
Table~\ref{tab:clusterstatistics} and Figures~\ref{fig:3324clusters}
and \ref{fig:iracclusters} suggest that the cluster may be
intermediate in YSO candidate properties, between the Gulf and the
Pelican clusters.  The relative fraction of objects is greatest in the
``flat'' class, but within Poisson counting statistics, the relative
fractions seen in each of the Class I/flat/Class II/Class III
categories is indistinguishable from the Gulf of Mexico cluster.

\begin{figure*}[tbp]
\epsscale{0.8}
\plotone{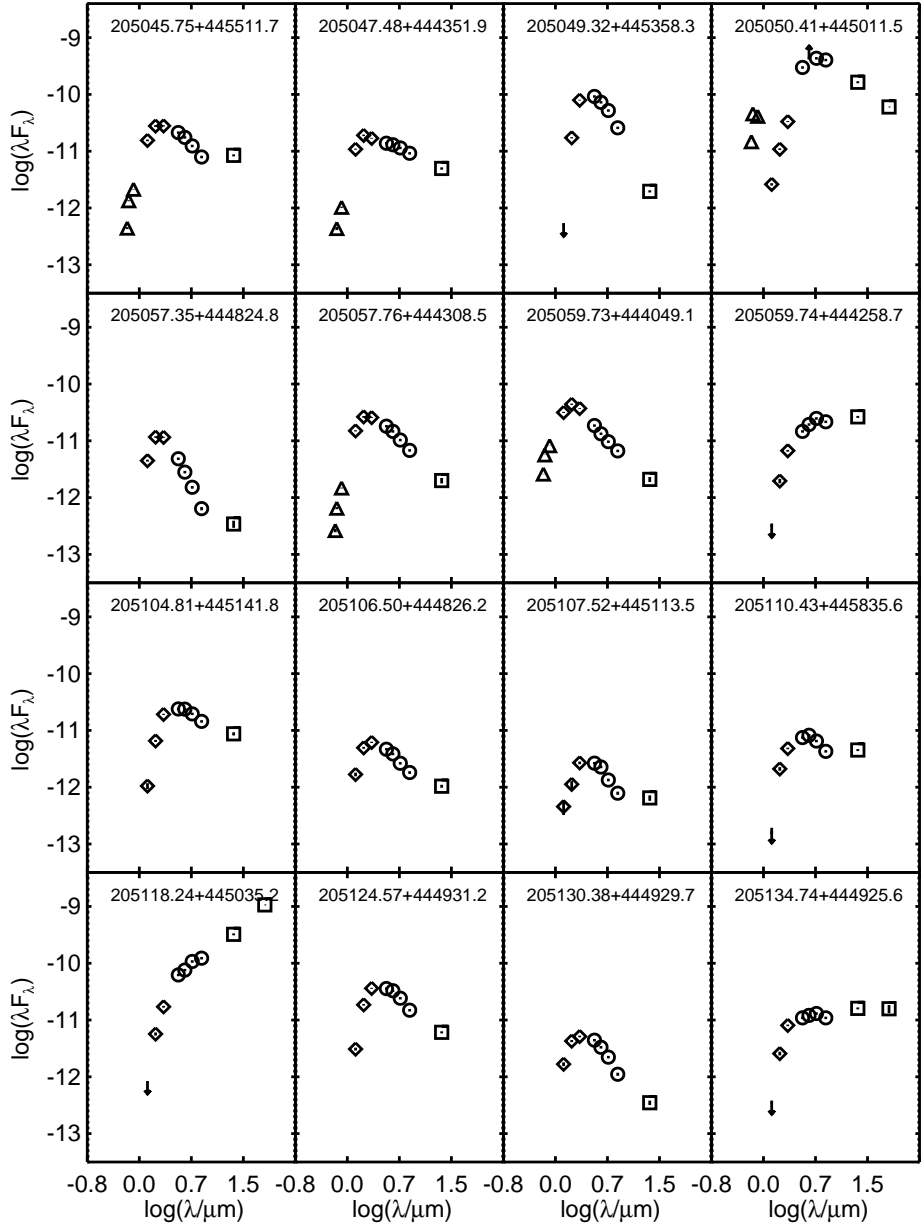}
\caption{
Sample of selected YSO candidate SEDs seen in the Pelican's Hat
cluster; $\lambda F_{\lambda}$ is in erg s$^{-1}$ cm$^{-2}$, and
$\lambda$ is in microns.  All of these new objects have IAU-compatible
names.   Vertical error bars are typically smaller than the points.
Symbols denote origin of photometry: $\times$={\em Vilnius} system,
triangles=IPHAS, $+$= Johnson (from the literature), diamonds=2MASS,
circles=IRAC, squares=MIPS; downward-pointing arrows are upper
limits.   }
\label{fig:pelhatseds}
\end{figure*}

Figure~\ref{fig:pelhatspace} shows the projected spatial distribution
of YSO candidates; note that there are no Class III objects (whereas
the other two clusters have a few), and that each category of object
seems to be smoothly distributed. This may be another indication of
youth, or this may mean that the actual cluster is larger than our
defined area. Figure~\ref{fig:pelhatseds} shows a sample of SEDs
selected out of our proposed cluster membership to represent the range
of colors and SED types found in this cluster.

There are $\sim$10 IRAC-selected but not MIPS-recovered YSO
candidates in this region, and they are almost all located on the
edges of our cluster definition.  It is difficult to assess if they
should be included in the cluster or not, and we have chosen not to
include them.

There are at least two objects nearby but not formally in the cluster
definition whose morphology in the 24 \mum\ image suggest youth; see
Figure~\ref{fig:pelhatimg}. About eight arcminutes to the northeast of
the north end of the dark lane (at 20:51:37, +45:05:15 or 312.9042,
+45.0875), there is a resolved patch of 24 \mum\ nebulosity, and
$\sim$5.5$\arcmin$ to the west of the north end of the dark lane (at
20:50:35 +44:57:16, or 312.6458, +44.95444), there is a resolved patch
of nebulosity that might be a bipolar nebula, and it is on the edge of
a larger circular feature $\sim$5.5$\arcmin$ in diameter. Since so
many of the objects in the NAN are young, we hesitate to add these
objects (and their apparently associated point sources) to the cluster
definition, but they are likely YSOs, possibly from the same local
episode of star formation that prompted the formation of objects in
the Pelican's Hat cluster.

\section{Ionizing source(s)}
\label{sec:ionizing}

Many previous authors have sought the ionizing source, or sources, for
the NAN; several articles have proposed various objects. Here we
summarize that literature and then consider the Spitzer properties of
several candidates. 

Herbig (1958) considers and rejects the following stars as the
exciting source of the nebula: 57 Cygni (too close, not enough
reddening), BD+44$^{\circ}$3627, BD+42$^{\circ}$3914, and MWC 1032 (HD
198931) (all projected onto cloud but behind heavy obscuration), HD
199579  (orientation of bright rims not correct, O6 V not bright
enough), $\alpha$ Cyg (Deneb; too far away), IV Cygni association
(orientation of bright rims not correct).  He concludes, agreeing with
Osterbrock (1957), that the source of excitation is likely behind the
Gulf of Mexico.  

The quest for the ionizing source of the complex continued in Neckel
\etal\ (1980). They dismiss the O6 star HD 199579 as having 2 or 3
times too faint flux to account for the observed emission.  They
suggest an object that now can be identified as 2MASS
J20535282+4424015. Their argument for this object being the exciting
source includes the fact that it is bright in the IR, that it is
18$\arcmin$ from the radio geometric center of the nebula, and
8$\arcmin$ in front of the strongest radio-bright rim.  They conclude
that if it is the exciting source, it must have a type earlier than
B1. 

Wendker \etal\ (1983) discuss the fact that HD 199579 must be a
background object; they assert that to match their long-wavelength
observations, there must be a group of (as of then unknown) eight O
stars powering the nebula.

Comer\'on \& Pasquali (2005) examined 19 bright preliminary candidates
and concluded that 2MASS J205551.25+435224.6, an O5 V (or earlier), is
likely to be the ionizing source of the nebula.  This object is also
\# 10 from Bally \& Scoville (1980), and is located near the geometric
center of the NAN based on radio observations.  However, they were
working on the assumption that the complex was at $\sim$610 pc, and
they conclude based on that distance that an O 5 star could be
responsible for the observed bulk properties of the NAN.  They observe
that there are few other comparably early candidates in the immediate
viscinity of this object, making this region distinctly different than
regions such as the Orion Nebula Cluster (ONC) where a cluster of one
late O star and several early B stars powers the nebula.  

Most recently, Strai\v{z}ys \& Laugalys (2008b) continue the search,
finding 13 stars resembling heavily reddened O stars in this region,
including the Comer\'on \& Pasqali (2005) source 2MASS
J205551.25+435224.6.   They confirm that it is an O5 V, and a
reasonable candidate for the exciting source of the nebula.  Of the
remaining 12 candidates, 4 of them could be O stars, contributing
to the ionizing flux in the region. The rest are dismissed as
background oxygen and carbon-rich AGB stars.

We can consider the SEDs of these objects based on the data we have
amassed. In our multi-wavelength database, we do not have optical data
for all of these objects, but for those that we do have, they are
highly reddened. Most are very bright at 2MASS and Spitzer bands, and
it could be that in reality all of the sources are highly reddened,
but the optical data are saturated in the available surveys so are
missing from our database. Several have at least one of their Spitzer
flux densities compromised due to saturation (or near-saturation). 
For the ones where enough of the Spitzer measurements are viable, only
one has any excess, MWC 1032 (HD 198931), and it is identified as a
YSO candidate in the lists above. Two of the objects from Strai\v{z}ys
\& Laugalys (2008b; their numbers 1 and 9) may have low-significance
excesses; they were dropped from our list of YSO candidates during the
vetting process above. 

While it is unsurprising that the SEDs of many if not all of these
candidate ionizing sources are highly reddened, it is interesting that
one or possibly three have an infrared excess. If the stars are indeed
young O or early B stars, it seems likely that they have already
ablated or otherwise dispersed their circumstellar material, so no
excess might be apparent.  If the stars are later type B stars, disks
could still be present. 

Looking at the larger scales depicted in Figure~\ref{fig:24mosaic}, 
the limb-brightened edges of the the MIPS-24 nebulosity appear to be
roughly symmetric about the Gulf of Mexico, suggesting that perhaps
the illuminating source really is behind the Gulf, and thus may be a
challenge to recover.  The Comer\'on \& Pasquali star (2MASS
J205551.25+435224.6) is located north of the Gulf of Mexico, very
roughly at the ``Bahamas,''  which is roughly consistent with this
illumination pattern.

\section{Conclusions}
\label{sec:concl}

We have presented Spitzer/MIPS observations of $\sim$7 square degrees
of the North American Nebula molecular cloud complex at 24, 70, and
160 \mum. There is large-scale and complex structure in the ISM
observed in all bands.  The 160 \mum\ emission follows the
near-IR-derived \av\ contours. About 4300 point sources are seen at 24
\mum, and $\sim$100 point sources are seen at 70 \mum.  We merged this
catalog to that derived from G09 for the Spitzer/IRAC data, as well as
extensive optical and near-IR data.

We have assembled a catalog of the previously-known objects in this
region, focusing on the $\sim$200 previously-identified YSOs and
candidates. (As supplemental material, we provide a list of these
objects not recovered by our search.)   We matched this catalog to
that derived from our Spitzer MIPS and IRAC data.  We used a series of
color criteria (plus other criteria, such as distance from the center
of the complex) to select objects as YSO candidates.  Using our
MIPS-focused criteria, we identified 1286 candidates, most of which
were independently identified using IRAC-focused criteria.  Some of
the IRAC-selected YSO candidates were not recovered using the MIPS
criteria because they were not detected at the less sensitive MIPS
bands. When the total number of unique YSO candidates as identified
from MIPS and IRAC are combined, we have identified 2076 YSO
candidates, about 10 times as many YSOs as were previously known here.
Most of our YSO candidates are Class II YSOs. 

Since the North American Nebula/Pelican Nebula region is in the
Galactic plane and is seen along a spiral arm, contamination rates
could be high. Our most likely lingering contaminant is AGB stars;
follow-up spectroscopy is needed to confirm the candidates as YSOs, or
identify them as contaminants. Additional observations such as X-ray
data and variability (monitoring) campaigns will also help in
eliminating contaminants.

A clustering analysis of the distribution of YSO candidates
suggests at least three clusters, named for their location in the
complex: the Gulf of Mexico, the Pelican, and the Pelican's Hat. 

The Gulf of Mexico cluster includes 283 MIPS-selected YSO candidates,
and surrounds a dramatic sinuous dust lane, which is dark even out to
70 \mum.  It is subject to the highest extinction (\av\ at least
$\sim$30) and has the widest range of infrared colors of the three
clusters, including the largest excesses and by far the most
point-source detections at 70 \mum. Just 3\% of the cluster members
were previously identified, and we have defined this cluster as about
10-100 times larger (in projected area) than was previously realized. 

The Pelican cluster is comparable in number of objects to the Gulf of
Mexico (247 MIPS-selected YSO candidates), but is subject to much less
reddening, and has a much higher fraction of less-embedded objects
(based on SED shape).  It includes a relatively high fraction (20\%)
of previously-identified objects, but the cluster boundaries are much
different than what was previously recognized.  

The Pelican's Hat cluster is an entirely new discovery, with no prior
identifications of YSOs. It surrounds a dark lane somewhat similar in
morphology to (though much smaller than) the Gulf of Mexico.  It
contains only 51 MIPS-selected YSO candidates, though it is close to
objects with interesting morphology in the 24 \mum\ image, bearing
further study.

We consider various ionizing source candidates, and conclude that some
of the candidates under consideration may have IR excesses.  The
illumination pattern is consistent with the ionizing star suggested by
Comer\'on \& Pasquali (2005), but it is also possible that some
ionizing source(s) are behind the Gulf of Mexico.

As additional material, we compared our source lists and images to
those available from IRAS, MSX, and AKARI, finding that the Spitzer
data resolves ISM and source confusion.  We also discuss some
candidate clusters within and projected onto the NAN which are
otherwise unremarkable in our data.  Some objects from the
superimposed cluster NGC 6997 may be debris disks.

\acknowledgements 

The authors wish to thank the following people: V.\ Strai\v{z}ys for a
helpful article referee report;  the c2d team for use of their SWIRE
reduction; V.\ Strai\v{z}ys and collaborators for providing
machine-readable tables (and coordinate checks) for several of their
publications prior to submission.

This research has made use of NASA's Astrophysics Data System (ADS)
Abstract Service, and of the SIMBAD database, operated at CDS,
Strasbourg, France.  This research has made use of data products from
the Two Micron All-Sky Survey (2MASS), which is a joint project of the
University of Massachusetts and the Infrared Processing and Analysis
Center, funded by the National Aeronautics and Space Administration
and the National Science Foundation.  These data were served by the
NASA/IPAC Infrared Science Archive, which is operated by the Jet
Propulsion Laboratory, California Institute of Technology, under
contract with the National Aeronautics and Space Administration.  This
research has made use of the Digitized Sky Surveys, which were
produced at the Space Telescope Science Institute under U.S.
Government grant NAG W-2166. The images of these surveys are based on
photographic data obtained using the Oschin Schmidt Telescope on
Palomar Mountain and the UK Schmidt Telescope. The plates were
processed into the present compressed digital form with the permission
of these institutions.
This paper makes use of data obtained as part of the INT Photometric H 
$\alpha$ Survey of the Northern Galactic Plane (IPHAS) carried
out at the Isaac Newton Telescope (INT). The INT is operated on the  
island of La Palma by the Isaac Newton Group in the Spanish  
Observatorio del Roque de los Muchachos of the Instituto de  
Astrofisica de Canarias. All IPHAS data, processed by the Cambridge  
Astronomical Survey Unit,
were obtained via the database and image access provided by Astrogrid.

The research described in this paper was partially carried out at the
Jet Propulsion Laboratory, California Institute of Technology, under
contract with the National Aeronautics and Space Administration.

\appendix

\clearpage

\section{Comparison to IRAS, MSX, and AKARI}
\label{sec:irasmsx}

Because Spitzer goes deeper and has higher resolution than prior
infrared missions such as IRAS and MSX, and the current mission AKARI,
by comparing the data, we can learn how well these other missions did
in identifying point sources in a region such as this with complex
nebulosity and high source densities, both leading to confusion.  

\subsection{IRAS}

\begin{deluxetable}{ll}
\tablecaption{Comparison to IRAS 25 \mum}
\label{tab:iras}
\tablewidth{0pt}
\tablehead{
\colhead{IRAS 25 \mum\ source} & \colhead{notes}  }
\startdata
20443+4441 & point source at 24 \mum\            \\
20445+4436 & point source at 24 \mum\            \\
20447+4441 & point source at 24 \mum\ (cluster?)   \\
20453+4441 & point source at 24 \mum\            \\
20458+4416 & point source at 24 \mum\            \\
20467+4340 & point source at 24 \mum\            \\
20468+4410 & nebulosity             \\
20469+4352 & nebulosity             \\
20469+4435 & point source at 24 \mum\            \\
20472+4338 & point source at 24 \mum\            \\
20472+4411 & nebulosity             \\
20474+4416 & nebulous knot         \\
20475+4431 & nebulosity/maybe point source at 24 \mum\   \\
20476+4422 & point source at 24 \mum\            \\
20479+4419 & nebulosity             \\
20479+4438 & nebulosity             \\
20481+4404 & point source at 24 \mum\            \\
20485+4407 & nebulosity             \\
20485+4423 & nebulosity             \\
20489+4410 & nebulosity/cluster of point sources at 24 \mum?     \\
20490+4351 & point source at 24 \mum\            \\
20490+4417 & bright nebulosity       \\
20495+4320 & nebulosity             \\
20496+4354 & point source at 24 \mum\            \\
20499+4337 & 2 point sources at 24 \mum\         \\
20511+4225 & nebulous point source at 24 \mum\         \\
20516+4304 & point source at 24 \mum\            \\
20524+4227 & point source at 24 \mum\            \\
\enddata
\end{deluxetable}

IRAS (the Infrared Astronomical Satellite) flew in 1983 and observed
at 12, 25, 60, and 100 \mum. The angular resolution of IRAS ranged
from $\sim$0.5$\arcmin$ at 12 \mum\ to $\sim$2$\arcmin$ at 100 \mum;
source confusion is clearly a problem in regions like the NAN, where
the source density varies from $\sim$20 per square arcmin at IRAC
bands to $\sim$0.2 per square arcmin at 24 \mum, and the nebular
emission is complex at essentially all bands. 

Previous results in Perseus from Rebull \etal\ (2007) comparing MIPS
and IRAS data suggest that the IRAS point source catalog (PSC;
Beichman \etal\ 1988) and faint source catalog (FSC; Moshir \etal\
1992) did reasonably well in identifying true point sources; while the
extended emission is complicated in Perseus (as it is in the NAN), the
source density is not as high as in the NAN. 

IRAS surveyed more than 96\% of the sky, but one of those missing sky
segments is in the NAN region.  There is a missing ``wedge'' of IRAS
imaging data along the Eastern edge of our map.  Apparently as a
result, the point source extraction routines were not run over much of
the NAN map.  There are no FSC sources found at all within 4 degrees
radius of our map center.  There are only PSC sources found in a
narrow strip in the Western edge of our map, from ecliptic longitudes
$\sim$336$\arcdeg$ to 337$\arcdeg$.

There are 69 PSC objects that overlap our MIPS map, 28 of which are
listed as solid detections (quality flag =3) at 25 \mum, and they
appear in Table~\ref{tab:iras}.  Thirteen of these resolve into
nebulosity when viewed at 24 \mum; the rest resolve into at least one
point source.  At 60 \mum, there are 28 solid detections (quality flag
=3), just three of which are detected as point sources in our 70 \mum\
map (IRAS 20472+4322, 20467+4340, and 20472+4338); the rest are
confused by  nebulosity.  The nebulosity is complex, and the overall
sky brightness is high, in this region.

While MIPS has much more sensitive detectors than IRAS, it also
samples a much smaller beamsize, so the achieved surface brightness
sensitivity is not tremendously improved over IRAS.  Most of the
large-scale structure seen in the MIPS maps can also be seen in the
IRAS maps.  Obviously the tremendous texture found in the
MIPS maps is lost in the IRAS maps.

\subsection{MSX}

The NAN region is included in the Midcourse Space Experiment (MSX)
survey of the Galactic plane (Price \etal\ 2001).  A U.S. Air Force
satellite, MSX flew in 1996-1997, and observed at 8.28, 12.13, 14.65,
and 21.3 \mum\ at a resolution of $\sim$18$\arcsec$.  As for the IRAS
maps, hints of the structure found in the ISM can be seen in the
lower-spatial-resolution MSX images.  The dark lane in the Gulf of
Mexico can clearly be seen in the 8.28 \mum\ image (and hints of it
can even be seen the 12.13 \mum\ image).  The brightest emission
regions discussed in \S\ref{sec:extemiss} can be seen in the MSX
images as well: the Pelican's neck, the Mexican Riviera, and the
bright resolved blobs.

About 900 sources are retrieved from the MSX point source catalog
(Egan \etal\ 1999) for this region.  Simply by overlaying them on the
Spitzer images, it can be seen that many are clearly confused by the
ridges and structure in the nebulosity, as well as point source
confusion itself, though many do correspond to real point sources seen
in the 24 \mum\ data.  It is somewhat difficult to compare the flux
densities quantitatively, as the Spitzer instruments saturate at about
the point where the MSX channels are operating at the lower limit of
their completeness.  There are $\sim$20 sources in common between MSX
Band A (8.28 \mum) and IRAC 8 \mum, most of which are reasonably
well-matched in measured flux densities (1$\sigma$ scatter for the
brighter sources is 17\%); about 20\% of the IRAC observations clearly
resolve source confusion in the MSX data.   There are $\sim$40 sources
in common between MSX Band E (21.34 \mum) and MIPS 24 \mum, the
brighter half of which are comparable in measured flux densities
(1$\sigma$ scatter for the brighter sources is 37\%); about 40\%
clearly resolve source confusion in the MSX data, most of which are
below the 2 Jy completeness limit for the MSX Band E data. 

\subsection{AKARI}

AKARI is a Japanese mission, and it surveyed the whole sky at 9, 18,
65, 90, 140, and 160 \mum\ during its cryogenic phase (2006-2007).
With a primary mirror of 68.5 cm, AKARI's resolution is slightly worse
than Spitzer's; the spatial resolution is $\sim$5$\arcsec$ at the
shortest bands and $\sim$90$\arcsec$ at the longest.  As of this
writing, images from this survey are not available, but catalogs are. 
We extracted $\sim$1100 sources in this region from the IRC 9 and 18
\mum\ all-sky point source catalog (Ishihara \etal\ 2010) and
$\sim$2100 sources in this region from the FIS 65, 90, 150, and 160
\mum\ all-sky bright source catalog (Yamamura \etal\ 2009).  Not all
of those sources are indicated to be high-quality detections at each
band; 96\% of the IRC catalog is a high-quality detection at 9 \mum,
whereas just 30\% are similar quality detections at 18 \mum. For the FIS
catalog, 13, 92, 31, and 15\% are high-quality detections at 65, 90,
140, and 160 \mum, respectively.

By empirical inspection of the Spitzer images, the IRC catalog is not
often confused by nebulosity; the overwhelming majority of IRC point
sources are point sources for Spitzer as well. Unsurprisingly, the
Spitzer data are substantially deeper; the AKARI catalogs only go to
$\sim$0.1 Jy at 9 \mum.  There are $\sim$640 objects in common between
the catalogs, and the flux densities are reasonably well-matched
between the AKARI 9 \mum\ and the Spitzer 8 \mum\ channels (1$\sigma$
scatter is 27\%), keeping in mind not only filter variation but
intrinsic stellar variability. 

In contrast, the FIS catalog seems more often to not recover point
sources seen in the Spitzer images. Without the AKARI images to
compare, it is difficult to understand many of the FIS catalog
detections in this confused and complex region. Only about 20 ($<$1\%)
of the FIS catalog detections correspond by eye to 70 \mum\ Spitzer
point or point-like sources. A significant fraction of the sources are
clearly confused by structure in the nebulosity (e.g., there are
``chains'' of sources along emission ridges).

\section{Clusters not part of the NAN}
\label{sec:appendixclusters}

There are some clusters identified in the literature that are likely
foreground objects, or are not recovered in the Spitzer data.  We
discuss these here.

\subsection{NGC 6997}
\label{sec:ngc6997}

\begin{figure*}[tbp]
\epsscale{1}
\plotone{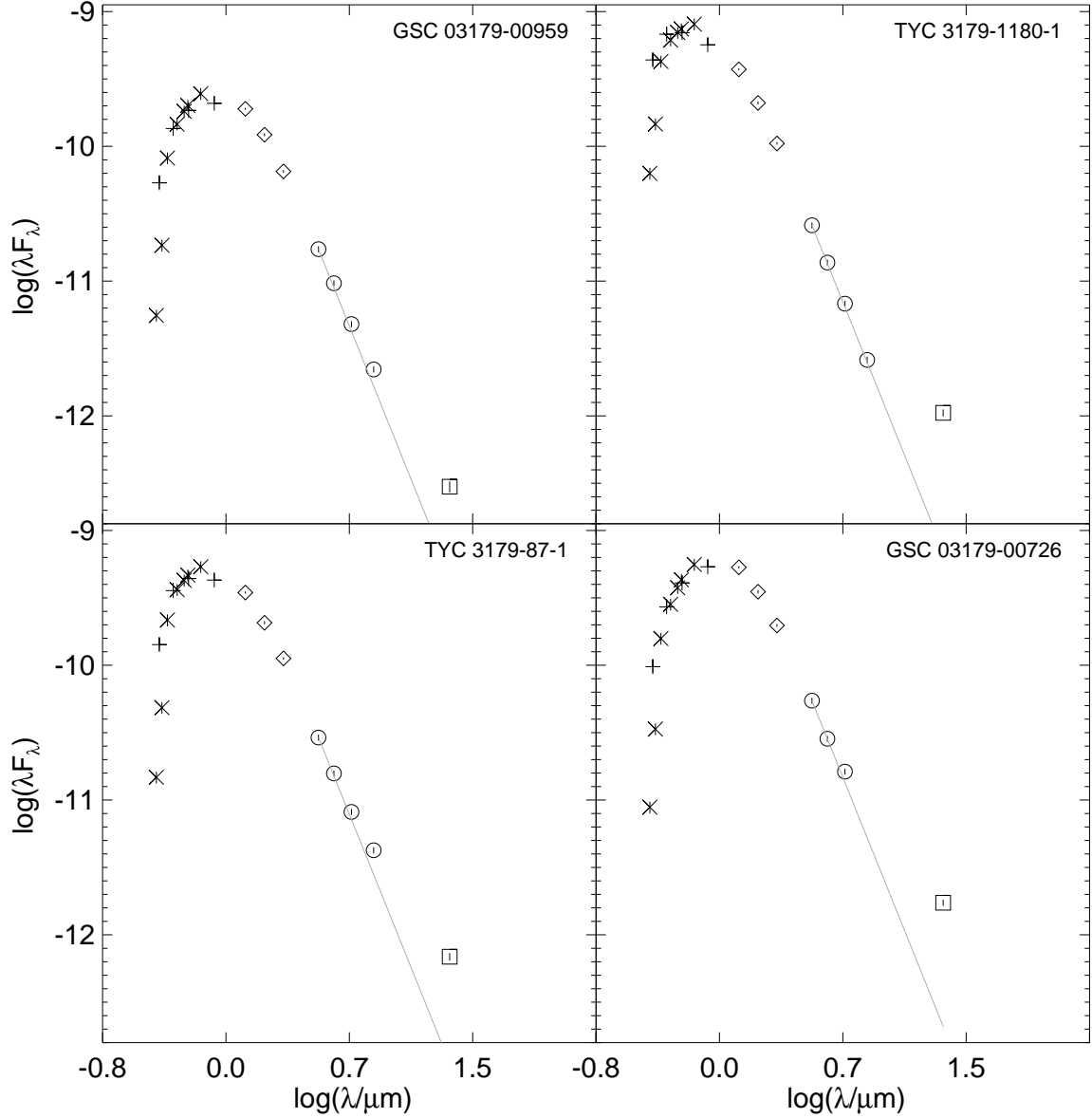}
\caption{SEDs for the four high-confidence YSO candidates which have
also been identified as NGC 6997 members; $\lambda F_{\lambda}$ is in
erg s$^{-1}$ cm$^{-2}$, and $\lambda$ is in microns.  Notation is as
in prior SED figures; the grey line from 3.6 \mum\ traces the flux
expected from a photosphere.  If these objects are really NGC 6997
members, and if the cluster is $\sim$350 Myr old, then these are
candidate debris disks, though the apparent IR excess could be due to
source confusion (including with a low-mass companion).}
\label{fig:n6997seds}
\end{figure*}

NGC 6997 has been identified historically as being a cluster
superimposed either in front of or behind the NAN complex.  Note that
this cluster has been commonly confused with NGC 6996; see Laugalys
\etal\ (2006b) and Corwin (2004).  NGC 6997 has been recently studied
by three papers in the literature.  Zdanavi\v{c}ius \& Strai\v{z}ys
(1990) identified 15 cluster members, placing the distance at 620 pc
(whereas the NAN is taken in that work to be at 550 pc) and the age at
$\sim10^8$ years.  Villanova \etal\ (2004) doubles the cluster spatial
size, finding $\sim$50 new probable or likely members, and placing the
cluster at $\sim$760 pc and $\sim$350 Myr, in front of their distance
for the NAN complex (which they take to be 1 kpc).  Laugalys \etal\
(2006b) finds $\sim$50 more members, places the cluster at 780-790 pc
(now again behind the NAN), and estimates the age at $3\times10^8$
years.  It is identified in Figure~\ref{fig:whereyso} above. 

For our discussion here, we do not attempt to redefine the cluster
boundaries; we simply accept as legitimate members the 112 objects
identified in these papers (and detected in our Spitzer survey) as
cluster members, and it does not matter if the cluster is in front of
or behind the NAN. On the whole, the stellar population of NGC 6997
indeed looks different than the rest of the cloud.  None of the stars
currently identified as NGC 6997 members have IRAC excesses,
consistent with the cluster being older than (and probably unrelated
to) the NAN complex.  Twelve of the 112 objects are seen at 24 \mum,
and none at 70 \mum.  Out of these 12 objects, four are identified as
high-confidence YSO candidates. The SEDs for the four high-confidence
YSO candidates appear in Figure~\ref{fig:n6997seds}; they are GSC
03179-00959 (205622.0+443537.2), TYC 3179-1180-1 (205624.6+443921.3),
TYC 3179-87-1 (205633.1+443810.3), and GSC 03179-00726
(205639.5+443821.6).  There are two more lower-confidence YSO
candidates (GSC 03179-01093=205633.3+443622.3 and GSC
03179-00980=205637.1+444009.2).  These latter two objects have much
lower-significance excesses at 24 \mum. All of these YSO candidates
appear in the complete tables of YSO candidates above.

If these stars are truly members of NGC 6997, then these observed
excesses could be only apparent excesses, due to source confusion or
an unresolved lower-mass companion, or they could be real excesses. 
If they are real, then they are likely debris disks.  The presence of
such debris disks is broadly consistent with the age of the cluster
being $\sim$350 Myr. Cieza \etal\ (2008) find, for the Hyades
($\sim$650 Myr), that none of the FGK stars have debris disks, but
that two of eleven A stars have debris disks. For these four
high-confidence YSO candidates in NGC 6997, the spectral types from
the literature are G7 V, B5 V, A3 V, and A7 III:, respectively; again,
this is broadly consistent with the literature.  A more complete study
of the membership of this cluster would allow a more complete estimate
of the disk fraction of this cluster.

We conclude that this cluster (as defined in the literature) is
sufficiently different in mid-infrared properties from the NAN that it
is unlikely to be related to the NAN, or at least much older than the
rest of the complex.

\subsection{Collinder 428}

Collinder 428 has been identified historically as being a cluster on
the far East side of the complex that is likely unrelated to the NAN
complex. Laugalys \etal\ (2007) concluded that the dust cloud in the
direction of Collinder 428 is at the same distance as the NAN complex
and that the star grouping is not a real cluster but rather a
relatively transparent window in the dust cloud, and therefore not a
separate object.  

The stars in this region are on the edge of our map, and none of them
are identified as YSO candidates using either IRAC or MIPS.  Such a
low disk fraction implies that, if Collinder 428 is a real cluster, it
is not part of the NAN complex.  This is consistent with the Laugalys
\etal\ (2007) interpretation that the stellar grouping is illusory due
to a window in the dust cloud, and not a separate entity.

\subsection{Remaining Cambr\'esy \etal\ (2002) clusters}

Cambr\'esy \etal\ (2002) identified 8 clusters based on 2MASS data. 
Clusters 1, 2, and 3a/b are part of the Gulf of Mexico, and cluster 5
is part of the Pelican cluster, as described above.  We now discuss
the remaining clusters. We do not define for further consideration any
additional clusters associated with any of these objects, for reasons
that will become apparent.  

Cluster 4 appears in the 24 \mum\ image as wispy faint ($\sim$40
MJy/sr) nebulosity, with the nearest very faint point source
2$\arcmin$ away (0.3 pc at 520 pc), and the nearest point source of a
brightness range comparable to the clusters above nearly
$\sim$7.5$\arcmin$ away (1.3 pc). 

Cluster 6 appears in the 24 \mum\ image as the heart of a roughly
arc-shaped patch of bright ($\lesssim$650 MJy/sr) feathery nebulosity
roughly  $11\arcmin \times 3\arcmin$ (1.7$\times$0.5 pc at 520 pc).  No
point sources can be discerned on top of the nebulosity, though some
faint point sources are seen close to the nebulosity.  

Cluster 7 appears in the 24 \mum\ image as a circular patch of bright
(250-650 MJy/sr) nebulosity $\sim 4 \arcmin$ (0.6 pc at 520 pc) in
diameter. No point sources can be detected on top of this nebulosity. 

Cluster 8, seen in 2MASS images as a tight, red clumping, appears as a
single very bright point source at 24 \mum, with no particular higher
density of surrounding point sources at 24 \mum.

\section{YSO candidates not recovered by our MIPS-based search}
\label{sec:iraconly}

There are two categories of objects that our primarily MIPS-based
search does not recover: objects seen to have an IRAC excess but not
detected at MIPS bands (likely to be primarily the lower mass members
of the cluster), and objects previously identified as YSO candidates
that do not have IR excesses (these were likely to have been
identified as YSOs using other methods than the search for IR
excesses).  Here, for completeness, we present flux density tables for
these objects. 

\subsection{IRAC-only YSOs}

In Table~\ref{tab:missingiracysos}, we list the Spitzer fluxes (and
cross-identifications) for those objects identified using the G09
methodology but not rediscovered using the MIPS-based selection
mechanisms here.

\subsection{Unrecovered YSO candidates}

In Table~\ref{tab:missingysos}, we list the Spitzer fluxes (and
cross-identifications) for those objects identified by the literature
as YSO candidates but not rediscovered using either the MIPS-based
selection mechanism that was the primary goal of this paper, or the
IRAC-based selection mechanism from G09 and reported in the prior
table.  These objects may or may not be legitimate NAN members; we
cannot determine this from our data set.

Note that 6 objects were identified in Witham \etal\ (2008) as bright
in H$\alpha$ in this region, but were not identified by us here as
having an IR excess. All six of these objects are undetected at MIPS
bands and on the edges of the IRAC maps such that they are only
detected at 2 of the 4 IRAC bands.  As such, they would not have been
identified by our YSO identification approaches. They are objects
204654.78+434806.6, 204703.24+434930.0, 204704.82+434911.2,
205934.89+452355.9, 205938.69+452414.5, and 210029.50+443655.4.

\include{monsterheaders_app}

\end{document}